\documentclass[11pt,a4paper]{article}
\usepackage{mathrsfs}
\usepackage{amsfonts}
\usepackage{graphicx,indentfirst}
\usepackage{subfigure}
\usepackage{amsmath,amssymb,amsthm}
\usepackage{natbib}
\usepackage{epstopdf}
\usepackage{fullpage}

 \numberwithin{equation}{section}
\allowdisplaybreaks[4] \theoremstyle{plain}
\newtheorem{thm}{Theorem}[section]
\newtheorem{lemma}{Lemma}[section]
\newtheorem{proposition}{Proposition}[section]
\newtheorem{corollary}{Corollary}[section]

\newtheorem{example}{Example}[section]
\theoremstyle{definition}
\newtheorem{definition}{Definition}[section]

\theoremstyle{remark}
\newtheorem{remark}{Remark}[section]

\makeatletter

\newcommand{\Rmnum}[1]{\expandafter\@slowromancap\romannumeral #1@}
\makeatother

\begin{document}
\begin{titlepage}
\title{\textbf{Time-Changed Ornstein-Uhlenbeck Processes And Their Applications In Commodity Derivative Models}\footnote{The authors thank Alexander Eydeland and the members of the Commodity Strategies Group at Morgan Stanley for providing data on implied volatilities in commodity futures options used in this paper and for their helpful comments.
This research was supported by the National Science
Foundation under grant DMS-0802720. }}
\author{
\textbf{Lingfei Li}\thanks{ Department of Industrial Engineering and
Management Sciences, McCormick School of Engineering and Applied
Sciences, Northwestern University, 2145 Sheridan Road, Evanston, IL
60208, E-mail: lingfeili2012@u.northwestern.edu.{}} \and
\textbf{Vadim Linetsky}\thanks{ Department of Industrial Engineering
and Management Sciences, McCormick School of Engineering and Applied
Sciences, Northwestern University, 2145 Sheridan Road, Evanston, IL
60208, Phone: (847) 491-2084, E-mail:
linetsky@iems.northwestern.edu.{}} }\date{August 27, 2011}
\maketitle

\begin{abstract}
This paper studies {\em subordinate Ornstein-Uhlenbeck (OU)
processes}, i.e., OU diffusions time changed by L\'{e}vy
subordinators. We construct their sample path decomposition,
show that they possess mean-reverting jumps, study their
equivalent measure transformations, and the spectral representation
of their transition semigroups in terms of Hermite expansions. As an application, we propose a new
class of commodity models with mean-reverting jumps based on
subordinate OU process. Further time changing by the integral of a
CIR process plus a deterministic function of time, we induce
stochastic volatility and time inhomogeneity, such as seasonality,
in the models. We obtain analytical solutions for commodity futures
options in terms of Hermite expansions. The models are consistent
with the initial futures curve, exhibit Samuelson's maturity effect,
and are flexible enough to capture a variety of implied
volatility smile patterns observed in commodities futures options.
\end{abstract}


\end{titlepage}

\section{Introduction}

The contribution of this paper is two-fold. The first part studies {\em subordinate Ornstein-Uhlenbeck (SubOU)
processes}. A SubOU process can be constructed by time changing an
OU diffusion by a L\'{e}vy subordinator. SubOU processes are Markov semimartingales
with mean-reverting jumps.
SubOU transition semigroups possess spectral representations in terms of Hermite expansions.
As an application, the second part of the paper develops a new class
of analytically tractable commodity models with mean-reverting jumps by modeling the
commodity spot price as the (scaled and compensated) exponential of a
SubOU process. To model stochastic volatility
and time inhomogeneity, such as seasonality, we further time change
SubOU processes by the integral of the sum of an independent CIR diffusion
and a deterministic function of time. The resulting models
have the following features: (1) mean-reverting jumps, (2) stochastic volatility, (3) time
inhomogeneity, (4) analytical
solutions for futures options in terms of Hermite
expansions, (5) consistency with the initial futures curve,
(6) Samuelson's maturity effect, and (6) flexibility to
capture a variety of implied volatility smile patterns observed
in commodity futures options.

The mathematical part of the paper contains a self-contained
presentation of SubOU processes. Section 2.1 defines SubOU
semigroups as Bochner's subordinates of OU semigroups and gives
explicit expressions for their infinitesimal generators based on the
application of R.S. Phillips' theorem. This material is classical
(see Schilling et al. (2010) for an excellent recent survey of
Bochner's subordination and Albeverio and Rudiger (2003), (2005) for
the treatment of SubOU semigroups in particular). Section 2.2
defines a class of SubOU Markov semimartingales, gives their local
characteristics, proves uniqueness of the associated martingale
problem, and proves the mean reversion property of their jumps.
While the material in this section follows from the general
semimartingale theory (our presentation follows Jacod and Shiryaev
(2003)), it has not been presented in the literature in this form.
Section 2.3 presents results on equivalent measure transformations
for SubOU processes. In particular, a class of locally equivalent
measure changes that transform one SubOU process into another SubOU
process is characterized, along with a detailed treatment of some
special cases important in applications. This section presents
original results that, to the best of our knowledge, have not
previously appeared in the literature. It serves as the basis for
financial applications, characterizing equivalent martingale
measures (EMMs) for this class of models.  Section 2.4 presents the
spectral decomposition of the SubOU semigroup in $L^2({\mathbb
R},{\mathfrak m})$, where  ${\mathfrak m}$ is the Gaussian measure,
in terms of Hermite expansions. The $L^2$ spectral theory of SubOU
semigroups has been previously given by Albeverio and Rudiger
(2003), (2005). We supplement it with pointwise convergence results
and truncation error bounds for the expansion that are important for
options pricing.

The second part of the paper provides the development of our commodity futures model. Section 3.1 defines the model for the commodity spot price as the exponential of a SubOU process scaled and compensated so that, under ${\mathbb Q}$, the mean spot price evolves along the fixed initial futures curve. We then explicitly solve for the futures dynamics under ${\mathbb Q}$ in the form of a martingale expansion with basis martingales associated with Hermite polynomials. Section 3.2 demonstrates Samuelson's maturity effect in commodity futures in this class of models. Section 3.3 derives explicit analytical solutions for futures options in terms of Hermite expansions. In section 4 we further time change SubOU
processes to induce stochastic volatility and time inhomogeneity and
study the resulting commodity futures models. In particular, we derive the futures price process, demonstrate Samuelson's maturity effect, and obtain solutions for futures options.
In section 5 we discuss efficient model
implementation based on recursions for Hermite polynomials
and present model calibration examples to futures options on a variety of commodities, including metals, energies and agriculturals.
Appendix A contains a number of results on the CIR
process needed in the development of models with stochastic
volatility. Proofs are collected in Appendix B.

In the rest of this introduction we discuss relationships of models developed in this paper to the literature. We start with a brief survey of the commodity derivatives modeling literature.
Mean reversion and jumps are two of the salient features of
commodities prices (see monographs \cite{EydelandWolyniec},
\cite{Geman}, and \cite{Geman08} for introduction to commodity and
energy derivatives markets and modeling). Mean reversion in
commodities markets is well documented in numerous empirical studies
in the literature (e.g., \cite{BSCC}, \cite{Pindyck},
\cite{CasassusDufresne}). To capture the mean reversion property,
the classical commodity models are based on OU diffusions. The
simplest such model is the exponential OU model of
\cite{Schwartz97}. In this model the commodity spot price is assumed
to follow the exponential of an OU process with constant long-run
mean level, rate of mean reversion, and volatility. While the OU
process itself lives on the whole real line, taking the exponential
leads to the positive process for the commodity spot price. The
geometric OU model plays the same role in commodity markets that the
geometric Brownian motion model plays in the equity markets, serving as the
simplest analytically tractable commodity derivatives pricing model.
Being inherently the spot price model, the futures curve is derived
endogenously in this model and, hence, does not generally match the
futures curve observed in the market. This situation is similar to
the \cite{Vasicek} model of the short interest rate, where the yield
curve is derived endogenously in the model and does not generally
match the market yield curve. Similar to how the Vasicek model is
extended to match an arbitrary market yield curve by making the
long-run mean level of the short rate time-dependent (e.g.,
\cite{HullWhite}), the exponential OU model can be extended
to match an arbitrary market-observed futures curve (e.g.,
\cite{ClelowStrickland}). In this model futures prices of all
maturities follow continuous martingales under ${\mathbb Q}$.

Along with mean reversion, discontinuous price movements (jumps) are
another salient feature of commodity and energy markets. While jumps
are a ubiquitous feature of all asset prices and financial
variables, from equities to foreign exchange to interest rates,
commodity and energy prices exhibit particularly large and frequent
jumps, perhaps more so than other asset classes (see, e.g.,
\cite{HilliardReis99}, \cite{Deng}, \cite{GemanRoncoroni} for
empirical evidence of jumps in commodity and energy prices). The
question then arises as to how to extend commodity models based on
mean-reverting OU diffusions to jumps. The first line of attack is
to add a jump component to the diffusive mean-reverting component to
form a jump-diffusion process similar to \cite{Merton} classical
jump-diffusion model widely used in equities. A variety of
jump-diffusion models along these lines have been introduced in
commodity markets (e.g., \cite{HilliardReis}, \cite{HilliardReis99},
\cite{Deng}, \cite{Yan}, \cite{2Benth}, \cite{GemanRoncoroni},
\cite{Andersen} and \cite{Crosby}). Virtually all of the
jump-diffusion models in the literature, with the exception of
\cite{GemanRoncoroni} and \cite{Andersen}, add state-independent
jumps to the mean-reverting diffusion. The resulting models exhibit
mean reversion  due to the OU drift, but do not have mean reversion
in their jump measure that remains state-independent. That is, upon
arrival, the direction  of the jump and the probability distribution
of its amplitude are independent of the current state of the
process. The drift acting upon the process between the jumps is
forced to account for all of the mean reversion in these models. A
model with mean reverting jumps would, in contrast, feature
state-dependent mean reverting jumps with the jump direction and the
jump amplitude dependent on the current state of the process.

In contrast to jump-diffusion models with state-independent jumps,
\cite{GemanRoncoroni} propose a jump-diffusion model with Poisson
jumps independent of the diffusion component but with jump direction
dependent on the pre-jump state of the process. They show that such
models capture some of the empirical properties of electricity price
data. However, analytical solutions for futures options have not
been obtained in their model. \cite{Andersen} considers
jump-diffusion processes with jumps driven by a continuous time
Markov chain whose states are interpreted as different market
regimes. Jumps in this framework are dependent on the regimes and,
hence, are state dependent. However, option pricing in this
regime-switching framework is generally highly non-trivial unless
some simplifying assumptions are made.

In this paper we take an alternative approach to the previous
literature on commodity and energy models with jumps. Instead of
adding state-independent jumps to the mean-reverting diffusion
process, we time change the mean-reverting OU diffusion with a
L\'{e}vy subordinator to yield a pure jump or a jump-diffusion
process (depending on whether or not the subordinator has a positive
drift) with state-dependent and mean reverting jumps. As such, our
models can be viewed as a commodity markets counterpart of the
time-changed L\'{e}vy process-based models in equity markets by
\cite{MadanCarrChang}, \cite{BarndorffNielsen},
\cite{GemanMadanYor}, \cite{CarrGemanMadanYorSVLP}, and
\cite{CarrWu}. However, since mean reversion is the crucial feature
of commodity markets, instead of time changing Brownian motion as in
those references, we time change OU diffusions and obtain pure jump
or jump-diffusion Markov semimartingales with state-dependent mean
reverting jumps. Similar to L\'{e}vy-based models in equity markets,
our models based on SubOU processes calibrate well to a variety of
implied volatility smiles in commodity markets when the maturity is
fixed.

To induce stochastic volatility (the need for stochastic volatility
in energy markets has been advocated by \cite{EydelandGeman}), we
further time change these jump processes with the integral of an
activity rate (stochastic volatility) that follows a CIR process.
This is similar to the approach of Carr et al. (2003) and Carr and
Wu (2004), but in contrast to those references we time change Markov
jump processes that are generally not L\'{e}vy processes. This
yields pure jump or jump-diffusion models with stochastic volatility
modulating jump amplitudes. To additionally introduce explicit time
dependence to capture the term structure of at-the-money (ATM)
volatilities observed in commodity futures options markets (e.g.,
the seasonality effects in volatility, as well as the sharply
declining term structure of ATM volatility often seen in some
commodity futures options), we add a purely deterministic function
of time to the CIR activity rate (turning it into the so-called
CIR++ process, e.g., \cite{BrigoMercurio}). Such models with
mean-reverting jumps, stochastic volatility, and time dependence can
be calibrated to the entire volatility surface across both the
strike and maturity dimensions.


To conclude this introduction, we discuss analytical and
computational aspects. While the time changed L\'{e}vy models of
\cite{CarrGemanMadanYorSVLP} lead to fast and efficient option
pricing by means of Fourier analysis (see \cite{CarrMadanFFT} for
the Fast Fourier Transform methodology and \cite{FengLinetksy08} and
\cite{FengLinetsky09} for the closely related Hilbert transform
methodology), our time changed OU models also lead to analytical
option pricing, but by different mathematical means. While in the
context of L\'{e}vy processes one exploits the explicit knowledge of
the characteristic function, in the context of OU processes we
exploit the explicit knowledge of the eigenfunction expansion of the
SubOU transition semigroup. The eigenfunction expansion method is a
powerful tool for pricing contingent claims written on symmetric
Markov processes (see \cite{LinetskyIJTAF} and \cite{Linetsky} for
surveys). It is particularly well suited to time changes since the
time variable enters the eigenfunction expansion of the transition
semigroup only through the exponentials $e^{-\lambda_n t}$ and,
after the time change with a L\'{e}vy subordinator, the
eigenfunction expansion has the same form as for the original
process, but with $e^{-\lambda_n t}$ replaced with
$e^{-\phi(\lambda_n) t}$, where $\phi(\lambda)$ is the Laplace
exponent of the subordinator. We note that the seminal paper by
\cite{Bochner} already contained this observation (see Eq.(11) in
\cite{Bochner}; further see Albeverio and Rudiger (2003), (2005) for
the mathematical development of subordination of symmetric Markov
processes). In Mathematical Finance, this observation has been
previously exploited by \cite{AlbaneseKuznetsov} in the context of
volatility smile modeling for equities, by
\cite{BoyarchenkoLevendorskii} in the context of interest rate
modeling, and by \cite{MendozaCarrLinetsky} in the context of
unified credit-equity modeling.

\section{Subordinate Ornstein-Uhlenbeck Processes}

\subsection{SubOU Semigroups}

We start with an OU semigroup $(\mathcal{P}_t)_{t\geq0}$ defined on
$\mathfrak{B}_b(\mathbb{R})$ (the space of bounded Borel measurable
functions), where
$\mathcal{P}_tf(x)=\int_{\mathbb{R}}f(y)p(t,x,y)dy$ with the OU
transition kernel:
\begin{equation}\label{eq:OUTPD}
p(t,x,y)=\frac{1}{\sqrt{\frac{\pi\sigma^2}{\kappa}(1-e^{-2\kappa
t})}}\exp\Big\{-\frac{\left(y-x+(x-\theta)(1-e^{-\kappa
t})\right)^2}{\frac{\sigma^2}{\kappa}(1-e^{-2\kappa t})}\Big\}.
\end{equation}
$p(t,x,y)$ is the transition density of an OU diffusion with the
rate of mean reversion $\kappa> 0$, long-run level $\theta\in
{\mathbb R}$, and volatility $\sigma>0$. $(\mathcal{P}_t)_{t\geq0}$
is a strongly continuous contraction semigroup on
$\mathfrak{B}_b(\mathbb{R})$. Restricted to $C_0(\mathbb{R})$ (the
space of continuous functions vanishing at infinity), it is a Feller
semigroup, and $C^\infty_c(\mathbb{R})$ is a core of the domain
$D(\mathcal{G})$ of its infinitesimal generator $\mathcal{G}$ acting
on $C^2_c(\mathbb{R})$ (the subscript $c$ stands for functions with
compact support) by
$\mathcal{G}f(x)=\kappa(\theta-x)f'(x)+\frac{1}{2}\sigma^2 f''(x)$
(c.f. \cite{DuffieFilipovicSchachermayer} Theorem 2.7).

Consider a vaguely continuous convolution semigroup $(q_t)_{t\geq0}$
of probability measures on $\mathbb{R}_+$ (c.f.
\cite{SchillingSongVondracek} Definition 5.1). For each $t$,
$q_t\left([0,\infty)\right)=1$ (we consider only conservative case
in this paper), and its Laplace transform is given by the
L\'{e}vy-Khintchine formula with the Laplace exponent $\phi(\lambda)$ defined for all $\lambda\geq 0$:
$$
\int_{[0,\infty)} e^{-\lambda s}q_t(ds)=e^{-t\phi(\lambda)},\quad
\phi(\lambda)=\gamma\lambda+\int_{[0,\infty)}(1-e^{-\lambda
s})\nu(ds)
$$
with drift $\gamma\geqslant0$ and L\'{e}vy measure $\nu$ satisfying
the integrability condition
$\int_{[0,\infty)}(s\wedge1)\nu(ds)<\infty$. $(q_t)_{t\geq0}$ is the
family of transition probabilities of a subordinator, i.e., a
non-negative L\'{e}vy process starting at the origin (c.f.
\cite{Bertoin} or  \cite{SchillingSongVondracek}).

We define a subordinate semigroup $(\mathcal{P}^{\phi}_t)_{t\geq0}$
on $\mathfrak{B}_b(\mathbb{R})$ as the Bochner integral:
$$
\mathcal{P}^{\phi}_tf(x):=\int_{[0,\infty)}\mathcal{P}_sf(x)q_t(ds).
$$
This procedure is called Bochner's subordination (c.f.
\cite{SchillingSongVondracek} Definition 12.2). From
\cite{SchillingSongVondracek} Proposition 12.1, the subordinate
semigroup $(\mathcal{P}^{\phi}_t)_{t\geq0}$ is also a strongly
continuous contraction semigroup on $\mathfrak{B}_b(\mathbb{R})$. We
call it the {\em SubOU semigroup} with generating tuple
$(\kappa,\theta,\sigma,\gamma,\nu)$. The superscript $\phi$ in
$(\mathcal{P}^{\phi}_t)_{t\geq0}$ signifies that it is constructed
by subordinating the semigroup $(\mathcal{P}_t)_{t\geq0}$ with the
convolution semigroup of a subordinator with the Laplace exponent
$\phi$.

From \cite{JacobVol1} Corollary 4.3.4, a Feller semigroup remains a
Feller semigroup after subordination. It implies that
$(\mathcal{P}^\phi_t)_{t\geq0}$ restricted to $C_0(\mathbb{R})$ is
Feller. Its infinitesimal generator is given by Phillips' Theorem
(\cite{Sato} Theorem 32.1). The assertion on its core comes from
\cite{Sato} Proposition 32.5 (ii) and the fact that
$C^\infty_c(\mathbb{R})$ is a core of $D(\mathcal{G})$. We summarize
these results in the following.

\begin{thm}\label{thm:SubOUgenerator}
(i) A SubOU semigroup with generating tuple
$(\kappa,\theta,\sigma,\gamma,\nu)$ is a Feller semigroup.

\noindent (ii)Let
$\mathcal{G}^{\phi}$ be its infinitesimal generator. Then
$C^\infty_c(\mathbb{R})$ is a core of $D(\mathcal{G}^{\phi})$,
$C^2_c(\mathbb{R})\subseteq D(\mathcal{G}^{\phi})$, and for any
$f\in C^2_c(\mathbb{R})$,
\begin{equation*}
\mathcal{G}^{\phi}f(x)=\frac{1}{2}\gamma\sigma^2f''(x)+b(x)f'(x)+\int_{y\neq0}\left(f(x+y)-f(x)-y{1}_{\{|y|\leqslant1\}}f'(x)\right)\Pi(x,dy),
\end{equation*}
with the state-dependent L\'{e}vy measure $\Pi(x,dy)=\pi(x,y)dy$
with density defined for all $y\neq 0$
\begin{equation}
\pi(x,y)=\int_{[0,\infty)} p(s,x,x+y)\nu(ds),
\end{equation}
where $p(t,x,y)$ is the OU transition density (2.1). The drift with
respect to the truncation function $y{1}_{\{|y|\leqslant1\}}$ is
$$
b(x)=\gamma\kappa(\theta-x)+\int_{[0,\infty)}\left(\int_{\{|y|\leqslant1\}}yp(s,x,x+y)dy\right)\nu(ds).
$$
\end{thm}

\begin{remark}
(i) On $C^\infty_c(\mathbb{R})$, $\mathcal{G}^{\phi}$ can be
represented as a pseudo-differential operator (PDO) (see for example
\cite{Schnurr} Corollary 1.21)
$\mathcal{G}^{\phi}f(x)=-p(x,D)f(x)=-\int_{\mathbb{R}}p(x,\xi)\hat{f}(\xi)e^{ix\xi}d\xi$,
where $\hat{f}(\xi)=\frac{1}{2\pi}\int_\mathbb{R}e^{-i\xi x}f(x)dx$
is the Fourier transform of $f(x)$, and $p(x,\xi)$ is called the
symbol of the PDO and is expressed as
$p(x,\xi)=\frac{1}{2}\gamma\sigma^2\xi^2-ib(x)\xi-\int_{y\neq0}\left(e^{i\xi y}-1-i\xi y{1}_{\{|y|\leqslant1\}}\right)\Pi(x,dy)$.
Note that $p(x,\xi)$ is a continuous negative definite function
(CNDF) for each $x$ (c.f. \cite{JacobVol1} Definition 3.6.5).

\noindent(ii) $\pi(x,y)$ satisfies the condition
$\int_{y\neq0}(y^2\wedge1)\pi(x,y)dy<\infty$ for each $x$. This is a
direct result from the representation theorem for CNDF. See
\cite{JacobVol1} Theorem 3.7.7.

\noindent(iii) The L\'{e}vy measure of the SubOU semigroup has
finite activity if and only if the L\'{e}vy measure of the
subordinator has finite activity, which is justified by
interchanging the order of integration in
$\int_{y\neq0}\int_{[0,\infty)} p(s,x,x+y)\nu(ds)dy$ by Tonelli's
Theorem.

\noindent(iv) In general, it is not true that we can interchange the
order of integration in \linebreak
$\int_{[0,\infty)}\int_{|y|\leq1}yp(s,x,x+y)dy\nu(ds)$. However, if
the L\'{e}vy density satisfies the integrability condition
$\int_{|y|\leqslant1}|y|\pi(x,y)dy<\infty$, then the interchange is
valid and the truncation is not needed.
It can be shown that this integrability condition is equivalent to
$\int_0^1\sqrt{s}\nu(ds)<\infty\ (\text{if}\ x\neq\theta)$ and
$\int_0^1\nu(ds)<\infty\ (\text{if}\ x=\theta)$.
In this case the generator
takes the simpler form on $C^2_c(\mathbb{R})$:
$$
\mathcal{G}^{\phi}f(x)=\frac{1}{2}\gamma\sigma^2
f''(x)+\gamma\kappa(\theta-x)f'(x)+\int_\mathbb{R}\left(f(x+y)-f(x)\right)\pi(x,y)dy.
$$
\end{remark}

\subsection{SubOU Processes as Markov Semimaringales}

\begin{definition}
{\em A time-homogeneous Markov process $(\Omega,\mathcal
{F},(\mathcal{F}_t)_{t\geq0},X,\mathbb{P}^x)_{x\in\mathbb{R}}$ with
state space $(\mathbb{R},\mathscr{B}(\mathbb{R}))$ is called a
subordinate OU (SubOU) process with generating tuple \linebreak
$(\kappa,\theta,\sigma,\gamma,\nu)$ if its semigroup is a SubOU
semigroup with the same generating tuple.}
\end{definition}
Since a SubOU semigroup is Feller, a SubOU process is a Feller
process. Every Feller process has a c\`{a}dl\`{a}g modification
(c.f. \cite{JacobVol3} Theorem 3.4.9 or \cite{RevuzYor} Theorem
\Rmnum{3}.2.7), so immediately we have the following

\begin{corollary}
Every SubOU process admits a c\`{a}dl\`{a}g modification.
\end{corollary}

We will always consider c\`{a}dl\`{a}g SubOU processes in this
paper. From now on, without explicit mention, we will assume that
$(X,(\mathbb{P}^x)_{x\in\mathbb{R}})$ is the canonical realization
of a given SubOU semigroup defined on
$(\Omega,\mathcal{F}^0,(\mathcal{F}^0_t)_{t\geq0})$, where
$\Omega=\mathbb{D}(\mathbb{R})$ (the Skorohod space of
c\`{a}dl\`{a}g functions with values in $\mathbb{R}$, c.f.
\cite{JacodShiryaev} Definition \Rmnum{6}.1.1),
$\mathcal{F}_t^0=\sigma(X_s,s\leq t)$, and
$\mathcal{F}^0=\bigvee_{t\geq0}\mathcal{F}_t^0$.

\cite{Schnurr} gives an excellent discussion on the connection
between c\`{a}dl\`{a}g Feller processes and semimartingales. The
Feller property of the SubOU process together with
$C^\infty_c(\mathbb{R})\subseteq D(\mathcal{G}^\phi)$ allows us to
conclude that it is a semimartingale w.r.t. every $\mathbb{P}^x$ with $x\in {\mathbb R}$
(c.f. \cite{Schnurr} Theorem 3.1). From \cite{Schnurr} Theorem 3.14,
the pseudo-differential operator representation of the infinitesimal
generator $\mathcal{G}^\phi$ gives us the triplet $(B,C,\Pi)$ of
semimartingale characteristics of the SubOU process. For the
definition of semimartingale characteristics see
\cite{JacodShiryaev} Chapter \Rmnum{2}.
\begin{thm}\label{thm:SubOUSemiMG}
(i) The SubOU process $(\Omega,\mathcal
{F}^0,(\mathcal{F}_t^0)_{t\geq0},X,\mathbb{P}^x)_{x\in\mathbb{R}}$
with generating tuple \linebreak $(\kappa,\theta,\sigma,\gamma,\nu)$
is a semimartingale w.r.t. every $\mathbb{P}^x$ and admits
semimartingale characteristics $(B,C,\Pi)$ w.r.t to the truncation
function $h(x)=x{1}_{\{|x|\leq1\}}$, where
$$
B_t(\omega)=\int_0^t\Big[\gamma\kappa(\theta-X_{s-}(\omega))+\int_0^\infty\int_{\{|y|\leq1\}}yp(u;X_{s-}(\omega),X_{s-}(\omega)+y)dy\nu(du)\Big]ds,
$$
$$
C_t(\omega)=\gamma\sigma^2t,\quad
\Pi(\omega,dt,dy)=\pi(X_{t-}(\omega),y)dtdy,
$$
where $\pi(x,y)$ is given in Theorem 2.1.

\noindent (ii) Denote by $\mu^X$ the integer-valued random measure associated with
the jumps of $X$ (c.f. \cite{JacodShiryaev} Proposition
\Rmnum{2}.1.16) and $X^c$ the continuous local martingale part of
$X$. Then $X$ has the following sample path decomposition (under the
starting point $x$):
\begin{equation}
X_t(\omega)=x+B_t(\omega)+X^c_t(\omega)+h(x)*(\mu^X-\Pi)_t(\omega)+(x-h(x))*\mu^X_t(\omega)
\end{equation}
with the quadratic variation of the continuous part
$[X^c,X^c]_t(\omega)=C_t(\omega)=\gamma\sigma^2t$ ($*$ denotes
integration w.r.t. a random measure).\vspace{0.3cm}

\noindent (iii) If $X'$ is an $\mathbb{R}$-valued semimartingale
defined on some filtered probability space \linebreak $(\Omega',
\mathcal{F}',(\mathcal{F}'_t)_{t\geq0},\mathbb{P}')$ with
$\mathbb{P'}(X'_0=x)=1$ and with the semimartingale characteristics
$(B',C',\Pi')$ given in (1), where $X$ is replaced by $X'$, then
$\mathbb{P}'\circ X'^{-1}=\mathbb{P}^x$.
\end{thm}

The proof of Part (iii) of Theorem \ref{thm:SubOUSemiMG} is given in
Appendix B. It essentially says the solution to the Martingale
Problem in the canonical space setting as defined in
\cite{JacodShiryaev} Definition \Rmnum{3}.2.4 is unique. This is a
key result for the study of locally equivalent measure changes for
SubOU processes in section \ref{subsect:EMCSubOU}.

\begin{remark}
(i) It is clear that the SubOU process is a jump-diffusion process
if $\gamma>0$ and a pure jump process if $\gamma=0$.

\noindent (ii) If $\int_0^1\sqrt{s}\nu(ds)<\infty$, then
$\int_{|y|\leqslant1}|y|\pi(x,y)dy<\infty$ for all $x\neq\theta$.
Hence $|h(x)|*\Pi\in\mathscr{A}^+_{loc}$, which implies
$h(x)*(\mu^X-\Pi)=h(x)*\mu^X-h(x)*\Pi$ (c.f. \cite{JacodShiryaev}
Proposition \Rmnum{2}.1.28) and
$$
X_t(\omega)=x+\int_0^t\gamma\kappa(\theta-X_{s-}(\omega))ds+X^c_t(\omega)+x*\mu^X_t(\omega).
$$
Hence, in this case, the jump part of the SubOU process is of finite
variation.
\end{remark}

A SubOU process is a process with
mean-reverting jumps. The mean reversion property of the
state-dependent SubOU L\'{e}vy measure $\Pi(x,\cdot)$ is
characterized in the following.

\begin{thm}\label{thm:mean-revertingjumpmeasure}
For any $y>0$, we have
\begin{itemize}
\item[(i)] If $x>\theta$, then $\pi(x,-y)>\pi(x,y)$, and $\Pi(x,(-\infty,-y))>\Pi(x,(y,\infty))$.
\item[(ii)] If $x<\theta$, then $\pi(x,-y)<\pi(x,y)$, and $\Pi(x,(-\infty,-y))<\Pi(x,(y,\infty))$.
\item[(iii)] If $x=\theta$, then $\pi(x,-y)=\pi(x,y)$, and
$\Pi(x,(-\infty,-y))=\Pi(x,(y,\infty))$.
\end{itemize}
\end{thm}

This theorem tells us that when the current state $x$ is above
(below) the long-run level $\theta$, a downward (upward) jump is
more likely to occur. When $x=\theta$, the intensity of downward and
upward jumps are equal. This mean-reverting nature of jumps makes
SubOU processes a natural candidate for modeling mean-reverting
prices and other financial variables. If $\gamma=0$, a SubOU process
is a pure jump process with mean-reverting jumps. If $\gamma>0$, it
is a jump-diffusion process with mean-reverting diffusion drift and
mean-reverting jumps.

\begin{figure}[htbp!]
\centering \subfigure[CPP
$x=1$]{\includegraphics[scale=0.25]{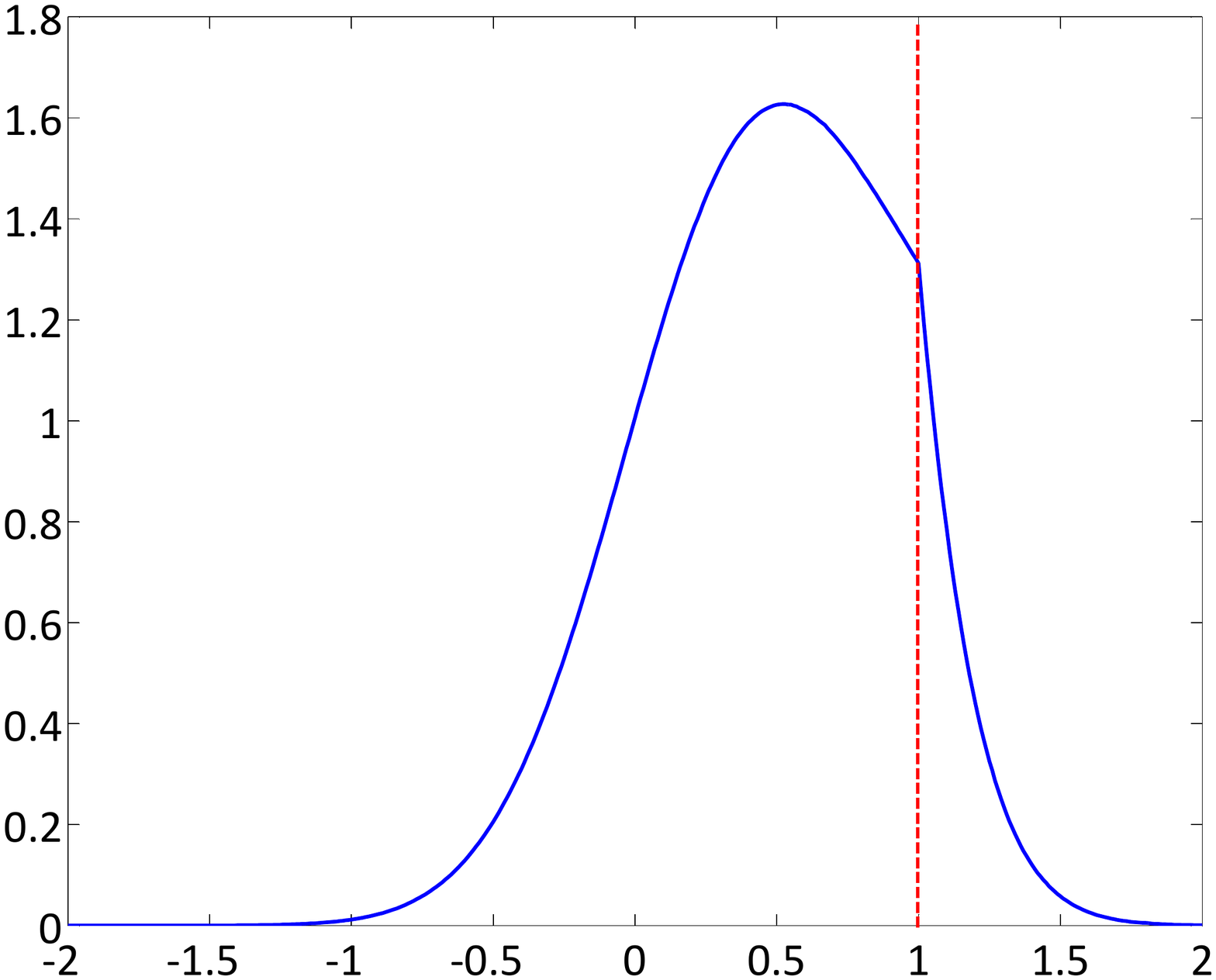}} \subfigure[CPP
$x=0.2$]{\includegraphics[scale=0.25]{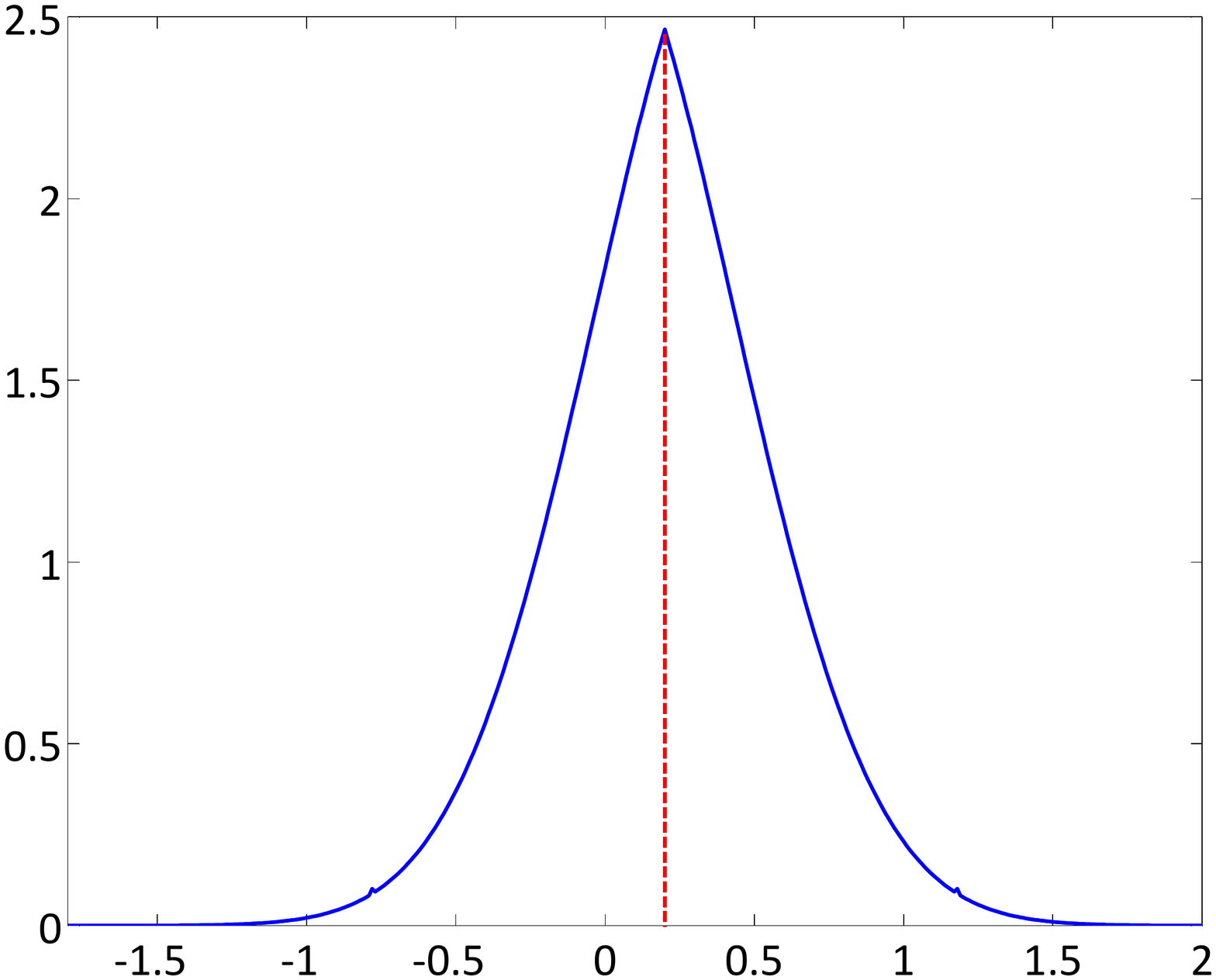}}\\
\subfigure[CPP $x=-1$]{\includegraphics[scale=0.25]{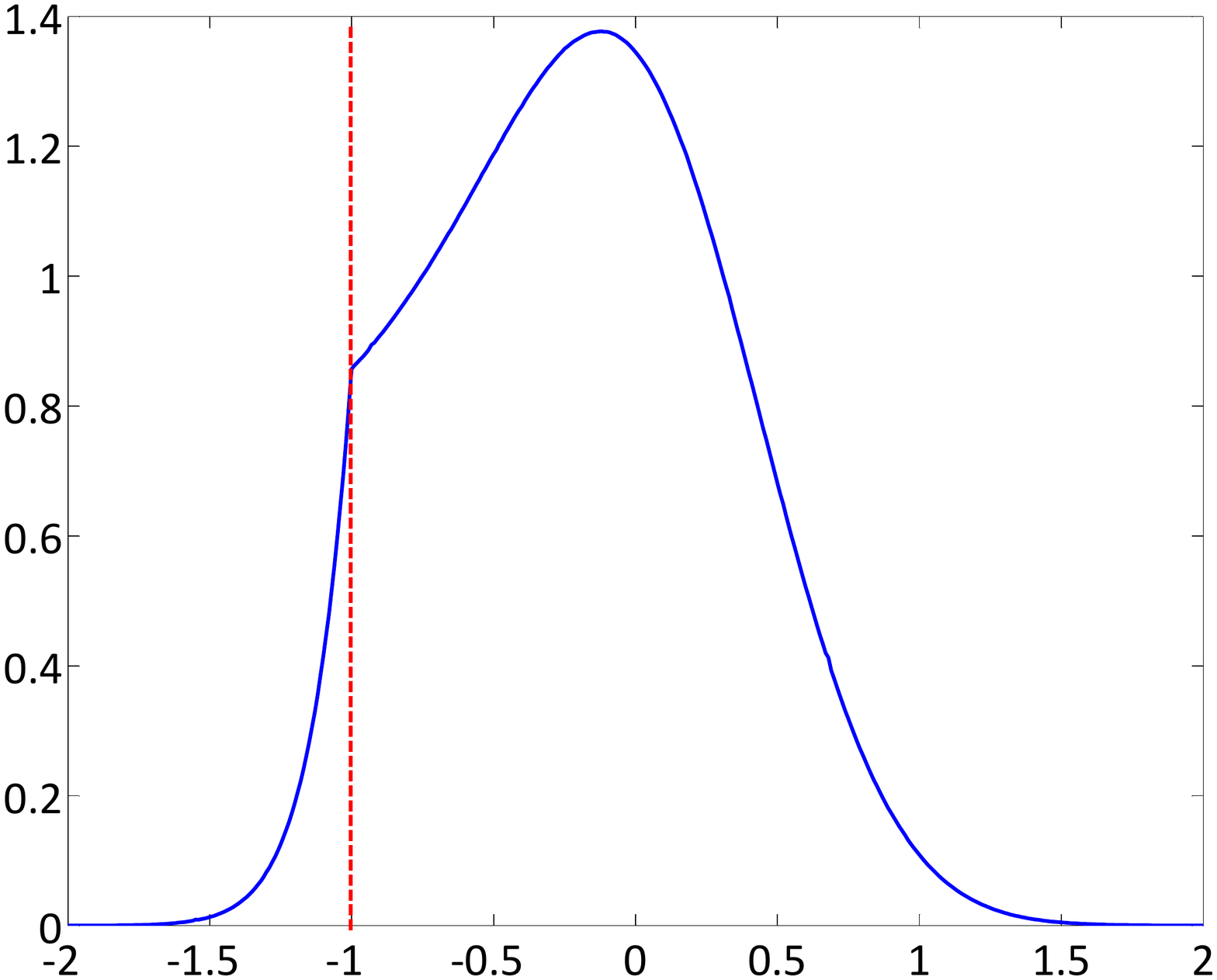}}
\subfigure[IG $x=1$]{\includegraphics[scale=0.25]{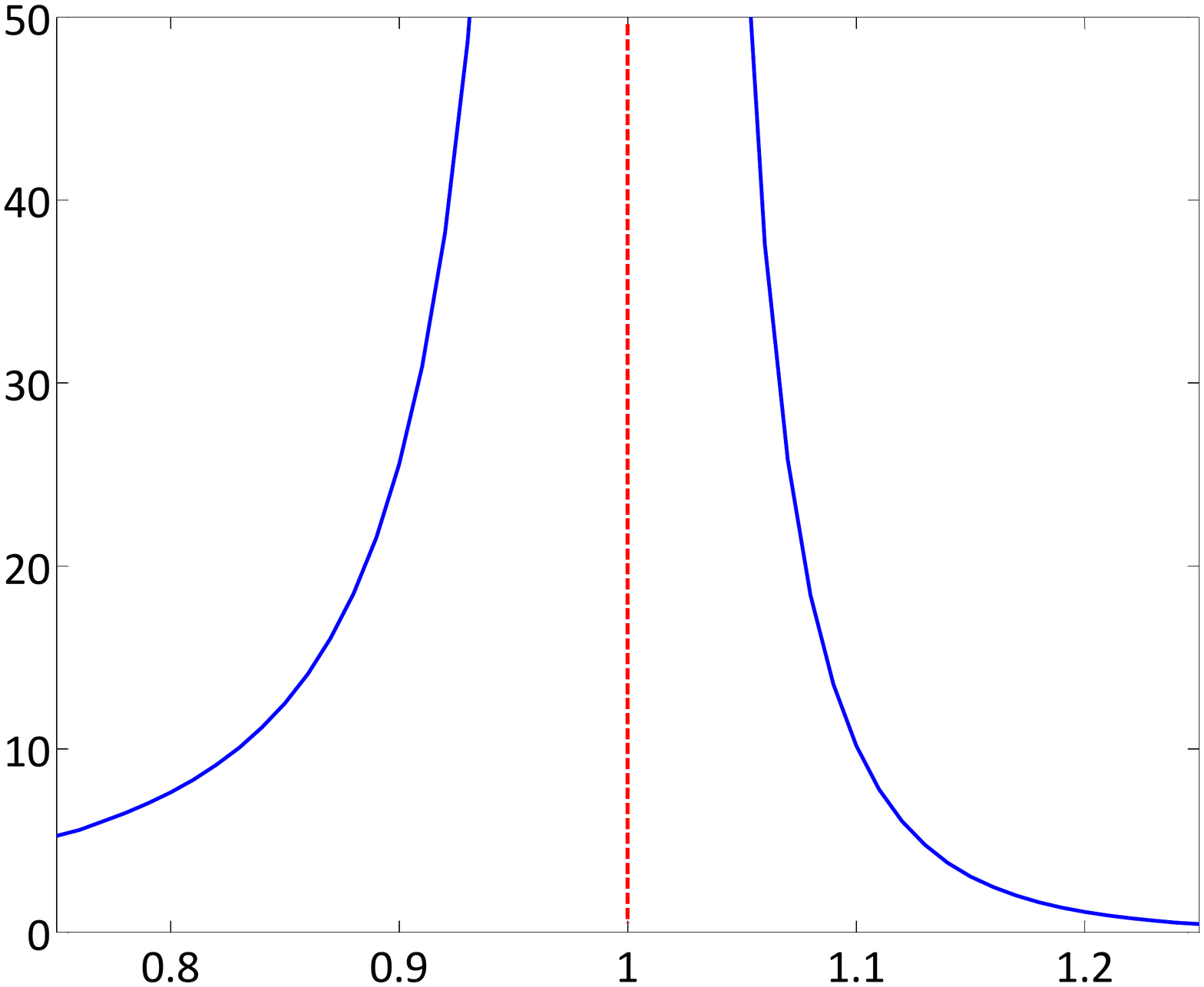}}\\
\subfigure[IG
$x=0.2$]{\includegraphics[scale=0.25]{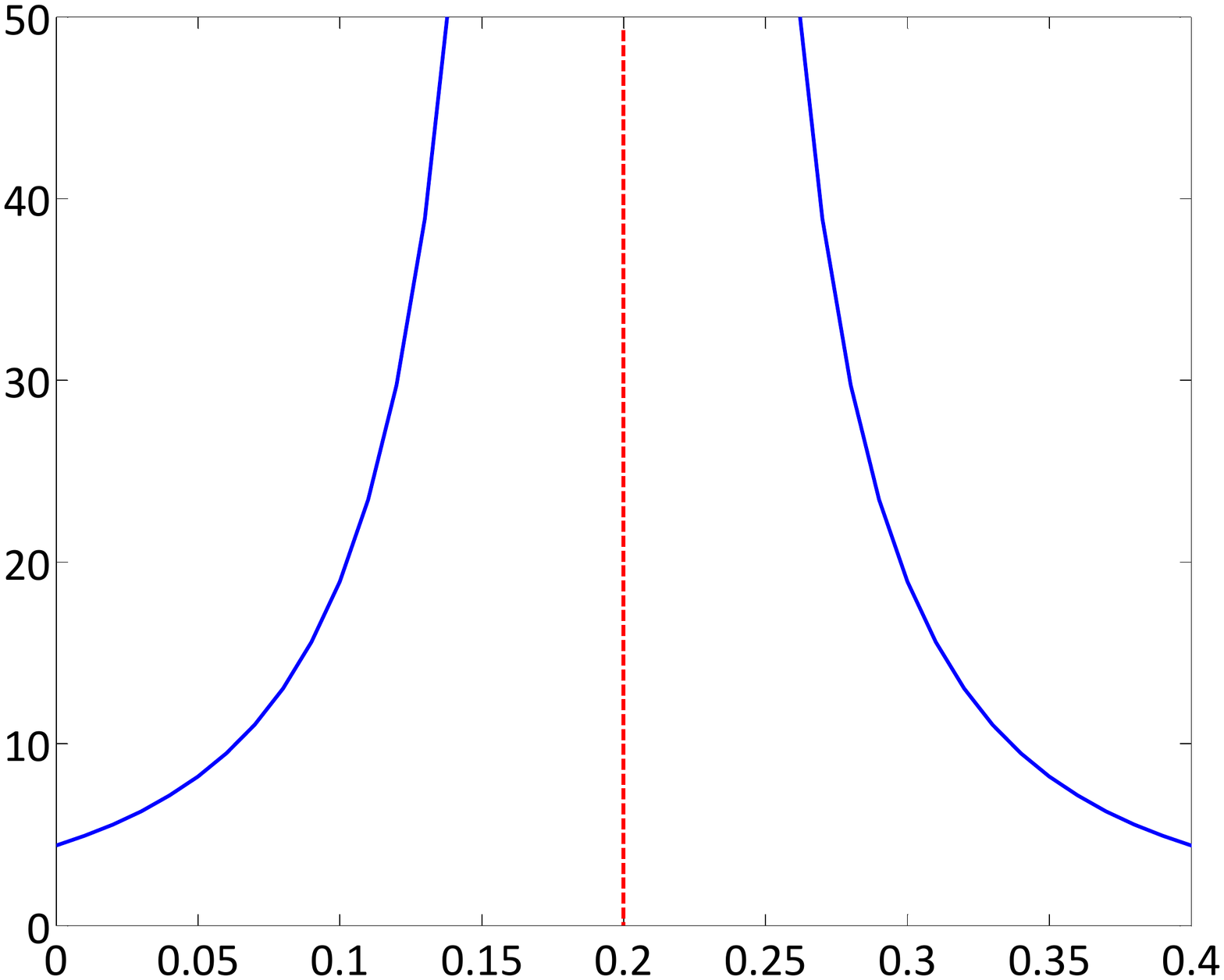}}
\subfigure[IG $x=-1$]{\includegraphics[scale=0.25]{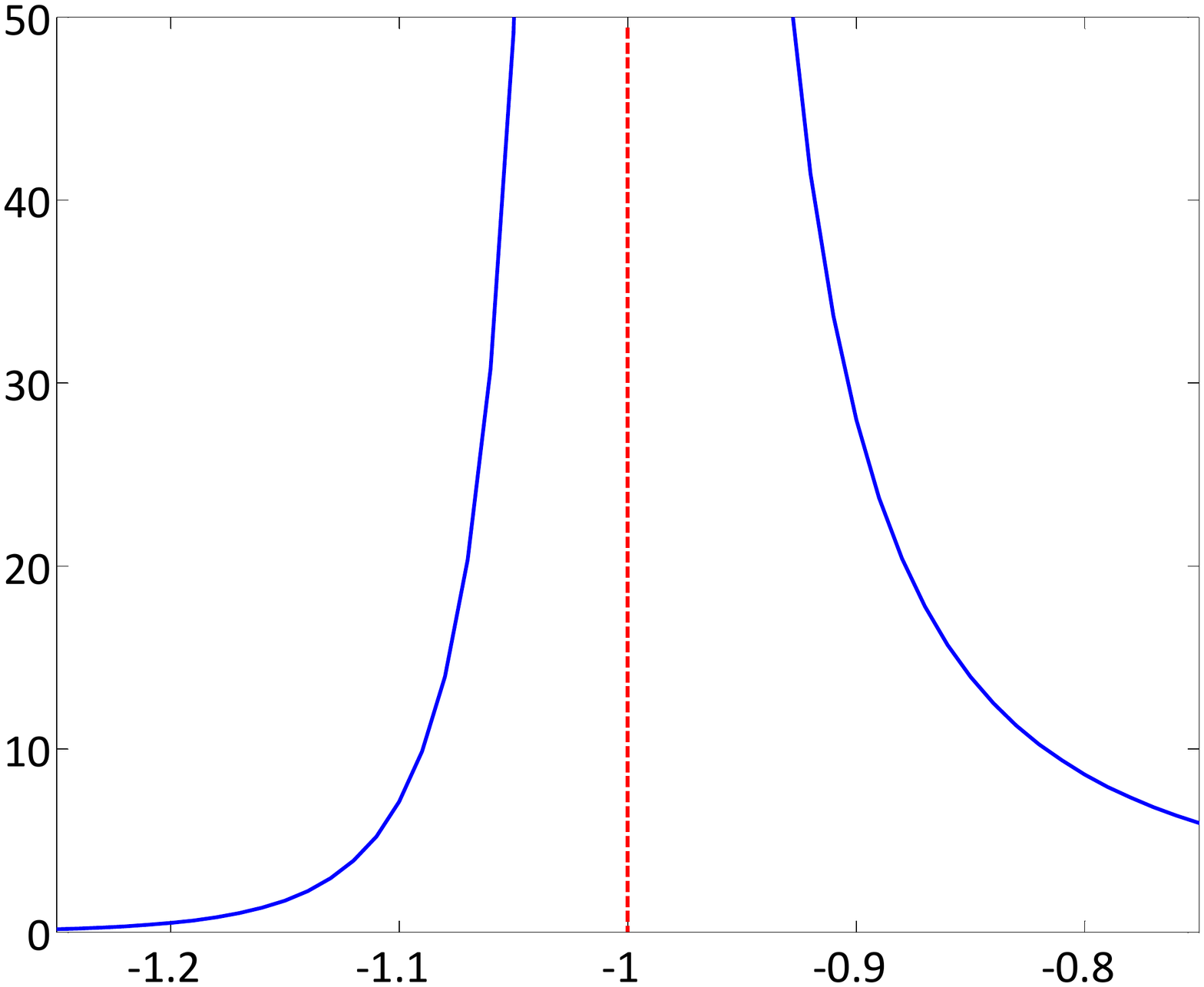}}
\caption{State-dependent L\'{e}vy densities of SubOU processes with
$\theta=0.2$, $\kappa=1$, and $\sigma=0.6$ when $\nu$ is the
L\'{e}vy measure of a compound Poisson process with exponential
jumps (arrival rate $\alpha=2$, reciprocal of mean jump size
$\eta=1$) and an inverse Gaussian process (mean rate $\mu=1$,
variance rate $\nu=1$) with $x=-1$, $0.2$, $1$. To emphasize the
value of the current state, the horizontal axis plots the post jump
state (not the jump size).}\label{fig:jumpdensity}
\end{figure}

Figure \ref{fig:jumpdensity} plots SubOU L\'{e}vy densities when
$\nu$ are L\'{e}vy measures of a compound Poisson process with
exponential jump sizes and an inverse Gaussian (IG) process.

\begin{remark}
\textbf{Time Change Interpretation of Bochner's Subordination} The semigroup $(\mathcal{P}_t)_{t\geq0}$ gives rise to
an OU diffusion process $X$. The vaguely continuous convolution
semigroup of probability measures $(q_t)_{t\geq0}$ gives rise to a
subordinator $T$. Assume that both $X$ and $T$ are defined on the
same probability space and are independent. Then the time changed or
subordinate process $X^\phi_t:=X_{T_t}$ is again a Markov process.
By independence of $X$ and $T$, the associated operator semigroup is
given by
$$\mathcal{P}^\phi_tf(x)=\mathbb{E}[f(X_{T_t})]=\int_{[0,\infty)}\mathbb{E}_x[f(X_s)]q_t(ds)=\int_{[0,\infty)}\mathcal{P}_sf(x)q_t(ds).$$
That is, $X^\phi_t$ is a SubOU process according to our definition,
and Bochner's subordination can be interpreted as a stochastic time
change with respect to an independent subordinator (cf. Schilling,
Song and Vondracek (2010) p.141).
\end{remark}
\begin{remark} SubOU Markov semimartingales admit a representation in terms of a Brownian motion and an independent Poisson random measure. Explicit expressions follow from Cinlar and Jacod (1981) Theorem 3.13 and are omitted due to space constraints.
\end{remark}

\subsection{Equivalent Measure Transformations for SubOU Processes}\label{subsect:EMCSubOU}

For building financial models based on SubOU processes, we are
interested in locally equivalent measure changes\footnote{Two
probability measures $\mathbb{P}_1$ and $\mathbb{P}_2$ on a filtered
probability space $(\Omega,\mathcal{F},(\mathcal{F}_t)_{t\geq0})$
are said to be locally equivalent, if
$\mathbb{P}_1|_{\mathcal{F}_t}\sim \mathbb{P}_2|_{\mathcal{F}_t}$
for each $t\geq0$, where $\mathbb{P}|_{\mathcal{F}_t}$ is the
restriction of measure $\mathbb{P}$ on the $\sigma$-filed
$\mathcal{F}_t$.} that transform a SubOU process with a given
generating tuple into another SubOU process with another
generating tuple. We can then build financial models with SubOU processes under both the physical and the risk-neutral measures, and determine how the generating tuple of the SubOU process changes under the measure change.

As before, $\Omega$ is the space of all
c\`{a}dl\`{a}g functions taking values in $\mathbb{R}$. In this
section we follow \cite{JacodShiryaev}. In order to use their
results, we use the right-continuous version of the filtration
$(\widetilde{\mathcal{F}}_{t})_{t\geq0}$ with
$\widetilde{\mathcal{F}}_{t}=\mathcal{F}^0_{t+}$ and
$\widetilde{\mathcal{F}}=\bigvee_{t\geq0}\widetilde{\mathcal{F}}_t=\mathcal{F}^0$.
Let $X$ be the canonical process. It is clear that if $X$ is a SubOU
process, it is also Markov and a SubOU process w.r.t.
$(\widetilde{\mathcal{F}})_{t\geq0}$. We fix the truncation function
$h(x)=x{1}_{\{x\leq1\}}$. Let $\mathbb{P}_0$ be a probability
measure on $(\Omega,\widetilde{\mathcal{F}}_{0})$ taken to be the
initial distribution. Following \cite{JacodShiryaev} Definition
\Rmnum{3}.2.4, we call a probability measure $\mathbb{P}$ on
$(\Omega,\widetilde{\mathcal{F}},(\widetilde{\mathcal{F}}_t)_{t\geq0})$
a solution to the martingale problem associated with
$(\widetilde{\mathcal{F}}_0,X)$ and $(\mathbb{P}_0;B,C,\nu)$, where
$(B,C,\nu)$ are given semimartingale characteristics, if the
following hold: (i) the restriction $\mathbb{P}|_{\widetilde{\mathcal{F}}_0}=\mathbb{P}_0$;
(ii) $X$ is a semimartingale on the stochastic basis $(\Omega,\widetilde{\mathcal{F}},(\widetilde{\mathcal{F}}_t)_{t\geq0},\mathbb{P})$ with characteristics $(B,C,\nu)$ relative to the truncation function $h$.
The following proposition is crucial in proving the necessary and
sufficient conditions for locally equivalent measure change.

\begin{proposition}
Let $(B,C,\Pi)$ be the SubOU semimartingale characteristics defined
in Theorem \ref{thm:SubOUSemiMG}. The solution to the martingale
problem $(\sigma(X_0),X|\mathbb{P}_0;B,C,\Pi)$ exists and is unique.
Moreover, local uniqueness holds.
\end{proposition}

See \cite{JacodShiryaev} Definition \Rmnum{3}.2.35 for the
definition of local uniqueness. The existence of the solution is
quite obvious. Given a SubOU semigroup with generating tuple
corresponding to the given SubOU semimartingale characteristics
$(B,C,\nu)$, we can construct a time-homogeneous universal Markov
process on the space of c\`{a}dl\`{a}g functions taking values in
$\mathbb{R}$. Such a process is a semimartingale with
characteristics $(B,C,\nu)$ by Theorem \ref{thm:SubOUSemiMG} under
every $\mathbb{P}^x$, and set
$\mathbb{P}(A)=\int\mathbb{P}^x(A)\mathbb{P}_0(dx)$ for any
$A\in\mathcal{F}$. The uniqueness follows from part (iii) of Theorem
\ref{thm:SubOUSemiMG}. The local uniqueness is a result of
uniqueness and the Markov property of the process by
\cite{JacodShiryaev} Theorem \Rmnum{3}.2.40. We then have the following.

\begin{thm}\label{thm:EMCSubOU}
Let $\mathbb{P}$ and $\mathbb{P}'$ be two probability measures on
$(\Omega,\widetilde{\mathcal{F}},(\widetilde{\mathcal{F}}_t)_{t\geq0})$
such that the canonical process is a SubOU process with generating
tuples $(\kappa,\theta,\sigma,\gamma,\nu)$ and
$(\kappa',\theta',\sigma',\gamma',\nu')$, respectively, and with
initial distributions $\mathbb{P}_0$ and $\mathbb{P}'_0$,
respectively. Then the following two statements are equivalent.
\begin{itemize}
\item[(1)] $\mathbb{P}$ and $\mathbb{P}'$ are locally equivalent, i.e., $\mathbb{P}|_{\widetilde{\mathcal{F}}_t}\sim\mathbb{P}'|_{\widetilde{\mathcal{F}}_t}$ for every
$t\geq0$.
\item[(2)] The following conditions are satisfied:\\
\noindent (i) $\mathbb{P}_0\sim\mathbb{P}'_0$; (ii) $\gamma\sigma^2=\gamma'\sigma'^2$;\\
\noindent (iii) For every $x\in\mathbb{R}$, the Hellinger
condition
$\int_{y\neq0}\big(\sqrt{\pi'(x,y)}-\sqrt{\pi(x,y)}\big)^2dy<\infty$
holds, where $\pi(x,\cdot)$ and $\pi'(x,\cdot)$ are defined as in
Theorem \ref{thm:SubOUgenerator}.
\end{itemize}
\item[(3)] Furthermore, suppose these conditions are satisfied. Define
$$\beta_t(\omega):=\frac{(\gamma'\kappa'\theta'-\gamma\kappa\theta)-(\gamma'\kappa'-\gamma\kappa)X_{t-}(\omega)}{\gamma\sigma^2}{1}_{\{\gamma\neq0\}},
\quad \text{and}\quad
Y(\omega,t,y):=\frac{\pi'(X_{t-}(\omega),y)}{\pi(X_{t-}(\omega),y)}.$$
Let $X^c$ and $\mu^X$ denote the continuous martingale part of $X$
and the jump measure associated with $X$. Then $N=\beta\cdot
X^c+(Y-1)*(\mu^X-\Pi)$ is a $\mathbb{P}$-local martingale, and the
Radon-Nikodym density process $D$ of $\mathbb{P}'$ w.r.t.
$\mathbb{P}$ equals to
 the Doleans-Dade stochastic exponential $\mathscr{E}(N)$ of $N$.
\end{thm}

\begin{remark}
Define $\varphi(x,y):=\ln\left(\pi'(x,y)/\pi(x,y)\right)$. The
Hellinger condition
$\int_{y\neq0}\big(\sqrt{\pi'(x,y)}-\sqrt{\pi(x,y)}\big)^2dy<\infty$
is equivalent to the following (similar to Remark 33.3 of
\cite{Sato}):
$$\int_{\{y:|\varphi(x,y)|\leq1\}}\varphi(x,y)^2\pi(x,y)dy<\infty,\ \int_{\{y:\varphi(x,y)>1\}}\pi'(x,y)dy<\infty,
\ \int_{\{y:\varphi(x,y)<-1\}}\pi(x,y)dy<\infty.$$ Intuitively, when
$\pi$ and $\pi'$ both have infinite activity, the Hellinger
condition says that the region where large perturbations of the jump
density occurs should not be arbitrarily close to the origin.
\end{remark}

\begin{remark}\label{remark:limitcaseSubOU}
The limiting case $\kappa=0$ corresponds to the subordinate Brownian
motion without drift. Theorem \ref{thm:EMCSubOU} still holds when
$\kappa=0$ and/or $\kappa'=0$. When $\kappa>0$ and $\kappa'=0$, it
characterizes locally equivalent measure transformations of SubOU
processes into subordinate Brownian motions without drift. When
$\kappa=\kappa'=0$, Theorem \ref{thm:EMCSubOU} reduce to the special
case of Theorem 33.1 of \cite{Sato} for L\'{e}vy processes
specialized to the case of subordinate Brownian motions.
\end{remark}

The general Hellinger condition is difficult to check. We wish to
derive restrictions it places on SubOU generating tuples that can be
transformed into each other under locally equivalent measure
changes. It can be easily shown that:
(i) If  both $\nu$ and $\nu'$ are L\'{e}vy measures of finite
activity subordinators, then the Hellinger condition is
automatically satisfied.
(ii) If $\nu$ is a L\'{e}vy measure of a finite
activity subordinator and $\nu'$ is a L\'{e}vy measure of an
infinite activity subordinator (or vice versa), then the Hellinger
condition is not satisfied. Thus, equivalent measure changes cannot
transform a SubOU process with a finite activity subordinator into a
SubOU process with an infinite activity subordinator, and vice
versa.

We now investigate the case when $\nu$ and $\nu'$
are L\'{e}vy measures of infinite activity subordinators. To verify
the Hellinger condition in this case, we need to study the
asymptotic behavior of the SubOU L\'{e}vy density $\pi(x,y)$ given in Eq.(2.2) as
$y\rightarrow 0$. The following proposition shows that it is
equivalent to the asymptotic behavior of the L\'{e}vy density of
some subordinated Brownian motion.

\begin{proposition}\label{prop:asymptotic behavior SubOU=SubBM}
Let $\pi(x,y)$ be the L\'{e}vy density of a SubOU process with
generating tuple $(\kappa,\theta,\sigma,\gamma,\nu)$. Suppose
$\pi(x,y)\rightarrow\infty$ as $y\rightarrow0$. For each fixed $x\in
{\mathbb R}$, let $\overline{\pi}_x(y)$ be the L\'{e}vy density of a
subordinate Brownian motion starting at $0$ with drift
$\kappa(\theta-x)$, volatility $\sigma$, and the same $\gamma$ and
$\nu$. Then $\lim_{y\rightarrow0}\pi(x,y)/\overline{\pi}_x(y)=1$.
\end{proposition}

We can further show that the asymptotics of the L\'{e}vy density of
subordinate Brownian motion does not depend on drift. We then have.
\begin{proposition}\label{prop:SubBM Asymptotic Independence of Drift}
The asymptotics of $\pi(x,y)$ as $y\rightarrow0$ does not depend on
$\kappa$, $\theta$, and $x$.
\end{proposition}
Hence, $\kappa$ and $\theta$ can be freely changed by locally
equivalent measure changes. In particular, $\kappa>0$ can be changed
to $\kappa=0$ by a locally equivalent measure change. The problem of
investigating the Hellinger condition now reduces to finding the
asymptotics of the L\'{e}vy density of a subordinate Brownian
motion. \cite{SongVondracek} is an excellent reference on the
potential theory of subordinate Brownian motions and provides many
examples of subordinators and asymptotics of the L\'{e}vy densities
of subordinate Brownian motions. If the L\'{e}vy measure $\nu$ of
the subordinator has a density $\nu(s)$, then, in general,
we have Proposition \ref{prop:SubBMComputeAsymptoticGen} to compute
the asymptotics of the L\'{e}vy density of the subordinate Brownian
motion
\begin{equation}\label{eq:SubBMJumpMeasure}
\overline{\pi}(y):=\int_0^{\infty}\frac{1}{\sqrt{2\pi\sigma^2s}}e^{-\frac{y^2}{2\sigma^2s}}\nu(s)ds
\end{equation}
as $y\rightarrow0$. Proposition \ref{prop:SubBMComputeAsymptoticGen}
gives the asymptotics under two different types of sufficient conditions. The
first sufficient condition is based on Lemma 3.3 of
\cite{SongVondracek}. The applicability of their Lemma 3.3 is not
restricted to L\'{e}vy densities of subordinators. However, in this
case some of their conditions are not necessary. Below we give a
more general result for this case. The second sufficient condition
is a restriction of the L\'{e}vy density of the subordinator to the class of completely
monotone functions\footnote{A completely monotone function
$f:(0,\infty)\mapsto\mathbb{R}$ is a $C^\infty$ function such that
$(-1)^nf^{(n)}(x)\geq0$ for $n=0,1,2,\cdots$.} (see, for example
\cite{SchillingSongVondracek} for its characterization and
properties). This result is proved in Theorem 2.6 in
\cite{KimSongVondracek}.


\begin{proposition}\label{prop:SubBMComputeAsymptoticGen}
Let $\nu(s)$ be the L\'{e}vy density of a subordinator. Suppose
there exist constants $c_0>0$ and $\frac{1}{2}<\beta<2$ and a
function $\ell:(0,\infty)\rightarrow(0,\infty)$ slowly varying at
infinity\footnote{A function $\ell$ defined in a neighborhood of
infinity is called slowly varying at infinity if
$\lim_{x\rightarrow\infty}\frac{\ell(ax)}{\ell(x)}=1$ for all
$a>0$.} such that
\begin{equation}\label{eq:SubordinatorAsymptoticLevyDensityAt0}
\nu(s)\sim\frac{c_0}{s^\beta\ell(\frac{1}{s})}\,\,\,\text{as}\,\,\,
s\rightarrow0.
\end{equation}
Let $\overline{\pi}(y)$ be defined as in
\eqref{eq:SubBMJumpMeasure}. Suppose one of the following two
conditions is satisfied:
\begin{itemize}
\item[(1)] Let $g:(0,\infty)\rightarrow(0,\infty)$ be a function such that
$\int_0^\infty s^{\beta-\frac{3}{2}}e^{-s}g(s)ds<\infty$. Assume
there is also some $\xi>0$ such that $f_{\ell,\xi}(y,s)\leq g(s)$
for all $y,s>0$, where the auxiliary function $f_{\ell,\xi}(y,s)$ is
defined by
$f_{\ell,\xi}(y,s):=\frac{\ell(\frac{1}{y})}{\ell(\frac{2\sigma^2s}{y})}$
if $y<\frac{s}{\xi}$ and $0$ otherwise for any function $\ell$
slowly varying at infinity and any $\xi>0$.

\item[(2)] $\nu(s)$ is a completely monotone function.
\end{itemize}
Then
$$
\overline{\pi}(y)\sim\dfrac{c_0\Gamma(\beta-\frac{1}{2})}{\sqrt{\pi}(2\sigma^2)^{1-\beta}}\dfrac{1}{|y|^{2\beta-1}\ell(\frac{1}{y^2})}\,\,\,\,\text{as}\,\,\,\,
y\rightarrow0.
$$
\end{proposition}

\begin{remark}
For slowly varying functions and regularly varying functions see
\cite{BinghamGoldieTeugels}. Every regularly varying
function\footnote{A function $f$ defined in a neighborhood of
infinity is called regularly varying at infinity with index
$\rho\in\mathbb{R}$ if $\lim_{x\rightarrow\infty}\frac{f(\lambda
x)}{f(x)}=\lambda^\rho$ for all $\lambda>0$. It is called regularly
varying at $0$ if $f(\frac{1}{x})$ is regularly varying at
$\infty$.} at zero can be written in the form
$\frac{1}{x^\beta\ell(\frac{1}{x})}$ for some real number $\beta$
and $\ell$ slowly varying at infinity (c.f.
\cite{BinghamGoldieTeugels} Theorem 1.4.1). Hence, the assumption on
the asymptotics \eqref{eq:SubordinatorAsymptoticLevyDensityAt0} is
very general. Also note that from \cite{BinghamGoldieTeugels}
Proposition 1.3.6,
$\frac{1}{s^\beta\ell(\frac{1}{s})}\rightarrow\infty$ as
$s\rightarrow0$, so we are dealing with subordinators whose L\'{e}vy
density tends to infinity at $0$.
\end{remark}

For a subordinator with L\'{e}vy density, if
\eqref{eq:SubordinatorAsymptoticLevyDensityAt0} is satisfied, there
is a close connection between the Blumenthal-Getoor (BG) index and
the parameter $\beta$ in Proposition 2.4 when $\beta\geq 1$. For any
subordinator with L\'{e}vy measure $\nu(ds)$ its BG index is defined
by $p:=\inf\{\alpha>0: \int_{|s|\leq1}s^\alpha\nu(ds)<\infty\}$.

\begin{proposition}\label{prop:SubordinatorBetaAndBGIndex}
(1) Suppose \eqref{eq:SubordinatorAsymptoticLevyDensityAt0} holds
with $\beta\geq1$. Then the BG index is equal to $\beta-1$.

\noindent(2) Suppose the conditions in Proposition
\ref{prop:SubBMComputeAsymptoticGen} are satisfied for two
subordinators with $\beta\geq1$ and $\beta'\geq1$. Then the
Hellinger condition implies their BG indexes are equal.
\end{proposition}

We now apply Proposition \ref{prop:SubBMComputeAsymptoticGen} to the
key example important in financial applications.
\begin{example}\textbf{Tempered Stable Subordinators.}
Consider the tempered stable family of L\'{e}vy measures $\nu(s)=C
s^{-1-p}e^{-\eta s}$, where $C>0$, $p<1$, $\eta>0$. The limiting
stable family has $\eta=0$ and $p\in (0,1)$. The tempered stable
cases with $p\geq0$ ($p<0$) give rise to subordinators with infinite
activity (finite activity). Important special cases are the Gamma
subordinator with $p=0$ (\cite{MadanCarrChang}), the Inverse
Gaussian (IG) subordinator with $p=\frac{1}{2}$
(\cite{BarndorffNielsen}), and the compound Poisson subordinator
with exponential jumps with $p=-1$  and $\eta>0$. For this family,
the Laplace exponent is given by the following.
\begin{equation}\label{eq:LaplaceExponentTSS}
\phi(\lambda)=
\begin{cases}
\gamma\lambda-C\Gamma(-p)[(\lambda+\eta)^p-\eta^p], &p\neq 0\\
\gamma\lambda+C\ln(1+\lambda/\eta), &p=0
\end{cases},
\end{equation}
where $\Gamma(\cdot)$ is the Gamma function.

For tempered stable subordinators with $p>-\frac{1}{2}$, it is clear
that Proposition \ref{prop:SubBMComputeAsymptoticGen} condition (1)
holds with $c_0=C$, $\beta=1+p$, $\ell(x)=1$, $g(s)=1$, and $\xi$
chosen arbitrarily. Condition (2) also holds because the L\'{e}vy
density of the subordinator is completely monotone. Hence we have
$$
\overline{\pi}(y)\sim\frac{C\Gamma(p+\frac{1}{2})(2\sigma^2)^p}{\sqrt{\pi}|y|^{2p+1}}\,\,\,\text{as}\,\,\,y\rightarrow0.
$$
From Proposition \ref{prop:asymptotic behavior SubOU=SubBM},
$\pi(x,y)$ has the same asymptotics. It is now straightforward to show that
Theorem \ref{thm:EMCSubOU} reduces to the following result for SubOU processes with
tempered stable subordinators with drift.
\end{example}
\begin{corollary}\label{cor:EMC Tempered Stable}
Consider the setting in Theorem \ref{thm:EMCSubOU}. Suppose $\nu$
and $\nu'$ belong to the tempered stable family with parameters
$(C,p,\eta)$ and $(C',p',\eta')$ with $p,p'\geq 0$, respectively.
Then $\mathbb{P}|_{\widetilde{\mathcal{F}}_t}\sim
\mathbb{P}'|_{\widetilde{\mathcal{F}}_t}$ for every $t\geq0$ if and
only if $\mathbb{P}_0\sim \mathbb{P}'_0$ and the following
equalities hold:
$$
\gamma\sigma^2=\gamma'\sigma'^2,\quad p=p',\quad
C\sigma^{2p}=C'\sigma'^{2p}.
$$
\end{corollary}
Thus, if we have SubOU processes with tempered stable subordinators with drift under both the physical and the risk-neutral measure, the $p$ parameter $p$ must remain the same under both measures, $C$ and $C'$ are related by $C'=C(\sigma/\sigma')^{2p}$, the subordinator drifts and the OU volatilities are related by $\gamma'\sigma'^2=\gamma\sigma^2$, and the OU drift parameters $\theta$ and $\kappa$ and $\theta'$ and $\kappa'$ can be arbitrarily changed.

For other examples of L\'{e}vy densities, where, e.g.,
$\ell(x)=(\ln(1+x))^\alpha$, one can also use Proposition
\ref{prop:SubBMComputeAsymptoticGen}. See
\cite{SongVondracek} section 2 for examples of subordinators and
section 3 for the asymptotics of the L\'{e}vy density of subordinate
Brownian motions. Replace $4$ in their formulas by
$2\sigma^2$ to coincide with our notation. Once the asymptotics of the L\'{e}vy density is
determined, the Hellinger condition can be reduced to a simple relationship for the parameters similar to Corollary 2.2 for SubOU processes with tempered stable L\'{e}vy densities.

We are also interested in the following question: if under some
measure $\mathbb{P}$ the semimartingale $X$ is a SubOU process with
generating tuple $(\kappa,\theta,\sigma,\gamma,\nu)$, characterize
all measures $\mathbb{P'}$ locally equivalent to $\mathbb{P}$. In
particular, we are interested in conditions on the semimartingale
characteristics of $X$ under $\mathbb{P'}$. The following result
answers this question.

\begin{thm}\label{thm:SubOUGenMeasureChange}
Let $\mathbb{P}$ and $\mathbb{P}'$ be two probability measures on
$(\Omega,\widetilde{\mathcal{F}},(\widetilde{\mathcal{F}}_t)_{t\geq0})$
with initial distributions $\mathbb{P}_0$ and $\mathbb{P}'_0$,
respectively. Suppose under $\mathbb{P}$, the canonical process $X$
is a SubOU process with generating tuple
$(\kappa,\theta,\sigma,\gamma,\nu)$ and local characteristics
$(B,C,\Pi)$. Suppose under $\mathbb{P}'$, $X$ is a semimartingale
with local characteristics $(B',C',\Pi')$. If $\mathbb{P}'$ and
$\mathbb{P}$ are locally equivalent, then there exists a nonnegative
predictable function $Y(\omega,t,x)$ and a predictable process
$\beta$ such that:
\begin{gather}
B'_t(\omega)=B_t(\omega)+\gamma\sigma^2\int_0^t\beta_s(\omega)ds+\int_{[0,t]\times\mathbb{R}}y{1}_{\{|y|\leq1\}}(Y(s,\omega,y)-1)\pi(X_{s-}(\omega),y)dyds,\notag\\
C'_t(\omega)=\gamma\sigma^2t, \
\Pi'(\omega,ds,dy)=Y(\omega,s,y)\Pi(\omega,ds,dy),\label{eq:SubOUGenMCCharacteristicsRelations}
\end{gather}
\begin{equation}\label{eq:SubOUGenMCBeta}
\int_0^t|\beta_s(\omega)|ds<\infty,\
\int_0^t\beta_s^2(\omega)ds<\infty,\
\end{equation}
\begin{equation}\label{eq:SubOUGenMCRandomMeasure}
\int_0^t\int_{y\neq0}|y{1}_{\{|y|\leq1\}}(Y(s,\omega,y)-1)|\pi(X_{s-}(\omega),y)dyds<\infty,
\end{equation}
\begin{equation}\label{eq:SubOUGenMCHellinger}
\int_0^t\int_{y\neq0}(\sqrt{Y(s,\omega,y)}-1)^2\pi(X_{s-}(\omega),y)dyds<\infty,\quad\text{(Hellinger
condition)}
\end{equation}
$\mathbb{P}'$ and $\mathbb{P}$-a.s. for all $t\geq0$. Define
$N=\beta\cdot X^c+(Y-1)*(\mu^X-\Pi)$. Then the density process $Z$
of $\mathbb{P}'$ w.r.t. $\mathbb{P}$ is the Doleans-Dade stochastic
exponential $\mathscr{E}(N)$ of $N$.
\end{thm}

\subsection{The Spectral Representation of the SubOU Semigroup}

The OU and SubOU processes are stationary with the Gaussian
stationary density
$$
\mathfrak{m}(x)=\sqrt{ \frac{\kappa}{\pi \sigma^2}}e^{-\frac{\kappa(\theta-x)^2}{\sigma^2}}.
$$
Consider the Hilbert space $L^2(\mathbb{R},\mathfrak{m})$ with the
inner product  $(f,g)=\int_{\mathbb{R}}f(x)g(x)\mathfrak{m}(x)dx$,
and denote by $\|\cdot\|$ the $L^2$-norm. The OU and SubOU
semigroups are both symmetric semigroups in
$L^2(\mathbb{R},\mathfrak{m})$, i.e.
$(\mathcal{P}_tf,g)=(f,\mathcal{P}_tg)$ and
$(\mathcal{P}^\phi_tf,g)=(f,\mathcal{P}^\phi_tg)$ for any $f,g\in
L^2(\mathbb{R},\mathfrak{m})$. Their spectral decompositions in
$L^2(\mathbb{R},\mathfrak{m})$ are available in closed form.

\begin{thm}\label{thm:spectralexpansion}
(1) The OU semigroup has the following eigenfunction expansion in ${L}^2(\mathbb{R},\mathfrak{m})$:
\begin{equation}\label{eq:OUsp}
\mathcal{P}_tf(x)=\sum_{n=0}^{\infty}e^{-\kappa
nt}f_n\varphi_n(x),\quad f\in {L}^2(\mathbb{R},\mathfrak{m}),\quad
t\geq 0,
\end{equation}
with the orthonormal eigenfunctions expressed in terms of Hermite
polynomials (see, e.g., Lebedev (1965))
\begin{equation}
\varphi_n(x)=\frac{1}{\sqrt{2^n
n!}}H_n\left(\frac{\sqrt{\kappa}}{\sigma}(x-\theta)\right),\quad
n=0,1,\cdots,
\end{equation}
and expansion coefficients $f_n=(f,\varphi_n)$.\\
\noindent(2) The SubOU semigroup has the following eigenfunction expansion in ${L}^2(\mathbb{R},\mathfrak{m})$:
\begin{equation}\label{eq:SubOUsp}
\mathcal{P}^{\phi}_tf(x)=\sum_{n=0}^{\infty}e^{-\phi(\kappa
n)t}f_n\varphi_n(x),\quad f\in {L}^2(\mathbb{R},\mathfrak{m}),\quad
t\geq 0
\end{equation}
with the same eigenfunctions and expansion coefficients as the OU
semigroup.
\end{thm}

General results for the spectral representation of one-dimensional
diffusions go back to the fundamental work of \cite{McKean}. For each $t$, the OU
transition semigroup operator $\mathcal{P}_t$ has a purely discrete
spectrum with eigenvalues $\{e^{-\kappa n t}, n=0,1,...\}$. The
explicit form of the eigenfunction expansion of the OU semigroup in
terms of Hermite polynomials is well known and can be found in many
references, including \cite{Wong}, \cite{KarlinTaylor},
\cite{Schoutens}, \cite{BakryMazet}, \cite{AlbeverioRudiger03},
\cite{AlbeverioRudiger05}, and \cite{GorovoiLinetsky} p.62.
The general spectral representation of the
transition semigroup of a symmetric Markov process can be found in
\cite{FukushimaOshimaTakeda}. Bochner subordination
replaces the eigenvalues $e^{-\lambda_n t}$ with
$e^{-\phi(\lambda_n)t}$, where $\phi$ is the Laplace exponent of the
subordinator, while the eigenfunctions remain the same. Thus the
eigenvalues of the SubOU semigroup operator $\mathcal{P}_t^\phi$ are
$\{e^{-\phi(\kappa n) t}, n=0,1,...\}$ with the same eigenfunctions.
The general spectral representation of the semigroup of a
subordinate symmetric Markov process can be found in \cite{Okura} and in
\cite{AlbeverioRudiger03} and \cite{AlbeverioRudiger05}, where
subordinate OU processes and their semigroups are studied in the
general setting of symmetric Markov processes. Applications in
finance can be found in \cite{Linetsky}, \cite{MendozaCarrLinetsky}
and \cite{MendozaLinetsky}.

For $t\geq 0$ the eigenfunction expansions on the RHS of
\eqref{eq:OUsp} and \eqref{eq:SubOUsp} for the OU and the SubOU
semigroup converge to $\mathcal{P}_tf$ and $\mathcal{P}_t^{\phi}f$
in the $L^2$-norm for any $f\in L^2(\mathbb{R},\mathfrak{m})$. In financial applications, we are interested in pointwise convergence, as we need to compute values at specific levels of the underlying variable.
For $t>0$ pointwise convergence results are available for OU and SubOU semigroups.
\begin{thm}\label{thm:PointwiseConvergenceHermiteExpansion}(1) The eigenfunction expansion (2.11) converges to ${\cal P}_tf(x)$ pointwise in $x$ for each $t>0$ and each $f\in L^2(\mathbb{R},\mathfrak{m})$.\\
\noindent(2) If either of the following condition is satisfied: (i)
$f(x)=\sum_{n=0}^\infty f_n\varphi_n(x)$ converges absolutely for
all $x\in\mathbb{R}$, or (ii) $\sum_{n=1}^{\infty}e^{-\phi(\kappa
n)t}n^{-1/4}<\infty$ for all $t>0$, then the eigenfunction expansion
(2.13) converges to ${\cal P}_t^\phi f(x)$ pointwise for all
$x\in\mathbb{R}$ for each $t>0$ and each $f\in
L^2(\mathbb{R},\mathfrak{m})$.
\end{thm}
The eigenfunction expansion (2.11) for the OU semigroup converges
pointwise without any further conditions for each $t>0$ and $f\in
L^2(\mathbb{R},\mathfrak{m})$. The eigenfunction (2.13) for the
SubOU semigroup converges pointwise for each $t>0$ and $f\in
L^2(\mathbb{R},\mathfrak{m})$ under the mild sufficient condition on
the Laplace exponent of the subordinator in (2) of Theorem 2.7. In
practice this condition is satisfied for all subordinators with
drift $\gamma>0$ due to the factor $e^{-\gamma \kappa t}$. In the
pure jump case $\gamma=0$, it is satisfied for all tempered stable
subordinators with $p>0$. Furthermore, for subordinators for which
it is not satisfied, while the eigenfunction expansion (2.13) is not
guaranteed to converge pointwise for each $t>0$ and each $f\in
L^2(\mathbb{R},\mathfrak{m})$, it may converge pointwise for some
$t>0$ and some functions $f$, depending on the rate of decay of the
coefficients $f_n$ as $n$ increases.

We also have the following expansions for OU and SubOU transition densities.
\begin{thm}
(1) The OU transition density (2.1) has the eigenfunction expansion
\begin{equation}\label{eq:OUpdfsp}
p(t,x,y)=\mathfrak{m}(y)\sum_{n=0}^{\infty}e^{-\kappa
nt}\varphi_n(x)\varphi_n(y)
\end{equation}
converging for all $t>0$ uniformly in $x,y$ on compacts.\\
\noindent(2) If the Laplace exponent of the subordinator satisfies
$\sum_{n=1}^\infty e^{-\phi(\kappa n)t}n^{-\frac{1}{2}}<\infty$ for
all $t>0$, the SubOU transition density has the eigenfunction
expansion
\begin{equation}\label{eq:SubOUpdfsp}
p^\phi(t,x,y)=\mathfrak{m}(y)\sum_{n=0}^{\infty}e^{-\phi(\kappa
n)t}\varphi_n(x)\varphi_n(y)
\end{equation}
converging for all $t>0$ uniformly in $x,y$ on compacts.
\end{thm}

In the numerical implementation one needs to truncate eigenfunction
expansions after a finite number of terms. Truncation error bounds
of the expansion (2.15) in the $L^2$ and the pointwise sense can be
easily derived. Here we present the pointwise error bound, as it is
of most interest in finance. $L^2$ bounds can be derived similarly.
\begin{thm}
Suppose that the Laplace exponent of the subordinator satisfies
$\sum_{n=0}^{\infty}e^{-\phi(\kappa n)t}<\infty$ for all $t>0$. Then
for any $f\in L^2(\mathbb{R},\mathfrak{m})$, the truncation error
has the following bound:
$$
\left|\sum_{n=M}^{\infty}e^{-\phi(\kappa
n)t}f_n\varphi_n(x)\right|\leq 1.0864\|f\|
e^{\frac{\kappa(x-\theta)^2}{2\sigma^2}}\sum_{n=M}^{\infty}e^{-\phi(\kappa
n)t}.
$$
\end{thm}
If $\gamma>0$, we can derive a particularly simple pointwise truncation error estimate:
$$
\left|\sum_{n=M}^{\infty}e^{-\phi(\kappa
n)t}f_n\varphi_n(x)\right|\leq
1.0864\|f\|e^{\frac{\kappa(x-\theta)^2}{2\sigma^2}}\frac{e^{-\gamma\kappa
M t}}{1-e^{-\gamma\kappa t}}.
$$
From these estimates it is clear that the convergence rate is
governed by the OU mean reversion rate $\kappa$ and time to maturity
$t$, as well as the Laplace exponent of the subordinator. The
greater the $\kappa$ and the longer the time to maturity, the faster
the convergence. In particular, if $\gamma>0$, the convergence is
exponential. In the pure jump case $\gamma=0$ with tempered stable
subordinators with $p>0$, the truncation error can similarly be
shown to be $O(e^{tC\Gamma(-p)(\kappa M)^p})$ with $\Gamma(-p)<0$
and $C>0$. We note that these error bounds are conservative since
they rely on the estimate $|f_n|\leq \|f\|$. Depending on the
properties of $f$, the coefficients $f_n$ may converge to zero at a
fast rate, resulting in faster convergence than is implied by these
estimates.

\section{Commodity Models With Mean-Reverting Jumps}

\subsection{Futures Dynamics}

We start with
$(\Omega,\widetilde{\mathcal{F}},(\widetilde{\mathcal{F}}_t)_{t\geq0})$
as in section 2.3 endowed with a probability measure $\mathbb{Q}$
and assume that, under $\mathbb{Q}$, the canonical process $X$ is a
SubOU process with generating tuple
$(\kappa,\theta,\sigma,\gamma,\nu)$ and starting point $X_0=x_0\in
{\mathbb R}$. Let $\{F(0,t),t\geq0\}$ be the initial futures curve
(a given deterministic function of time). We take $\mathbb{Q}$ to be
the risk-neutral pricing measure chosen by the market and model the
commodity spot price $S_t$ under $\mathbb{Q}$ as the (scaled)
exponential of the SubOU process $X$:
\begin{equation}\label{eq:SubOUSpotPriceRiskNeutralMeasure}
S_t=F(0,t)e^{X_t-G(t)}.
\end{equation}
The function $G(t)$ is selected so that the expectation of the spot
price under $\mathbb{Q}$ is equal to the initial futures price,
$\mathbb{E}^{\mathbb{Q}}[S_t]=F(0,t)$, which implies $G(t)=\ln
\mathbb{E}^{\mathbb{Q}}[e^{X_t}]$.

To compute futures price dynamics, we need the following.
\begin{lemma}\label{lemma:expExpansion}
The expansion of the exponential function in the eigenfunction basis
(2.13) reads:
\begin{equation}
e^x=\sum_{n=0}^{\infty}f_n\varphi_n(x),\quad
f_n=e^{\theta+\frac{\sigma^2}{4\kappa}}\frac{1}{\sqrt{n!}}\left(\frac{\sigma}{\sqrt{2\kappa}}\right)^n.
\end{equation}
The expansion converges absolutely for each $x\in {\mathbb R}$.
\end{lemma}

We now compute the futures price process
$\{F(s,t)=\mathbb{E}^{\mathbb{Q}}[S_t|\widetilde{\mathcal{F}}_s],s\in[0,t]\}$
for each fixed maturity $t\geq0$ using Lemma
\ref{lemma:expExpansion}.
\begin{thm}\label{thm:SubOUfuturepricing}
(1) The function $G(t)$ in the model (3.1) is given by:
\begin{equation}\label{eq:LSOUspexp}
e^{G(t)}=\mathbb{E}^{\mathbb{Q}}[e^{X_{t}}]=\sum_{n=0}^{\infty}e^{-\phi(\kappa
n)t}f_n\varphi_n(x_0),
\end{equation}
where $f_n$ are given in (3.2), and the expansion converges
absolutely for each $x_0\in {\mathbb R}$, all $t\geq 0$ and any
Laplace exponent $\phi$.

\noindent(2) For each fixed maturity time $t>0$, the futures price
$F(s,t)$ is a martingale on $[0,t]$ given by:
\begin{equation}\label{eq:SubOUfuturesprocess}
F(s,t)=F(0,t)e^{-G(t)} \sum_{n=0}^{\infty}e^{-\phi(\kappa
n)(t-s)}f_n \varphi_n(X_s),\quad s\in [0,t].
\end{equation}
\end{thm}
At time zero, $s=0$, \eqref{eq:SubOUfuturesprocess} reduces to the
identity $F(0,t)=F(0,t)$. At maturity, $s=t$, the futures price is
equal to the spot price and \eqref{eq:SubOUfuturesprocess} reduces
to \eqref{eq:SubOUSpotPriceRiskNeutralMeasure} due to Eq.(3.2).
Eq.\eqref{eq:SubOUfuturesprocess} gives a martingale expansion for
the futures price. Note that for each $n$ the process
$\{e^{\phi(\kappa n)s}\varphi_n(X_s),s\geq 0\}$ is a martingale due
to the eigenfunction property:
$$
{\mathbb E}^{\mathbb{Q}}[\varphi_n(X_s)|X_t]=e^{-\phi(\kappa
n)(s-t)} \varphi_n(X_t).
$$
Thus, Eq.(3.4) represents the futures price process as an expansion
in martingales associated with the eigenfunctions of the SubOU
semigroup.

Since the process $X$ can be expressed in terms of the spot price
process $S$ and the initial futures curve by inverting (3.1),
\begin{equation}\label{eq:SubOUInTermsOfSpot}
X_s=\ln\left(S_s/F(0,s))\right)+G(s),
\end{equation}
Eq.\eqref{eq:SubOUfuturesprocess} expresses the dynamics of the
futures price in terms of the spot price dynamics and the initial
futures curve. Alternatively, we can view
Eq.\eqref{eq:SubOUfuturesprocess} as the process for the futures
price driven by the SubOU process $X$ without any reference to the spot
price $S$. In this interpretation, our model can be viewed as the
model for the evolution of the futures curve, rather than the spot
price model. Eq.\eqref{eq:SubOUfuturesprocess} directly defines the
martingale futures dynamics. The spot dynamics (3.1) then follows as
the limiting case.

\begin{remark} {\bf The Case without Time Change.} When $X_t$ is an OU rather than SubOU process, our model reduces to
the standard exponential OU model:
$$
S_t=F(0,t)e^{X_t-x_0e^{-\kappa t}-\theta(1-e^{-\kappa
t})-\frac{\sigma^2}{4\kappa}(1-e^{-2\kappa t})}.
$$
By applying It\^{o}'s formula, we obtain the spot price SDE:
$dS_t=\kappa(\Theta(t)-\ln{S_t})S_t dt+\sigma S_t dB_t$ with
$\Theta(t)=\frac{1}{\kappa}\left(\frac{d}{dt}\ln{F(0,t)}+\frac{\sigma^2}{4\kappa}(1-e^{-2\kappa
t})\right)+\ln{F(0,t)}$. This is essentially the same SDE as the
Model 1 in \cite{Schwartz97} but with the long run level $\Theta(t)$
taken to be a deterministic function of time completely determined
by the initial futures curve. Using the generating function of
Hermite polynomials (\cite{Lebedev} p.60),
$\sum_{n=0}^{\infty}\frac{w^n}{n!}H_n(z)=e^{2zw-w^2}$, when $X_t$ is
an OU process (i.e., $\phi(\lambda)=\lambda$),
Eq.\eqref{eq:SubOUfuturesprocess} reduces to:
\begin{align*}\label{eq:futuresprocessNoTimeChange}
F(s,t)&=F(0,t)\exp\left\{X_se^{-\kappa(t-s)}-x_0e^{-\kappa t}-\theta(e^{-\kappa(t-s)}-e^{-\kappa t})-\frac{\sigma^2}{4\kappa}(e^{-2\kappa(t-s)}-e^{-2\kappa t})\right\}\\
&=F(0,t)\left(\frac{S_s}{F(0,s)}\right)^{\exp\{-\kappa(t-s)\}}
\exp\left\{-\frac{\sigma^2}{4\kappa}e^{-\kappa t}(e^{2\kappa
s}-1)(e^{-\kappa t}-e^{-\kappa s})\right\}.
\end{align*}
This expression for the futures price dynamics in terms of the
initial futures curve and the spot price dynamics in the OU model
can be found in Clewlow and Strickland (1999), Eq.(2.5). Using
It\^{o}'s formula, one can show that
\begin{equation}\label{eq:SDEFutPriceNoTimeChange}
dF(s,t)=\sigma e^{-\kappa(t-s)}F(s,t)dB_s,\quad s\in[0,t].
\end{equation}
\end{remark}


We now discuss futures dynamics under the physical measure
$\mathbb{P}$. The form for the futures process is still given by
\eqref{eq:SubOUfuturesprocess}. However, the law of $X$ changes
under an equivalent measure change. Let
$(\overline{B}^P,\overline{C}^P,\overline{\Pi}^P)$ be the
semimartingale characteristics of $X$ under $\mathbb{P}$. Theorem
\ref{thm:SubOUGenMeasureChange} gives the general conditions on the
semimartingale characteristics of
$(\overline{B}^P,\overline{C}^P,\overline{\Pi}^P)$. Any
semimartingale satisfying these conditions can be chosen as a
candidate driver for the commodity model under  $\mathbb{P}$ that
leads to the model driven by the given SubOU process with generating
tuple $(\kappa,\theta,\sigma,\gamma,\nu)$ under $\mathbb{Q}$. In
order to retain analytical tractability under $\mathbb{P}$, we are
interested in equivalent measure transformations that transform a
given SubOU process into another SubOU process plus possibly a
deterministic function of time. Using Theorem \ref{thm:EMCSubOU} and
Theorem \ref{thm:SubOUGenMeasureChange}, we obtain the following
result.
\begin{thm}\label{thm:EMCSubOUvsSubOUDrift}
Consider the canonical process $X$ on
$(\Omega,\widetilde{F},(\widetilde{F}_t)_{t\geq0})$. Suppose under
measure $\mathbb{Q}$ with $\mathbb{Q}(X_0=x_0)=1$ the canonical
process $X$ is a SubOU process with generating tuple
$(\kappa,\theta,\sigma,\gamma,\nu)$ and under measure $\mathbb{P}$
with $\mathbb{P}(X_0=x_0)=1$ it is a SubOU process with generating
tuple $(\kappa_P,\theta_P,\sigma_P,\gamma_P,\nu_P)$ plus a
deterministic function $H(t)$. Then $\mathbb{Q}$ and $\mathbb{P}$
are locally equivalent if and only if:
\begin{itemize}
\item[(1)] $H$ is absolutely continuous with $H(0)=0$ if $\gamma>0$, and $H(t)=0$
for all $t$ if $\gamma=0$.
\item[(2)] $\gamma_P\sigma_P^2=\gamma\sigma^2$.
\item[(3)] the Hellinger condition $\int_{y\neq0}\big(\sqrt{\pi^P(x,y)}-\sqrt{\pi(x,y)}\big)^2dy<\infty$ is satisfied.
\end{itemize}
\end{thm}
The Hellinger condition (3) can be simplified using Proposition
\ref{prop:SubBMComputeAsymptoticGen}. For example, in the case where
the L\'{e}vy measures $\nu^P$ and $\nu^Q$ are both those of tempered
stable subordinators, the Hellinger condition (3) reduces to the
conditions presented in Corollary \ref{cor:EMC Tempered Stable}.

If $X$ under $\mathbb{P}$ is specified to be a SubOU process plus
some deterministic drift given by the function $H(t)$, the model
parameters can be estimated from the time series of futures prices
by filtering methods. In this case the transition density of the
underlying SubOU process is known explicitly and given by
\eqref{eq:SubOUpdfsp}. The pure OU diffusion based model has been
estimated by \cite{Schwartz97}. In that case, the noise term is
Gaussian and the standard Kalman filter can be used. In our SubOU
case, the noise term for the transition equation is not Gaussian,
and the particle filter algorithm (or the extended particle filter
or the unscented particle filter) can be used since we know the
transition density of $X$ in closed form (see \cite{Haykin} and
\cite{JavaheriLautierGalli}).

\subsection{The Maturity Effect}

The \emph{maturity effect} (also known as the \emph{Samuelson
hypothesis}, see \cite{Samuelson}) in the commodities futures
markets is the well-known increase in commodity futures price
volatility as the futures contract approaches maturity. The maturity
effect implies that long term futures are less volatile than short
term futures, and is well documented in the empirical literature
(see \cite{BSCC}, \cite{KalevDuong} and references therein). The
maturity effect is obviously present in the pure OU model
\eqref{eq:SDEFutPriceNoTimeChange}, where futures volatility $\sigma
e^{-\kappa \tau}$ decays exponentially as time to maturity
$\tau=t-s$ increases, with mean reversion rate controlling
the rate of decay. Here we investigate the maturity effect in our
SubOU model.

We start with characterizing futures volatililty in the general
semimartingale setting. For a futures contract with maturity time
$t$, define $r^t_s=\ln{\frac{F(s,t)}{F(0,t)}},$ $s\in [0,t]$, the
cumulative continuously compounded return process over the time
interval $[0,s]$ with $s\leq t$. Since $F(s,t)$ is a semimartingale,
$r^t$ is also a semimartingale. We measure volatility of the futures
return process $r^t$ experienced over the time interval $[0,s]$ by
its quadratic variation (QV) $[r^t,r^t]_s$ (the square-bracket
process). This definition of volatility has been widely used in the
econometric literature (see \cite{AndersenBollerslevDiebold}). With
this definition, the maturity effect can be mathematically defined
as follows.
\begin{definition}
{\em A futures model is said to exhibit the maturity effect almost surely
if}
$$
\mathbb{P}([r^{t_1},r^{t_1}]_s>[r^{t_2},r^{t_2}]_s)=1\quad \text{for
any}\quad 0<s<t_1<t_2.
$$
\end{definition}

\begin{remark}\label{remark:EquivalenceQVPhysicalRiskNeutral}
If $\mathbb{P}$ and $\mathbb{Q}$ are locally equivalent, then the QV
of a semimartingale under $\mathbb{P}$ is a version of the QV under
$\mathbb{Q}$ (\cite{JacodShiryaev} Theorem \Rmnum{3}.3.13).  Hence,
if the maturity effect is present in the futures dynamics under the
physical measure, it is also present under the risk-neutral measure.
We will compute $[r^t,r^t]_s$ under $\mathbb{Q}$.
\end{remark}

\begin{remark}
In the pure OU model \eqref{eq:SDEFutPriceNoTimeChange} the QV of
futures return process is a deterministic function
$[r^t,r^t]_s=\frac{\sigma^2}{2\kappa}e^{-2\kappa t}(e^{2\kappa
s}-1)$ decreasing in $t$ for each fixed $s$, $0<s<t$.
\end{remark}

Note that $[r^t,r^t]_s=[r^{tc},r^{tc}]_s+\sum_{u\leq s}(\Delta
r^t_u)^2$, where $r^{tc}$ denotes the continuous martingale part of
the process $r^{t}$. From Eq.\eqref{eq:SubOUfuturesprocess},
$r_s^t=-G(t)+\theta+\frac{\sigma^2}{4\kappa}+\ln{g(X_s,s,t)},$ where
the function $g$ is:
$$
g(x,s,t):=\sum_{n=0}^{\infty}e^{-\phi(\kappa
n)(t-s)}\Big(\frac{\sigma}{2\sqrt{\kappa}}\Big)^n\frac{1}{n!}H_n\Big(\frac{\sqrt{\kappa}}{\sigma}(x-\theta)\Big).
$$
Since we know the semimartingale characteristics of the SubOU
process $X$, from \cite{Kallsen} Proposition 2.5 we know that
$[r^{tc},r^{tc}]_s=\gamma\sigma^2\int_0^s\Big(\frac{\partial\ln{g(X_{u-},u,t)}}{\partial{x}}\Big)^2du=\gamma\sigma^2\int_0^s\Big(\frac{g_x(X_{u-},u,t)}{g(X_{u-},u,t)}\Big)^2du$ and
$(\Delta
r^t_u)^2=(r^t_u-r^t_{u-})^2=(\ln{g(X_{u},u,t)}-\ln{g(X_{u-},u-,t)})^2
=\Big(\int_{X_{u}\wedge X_{u-}}^{X_{u}\vee
X_{u-}}\frac{g_x(x,u,t)}{g(x,u,t)}dx\Big)^2.$
Therefore,
$[r^t,r^t]_s=\gamma\sigma^2\int_0^s\Big(\frac{g_x(X_{u-},u,t)}{g(X_{u-},u,t)}\Big)^2du+\sum_{u\leq
s}\Big(\int_{X_{u}\wedge X_{u-}}^{X_{u}\vee
X_{u-}}\frac{g_x(x,u,t)}{g(x,u,t)}dx\Big)^2.$
Note that $g(x,u,t)>0$ and $g_x(x,u,t)>0$ for each $x$, and
$g(x,u,t)$ depends only on $t-u$. We thus have the following result.
\begin{thm}\label{thm:SubOUSHSufficientConditions}
If $g_x(x,0,t)/g(x,0,t)$ is decreasing in $t$ for each $x$,
then the maturity effect holds in the SubOU model.
\end{thm}
While the condition in Theorem 3.3 is hard to check analytically since the
function $g$ is given by the Hermite expansion, it can be easily
checked numerically. We carried out extensive numerical testing for
a wide range of parameter scenarios in pure jump ($\gamma=0$) and
jump-diffusion ($\gamma>0$) cases and verified that it was indeed
satisfied in all the cases. We thus conjecture that the condition in
Theorem 3.3 is satisfied, and the maturity effect holds for our
SubOU models.

\begin{figure}[h]
\centering
\includegraphics[scale=0.3]{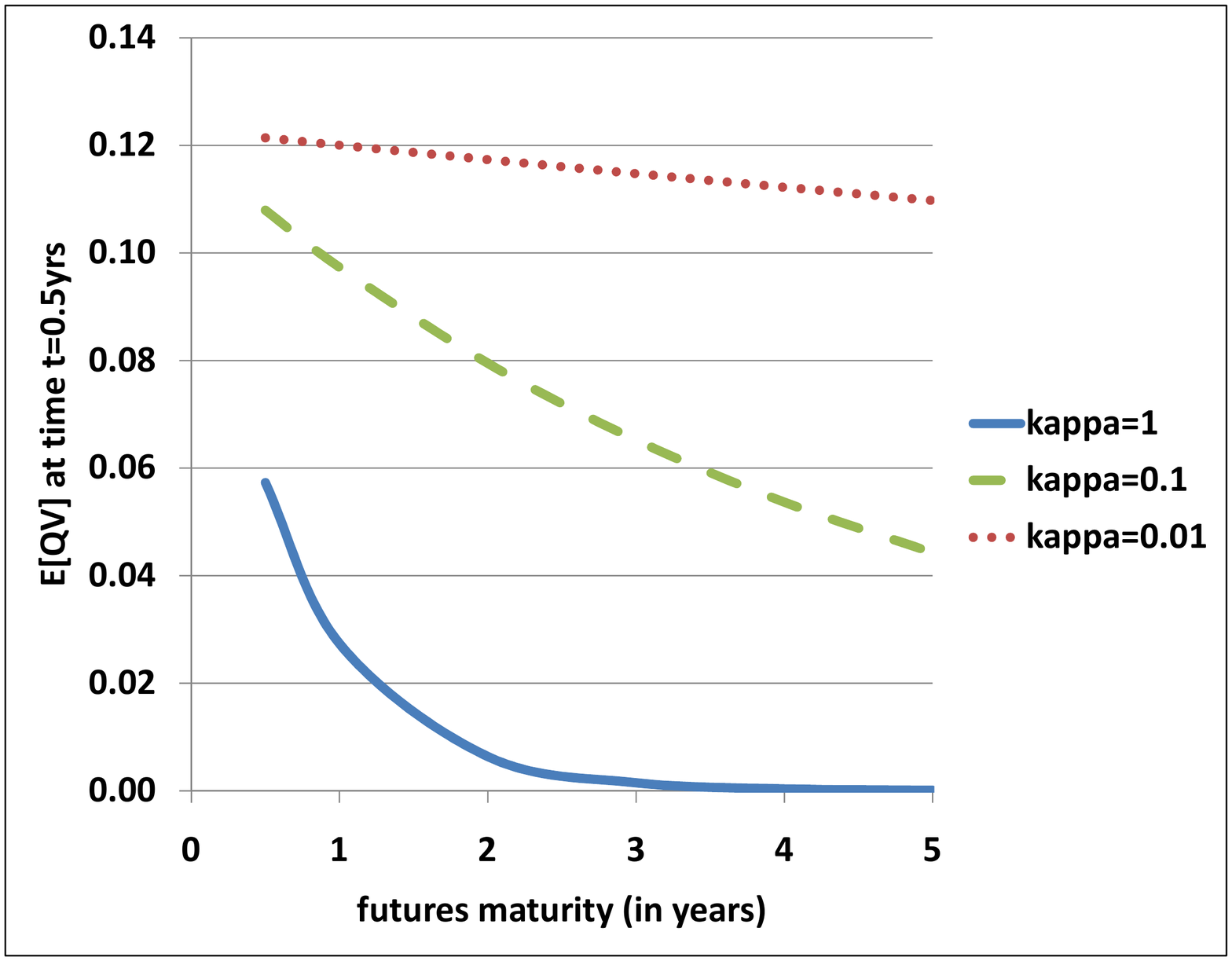}
\caption{$\mathbb{E}^{\mathbb{Q}}\{[r^t,r^t]_{0.5}\}$ as a function
of futures maturity $t$ for a SubOU Process with Inverse Gaussian
subordinator (with parameters $\kappa=0.01, 0.1, 1$, $\theta=0$,
$\sigma=0.5$, $\gamma=0$, mean rate $\mu=1$, variance rate
$\nu=1$).}\label{fig:EQVComparison}
\end{figure}

Figure \ref{fig:EQVComparison} illustrates the maturity effect as
follows. We simulated 10,000 sample paths on the time interval
$[0,1/2]$ of pure jump ($\gamma=0$) SubOU processes $X$ with
parameters $\theta=0$, $\sigma=0.5$, with the Inverse Gaussian
subordinator with  mean rate $\mu=1$ and variance rate $\nu=1$, and
with $\kappa=0.01, 0.1,$ and $1$. We then constructed 10,000 sample
paths of the futures price processes with maturities 1/2, 1, 2, 3, 4
and 5 years for each of the underlying SubOU processes using the
model relationship \eqref{eq:SubOUfuturesprocess} under
$\mathbb{Q}$, estimated realized quadratic variations of futures
returns on each sample path (the quadratic variation is the same
under $\mathbb{P}$ and under $\mathbb{Q}$), and verified that
$[r^{t_1},r^{t_1}]_{0.5}
>[r^{t_2},r^{t_2}]_{0.5}$ for $t_1<t_2$ on each sample path. Figure
\ref{fig:EQVComparison} plots the estimated mean of the quadratic
variation of futures returns as functions of futures contract
maturity for the three values of the rate of mean reversion
$\kappa=0.01, 0.1,$ and $1$.  The maturity effect is clearly seen in
the plot. As in the pure diffusion OU model,  $\kappa$ controls the maturity effect in pure jump and
jump-diffusion SubOU models.

\begin{figure}
\centering
\subfigure[the SubOU
process]{\includegraphics[scale=0.25]{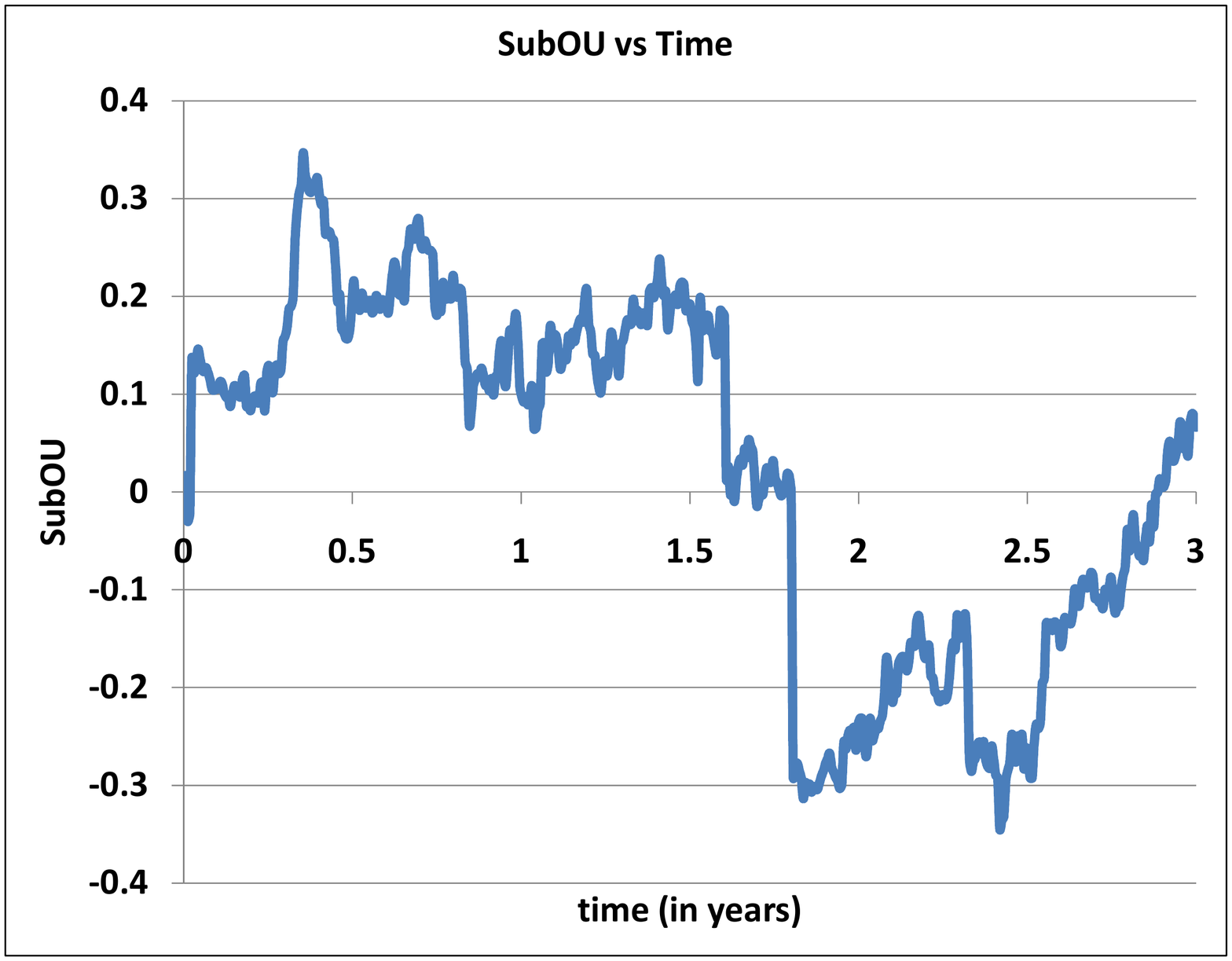}}\hspace{-1cm}
\subfigure[the futures
process]{\includegraphics[scale=0.25]{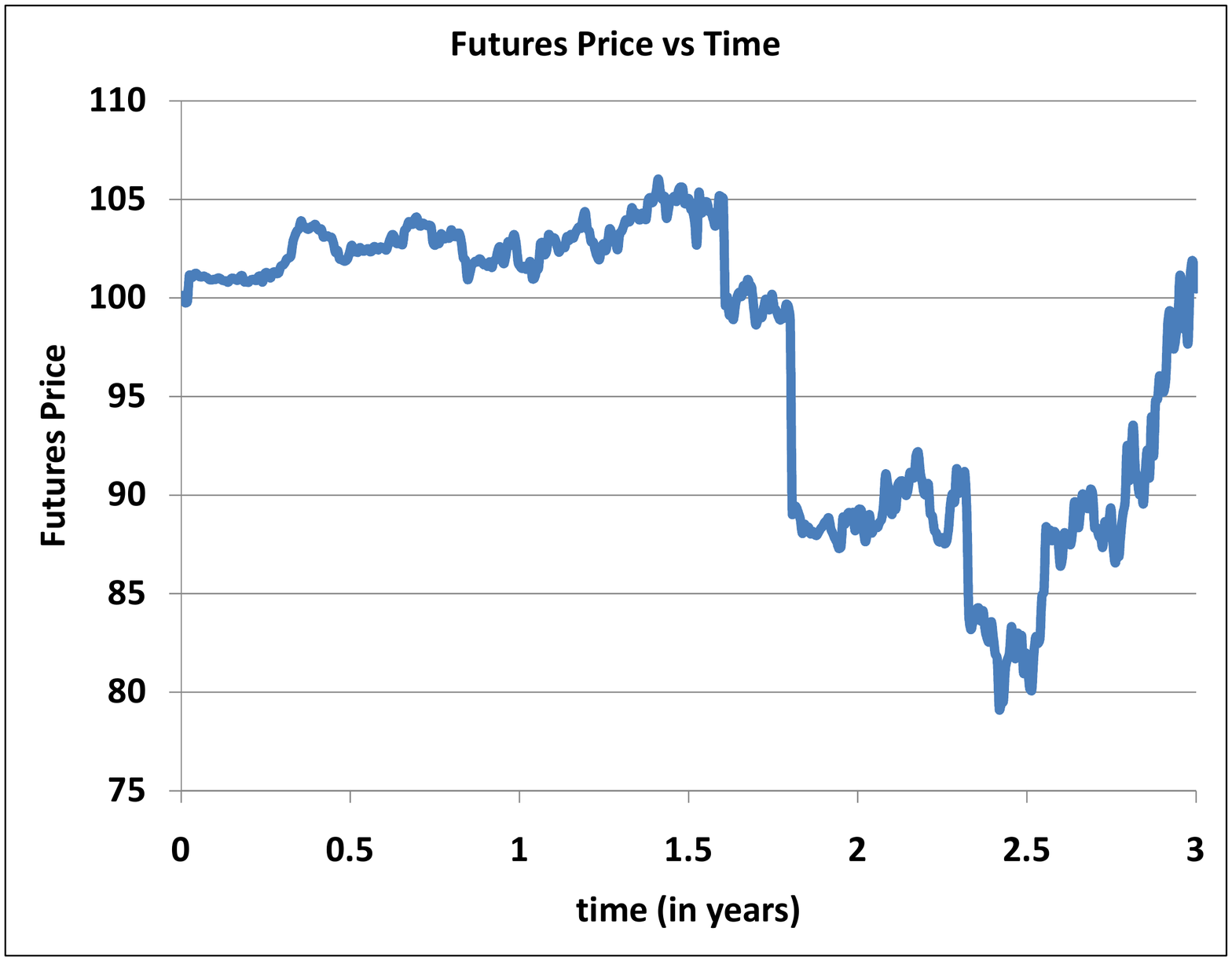}} \caption{A
Sample Path of a SubOU Process and the corresponding futures price
process with the Inverse Gaussian subordinator (with parameters
$\kappa=1$, $\theta=0$, $\sigma=0.5$, $\gamma=0.1$, mean rate
$\mu=1$, variance rate $\nu=1$).}\label{fig:SubOUFutVsTime}
\end{figure}

To further illustrate, Figure \ref{fig:SubOUFutVsTime} plots a
sample path of the driving SubOU process in the jump-diffusion case
and the corresponding futures price process with 3 years to maturity
at time zero. The maturity effect is clearly seen in the sample path
dynamics, as the futures price experiences low realized volatility
far away from maturity, and the realized volatility substantially
increases as the futures contract approaches maturity.

\begin{remark}\label{remark:WhyModelMeanReversionUnderRiskNeutralMeasure}
It is important to note that the rate of mean reversion $\kappa$
that enters the expression for the diffusion volatility in the pure
diffusion OU case \eqref{eq:SDEFutPriceNoTimeChange} and in the
quadratic variation process through the functional form
\eqref{eq:SubOUfuturesprocess} of the dependence of the futures
price on the SubOU process in the SubOU model is the {\em rate of
mean reversion under the risk-neutral pricing measure} $\mathbb{Q}$.
It is the {\em risk-neutral rate of mean reversion} that controls
the maturity effect. That is, the presence of the maturity effect in
the futures time series under the physical measure $\mathbb{P}$ is
governed by the rate of mean reversion under the pricing measure. If
there is no mean reversion under the pricing measure $\mathbb{Q}$,
i.e., $X$ is taken to be a subordinate Brownian motion under
$\mathbb{Q}$ rather than a subordinate OU process, there is no
maturity effect under $\mathbb{P}$. Thus, the presence of the
maturity effect under $\mathbb{P}$ requires $X$ to be a SubOU
process under $\mathbb{Q}$, as futures models built on subordinate
Brownian motions (L\'{e}vy processes) do not possess the maturity
effect. In contrast, SubOU models are capable of modeling the
maturity effect.
\end{remark}




\subsection{Futures Options Pricing}

We consider pricing European put and call options on a futures
contract. Suppose the strike price is $K$. The underlying futures
contract matures at time $t^*$ and the option expires at $t<t^*$.
The time $\tau=t^*-t$ varies across commodities, ranging from
several days for natural gas to one month for gold.

Here we only consider pricing the put option. The call option price
is given by the put-call parity. Alternatively, a similar
eigenfunction expansion can be obtained for the call pricing
function, and the put-call parity can be verified directly. The put
payoff at expiration $t$ is $(K-F(t,t^*))^+,$ where $F(t,t^*)$ is
the $t^*$-maturity futures price at time $t$. In our model it is
related to $X_t$ by \eqref{eq:SubOUfuturesprocess}. It is convenient
to write the payoff function as follows:
$$(K-F(x,t,t^*))^+=(K-F(x,t,t^*)){\bf 1}_{\{x<x^*\}},$$
where $x^*$ is the unique solution of the equation $F(x,t,t^*)=K$,
and $F(x,t,t^*)$ is the $t^*$-maturity futures price at time $t$ as
a function of the state variable $X_t=x$ given by
\eqref{eq:SubOUfuturesprocess}. Since $F(x,t,t^*)$ is a strictly
increasing function of $x$, the solution to this equation is unique
and can be easily computed numerically using bisection or any other
root bracketing algorithm. To price the put option at time zero, we
thus need to first find $x^*$ corresponding to the strike price $K$
and then compute the expectation in:
$$P(t,t^*,F(0,t^*),K)=B(0,t)\mathbb{E}^{\mathbb{Q}}\Big[(K-F(X_t,t,t^*)){\bf
1}_{\{X_t<x^*\}}\Big],$$ where $B(0,t)$ is the risk-free discount
factor from the option expiration $t$ to time zero.
\begin{thm}\label{thm:SubOUOptionPricing}
Let $x^*$ be the unique solution of the equation $F(x,t,t^*)=K$ and
define $w^*:=\frac{\sqrt{\kappa}}{\sigma}(x^*-\theta)$,
$\tau:=t^*-t$, $\alpha:=\frac{\sigma}{2\sqrt{\kappa}}$ and
$F:=F(0,t^*)$. Suppose the Laplace exponent of the subordinator
satisfies $\sum_{n=1}^\infty e^{-\phi(\kappa
n)t}n^{-\frac{1}{4}}<\infty$. Then the put price has the absolutely
convergent eigenfunction expansion:
\begin{equation}\label{eq:SubOUPutPrice}
P(t,t^*,K,F)=B(0,t)\sum_{n=0}^\infty e^{-\phi(\kappa n)t}
p_n(t,t^*,w^*,F)\varphi_n(x_0),
\end{equation}
\begin{equation}\label{eq:SubOUPutExpansionCoeff}
p_n(t,t^*,w^*,F)=\frac{1}{\sqrt{\pi2^nn!}}\Big\{ K b_n(w^*)
-Fe^{\theta +\frac{\sigma^2}{4\kappa} -G(t^*)} \sum_{m=0}^\infty
e^{-\phi(\kappa m)\tau}\frac{\alpha^m}{m!}a_{n,m}(w^*)\Big\},
\end{equation}
\begin{align}
b_n(w)&=\int_{-\infty}^{w}H_n(x)e^{-x^2}dx=\begin{cases}
\sqrt{\pi}\Phi(\sqrt{2}w), &n=0,\\
-H_{n-1}(w)e^{-w^2}, &n=1,2,\cdots
\end{cases},\label{eq:bCoeff}\\
a_{n,m}(w)&=\int_{-\infty}^{w}H_m(x)H_n(x)e^{-x^2}dx=\sum_{k=0}^{min(n,m)}\binom{m}{k}\binom{n}{k}2^kk!b_{n+m-2k}(w).\label{eq:aCoeff}
\end{align}
The call price is given by the put-call parity
$C(t,t^*,K,F)=B(0,t)(F-K)+P(t,t^*,K,F)$.
\end{thm}

\begin{remark}
The option written on the spot price is obtained by setting $t=t^*$
in \eqref{eq:SubOUPutExpansionCoeff}.
\end{remark}

\begin{remark} {\bf The Case Without Time Change.} In the pure diffusion OU model,
the option pricing formulas collapse to the Black-Scholes type
formulas for the exponential OU diffusion model obtained by
\cite{ClelowStrickland}:
$$P(t,t^*,K,F)=B(0,t)\left[K\Phi(-d_{-})-F\Phi(-d_{+})\right],\
C(t,t^*,K,F)=B(0,t)\left[F\Phi(d_{+})-K\Phi(d_{-})\right],$$
$$
d_{-}=\frac{\ln{\left(\frac{F}{K}\right)}-\frac{\sigma^2}{4\kappa}e^{-2\kappa\tau}(1-e^{-2\kappa
t})}{\frac{\sigma}{\sqrt{2\kappa}}e^{-\kappa\tau}\sqrt{1-e^{-2\kappa
t}}}, \quad
d_{+}=d_{-}+\frac{\sigma}{\sqrt{2\kappa}}e^{-\kappa\tau}\sqrt{1-e^{-2\kappa
t}}.
$$
\end{remark}

\section{Stochastic Volatility and Time Inhomogeneity}

Models based on SubOU processes described in the previous section
can be calibrated to fit a variety of volatility smile patterns
observed in commodity options markets. However, they are generally
not flexible enough in order to fit the entire volatility surface
across different maturities. In this section we study a further
extension of SubOU models to introduce stochastic volatility and
time inhomogeneity, such as seasonality in options' implied
volatility typical for some commodities, such as natural gas.


We consider absolutely continuous time changes of the form
\begin{equation}\label{eq:ACTimeChange}
T_t=\int_0^t\Big(a(u)+Z_u\Big)du,
\end{equation}
where $a(t)\geq 0$ is a deterministic function of time and $Z$ is a
CIR diffusion solving the SDE
$$dZ_t=\kappa_Z(\theta_Z-Z_t)dt+\sigma_Z\sqrt{Z_t}dB_t,\ Z_0=z_0,$$
with parameters assumed to satisfy the Feller condition,
$2\theta_Z\kappa_Z/\sigma_Z^2\geq 1$ to ensure that zero is
an inaccessible boundary.

The {\em activity rate} process $a(t)+Z_t$ has the form of the
so-called CIR++ process well known in the interest rate modeling
literature (e.g., \cite{BrigoMercurio}). The advantage of
 the CIR process is in its analytical tractability. Its
transition probability density, the Laplace transform of its
integral, and the Laplace transform conditional on the terminal
state of the process are all known in closed form. The relevant
results are collected in Appendix \ref{append:CIRResults}.

Define the process $S$ to be the inverse of $T$,
$S_t:=\inf\{u\geq0:T_u>t\}.$ Since $T$ is a strictly increasing
continuous process, so is $S$. It is also clear that
$T_t=\inf\{u\geq0:S_u>t\}$.

Assume that on some complete probability space
$(\Omega,\mathcal{F},\mathbb{P})$ we have a c\`{a}dl\`{a}g SubOU
process $X$ with generating tuple
$(\kappa,\theta,\sigma,\gamma,\nu)$, $X_0=x_0$ and an independent
absolutely continuous time change $T$ of the form in
\eqref{eq:ACTimeChange}. Let $(\mathcal{F}_t)_{t\geq0}$ be the
smallest right-continuous complete filtration generated by the
processes $X_t$, $Z_{S_t}$ and $S_t$. Then $T_t$ is a stopping time
w.r.t. $(\mathcal{F}_t)_{t\geq0}$ for every $t$, and we can define
the time changed filtration $\mathcal{G}_t:=\mathcal{F}_{T_t}$. It
is clear that $T$ and $Z$ are adapted to $(\mathcal{G}_t)_{t\geq0}$.
Define a new process $Y$ by $Y_t:=X_{T_t}$, with $Y_0=y_0=x_0$. From
\cite{Jacod} Corollary 10.12, $Y$ is a
$(\mathcal{G}_t)$-semimartingale, and from \cite{KallsenShiryaev}
Lemma 5 it admits the following local characteristics
$(\overline{B},\overline{C},\overline{\Pi})$:
$$
\overline{B}_t(\omega)=\int_0^t(a(s)+Z_s(\omega))\Big[\gamma\kappa(\theta-Y_{s-}(\omega))+\int_0^\infty\int_{|x|\leq1}xp(u;Y_{s-}(\omega),Y_{s-}(\omega)+x)dx\nu(du)\Big]ds,
$$
$$
\overline{C}_t(\omega)=\gamma\sigma^2\int_0^t(a(s)+Z_s(\omega))ds,\quad
\overline{\Pi}(\omega,dt,dx)=(a(t)+Z_t(\omega))\pi(Y_{t-}(\omega),x)dx,
$$
where $\pi(\cdot,\cdot)$ is defined in Theorem
\ref{thm:SubOUgenerator} and here $x$ is interpreted as the jump
size. From these expressions we see that the role of the absolutely
continuous time change is to scale all the local characteristics of
the SubOU process with the stochastic activity rate or stochastic
volatility. The bivariate process $(Y,Z)$ is also a
$(\mathcal{G}_t)$-semimartingale. We have the following result on
its cross-variation process.
\begin{proposition}\label{prop:cross-variation}
The cross-variation process $[Y^c,Z^c]_t=0$, where $Y^c$ and $Z^c$
are the continuous local martingale parts of $Y$ and $Z$
respectively.
\end{proposition}

It is clear that $(Y,Z)$ is also a Markov process w.r.t. the
filtration $(\mathcal{G}_t)_{t\geq0}$. Given $\mathcal{G}_t$, the
distribution of $Y_{t+s}$ depends only on $T_{t+s}-T_t$ and $Y_t$,
and $T_{t+s}-T_t$ depends only on $Z_t$. The distribution of
$Z_{t+s}$ depends only on $Z_t$. Thus, conditional expectations of
the form $\mathbb{E}[f(Y_t)|\mathcal{G}_s]$ reduce to
$\mathbb{E}[f(Y_t)|Y_s,Z_s]$ by the Markov property. Using
conditioning and the spectral representation of the SubOU semigroup,
such expectations can be computed in terms of eigenfunction
expansions.
\begin{thm}\label{thm:TI-CIR-SubOUSpectralExpansion}
For $f\in L^2(\mathbb{R},\mathfrak{m})$, suppose one of the
following two conditions is satisfied:
\begin{itemize}
\item[(1)] The eigenfunction expansion $f(x)=\sum_{n=0}^\infty f_n\varphi_n(x)$, where $f_n=(f,\varphi_n),$ converges
absolutely for each $x$.
\item[(2)] $\sum_{n=0}^\infty e^{-\phi(\kappa n)\int_s^t a(u)du}
\mathcal{L}_{CIR}\left(t-s,\phi(\kappa
n)\left|z\right.\right)n^{-\frac{1}{4}}<\infty$ for some $z>0$ (and
hence for all $z$; it is straightforward to show this using
\eqref{eq:AsymptoticLaplaceCIR}), where the Laplace transform
$\mathcal{L}_{CIR}\left(t,\cdot\left|z\right.\right)$ is given in
Appendix A.
\end{itemize}
Then
$\mathbb{E}[f(Y_t)|Y_s,Z_s]=\sum_{n=0}^\infty e^{-\phi(\kappa
n)\int_s^t a(u)du} \mathcal{L}_{CIR}\Big(t-s,\phi(\kappa
n)\Big|Z_s\Big) f_n\varphi_n(Y_s).$
\end{thm}


We can now introduce stochastic volatility and time inhomogeneity in
commodity models. Let $Y_t = X_{T_t}$ be the time changed SubOU
process as above. Under the risk-neutral pricing measure
$\mathbb{Q}$ chosen by the market, we model the spot price as
follows:
\begin{equation}\label{eq:TI-CIR-SubOUSpotPrice}
S_t=F(0,t)e^{Y_t-G(t)},
\end{equation}
where the function $G(t)$ is selected so that
$e^{G(t)}=\mathbb{E}^{\mathbb Q}[e^{Y_t}]$. Applying Theorem
\ref{thm:TI-CIR-SubOUSpectralExpansion} to the exponential function,
we obtain the futures price process.
\begin{thm}\label{Thm:TI-CIR-SubOUfuturepricing}
(1) $e^{G(t)}=\sum_{n=0}^{\infty}e^{-\phi(\kappa n)\int_0^t
a(u)du}\mathcal{L}_{CIR}\Big(t,\phi(\kappa n)\Big|z_0\Big)
f_n\varphi_n(y_0)$, where $f_n$ are given in Lemma
\ref{lemma:expExpansion}. The expansion converges absolutely for all
$z_0>0$, $y_0\in\mathbb{R}$, and any Laplace exponent $\phi$.

\noindent(2) For each $t>0$, the futures price is a martingale on
$[0,t]$ given by:
\begin{equation}\label{eq:TI-CIR-SubOUfuturesprocess}
F(s,t)=F(0,t)e^{-G(t)} \sum_{n=0}^{\infty} e^{-\phi(\kappa
n)\int_s^t a(u)du}\mathcal{L}_{CIR}\Big(t-s,\phi(\kappa
n)\Big|Z_s\Big)f_n \varphi_n(Y_s),\quad s\in [0,t].
\end{equation}
\end{thm}


To investigate the maturity effect, we need to compute the QV
process $[r^t,r^t]_s$, which is more involved in this case due to
the extra state variable $Z$. From
\eqref{eq:TI-CIR-SubOUfuturesprocess},
$r_s^t=-G(t)+\theta+\frac{\sigma^2}{4\kappa}+\ln{g(Y_s,Z_s,s,t)}$,
where
$$
g(y,z,s,t)=\sum_{n=0}^{\infty}e^{-\phi(\kappa n)\int_s^t
a(u)du}\mathcal{L}_{CIR}\Big(t-s,\phi(\kappa
n)\Big|z\Big)\Big(\frac{\sigma}{2\sqrt{\kappa}}\Big)^n\frac{1}{n!}H_n\Big(\frac{\sqrt{\kappa}}{\sigma}(y-\theta)\Big).
$$
Again we use \cite{Kallsen} Proposition 2.5 to compute
$[r^{tc},r^{tc}]_s$ from the local characteristics of the
semimartingale $(Y,Z)$. Since the cross-variation is zero by
Proposition \ref{prop:cross-variation}, we do not have cross
derivative terms and obtain:
$$
[r^t,r^t]_s=[r^{tc},r^{tc}]_s+
\gamma\sigma^2\int_0^s(a(u)+Z_u)\Big(\frac{g_y(Y_{u-},Z_u,u,t)}{g(Y_{u-},Z_u,u,t)}\Big)^2du+\sigma_Z^2\int_0^sZ_u\Big(\frac{g_z(Y_{u-},Z_u,u,t)}{g(Y_{u-},Z_u,u,t)}\Big)^2du
$$
$$
+\sum_{u\leq s}\Big(\int_{Y_{u}\wedge Y_{u-}}^{Y_{u}\vee
Y_{u-}}\frac{g_y(y,z,u,t)}{g(y,z,u,t)}dy\Big)^2.
$$
Note that $g$ and $g_y$ are positive, but $g_z$ is not necessarily
so, and $g(y,z,u,t)$ depends on $u$ and $t$ only through $t-u$. It
is thus clear that we have the following:
\begin{thm}\label{thm:SubOUSHSufficientConditions}
If $\frac{g_y(y,z,0,t)}{g(y,z,0,t)}$ and
$\Big(\frac{g_z(y,z,0,t)}{g(y,z,0,t)}\Big)^2$ are decreasing in $t$
for any $(y,z)$, then the maturity effect holds.
\end{thm}
As in the SubOU case in section 3.3, this condition is hard to check
analytically, but can be easily verified numerically. We have
conducted extensive numerical experiments and verified this
condition for all parameter specifications we have tested.


For the model with stochastic volatility the option pricing formula
is more involved since the futures price $F(t,t^*)$ at  expiration
of the option $t$ is now determined by the values of two state
variables $Y_t$ and $Z_t$ at that time, $F(t,t^*)=F(Y_t,Z_t,t,t^*)$.
We condition on the state of the CIR process $Z_t$ at time $t$ and
reduce the problem to the SubOU case. One then has to use the
conditional Laplace transform \eqref{eq:ConLaplaceCIR} instead of
\eqref{eq:LaplaceCIR}, since we have conditioned on $Z_t$. Hence,
the pricing formula is expressed as an integral with respect to the
transition density of the CIR process \eqref{eq:CIRTPD}. An
additional subtlety is that $y^*$ now depends on $z$. Namely, for
each fixed $z>0$, there exists a unique $y^*=y^*(z)$ such that
$F(y^*,z,t,t^*)=K.$ Then the put payoff function can be rewritten as
$(K-F(y,z,t,t^*))^+=(K-F(y,z,t,t^*)){\bf 1}_{\{y<y^*(z)\}}$.

\begin{thm}\label{Thm:TI-CIR-SubOUoptionpricing}
For each fixed $z>0$, let $y^*(z)$ denote the unique solution of the
equation
$F(y,z,t,t^*)=K,$
where $F(y,z,t,t^*)$ is the futures pricing function
\eqref{eq:TI-CIR-SubOUfuturesprocess}. Define
$w^*(z):=\frac{\sqrt{\kappa}}{\sigma}(y^*(z)-\theta)$,
$\tau:=t^*-t$, $\alpha:=\frac{\sigma}{2\sqrt{\kappa}}$ and
$F:=F(0,t^*)$. Suppose condition (2) of Theorem
\ref{thm:TI-CIR-SubOUSpectralExpansion} and the following condition
are satisfied:
\begin{equation}\label{eq:TI-CIR-SubConLaplaceTransformFinite}
\sum_{n=0}^\infty e^{-\phi(\kappa n)\int_0^t a(u)du}
\mathcal{L}_{CIR}\left(t,\phi(\lambda)\left|z_0,z\right.\right)n^{-\frac{1}{4}}<\infty
\ \text{for some}\ z\ \text{and hence for all $z$}.\end{equation}
(It is easy to show this using \eqref{eq:AsymptoticConLaplaceCIR}.)
Then the put price is given by:
\begin{gather}\label{eq:TI-CIR-SubOUputprice}
P(t,t^*,K,F)=B(0,t)\\
\times \int_0^\infty \left\{\sum_{n=0}^\infty e^{-\phi(\kappa
n)\int_0^t a(u)du}\mathcal{L}_{CIR}(t,\phi(\kappa
n)|z_0,z_t)p_n(t,t^*,w^*(z_t),F)\varphi_n(y_0)\right\}p_{CIR}(t,z_0,z_t)dz_t\notag,
\end{gather}
where $p_{CIR}(t,z_0,z_t)$ is the CIR transition density
\eqref{eq:CIRTPD} and
\begin{gather}
p_n(t,t^*,w^*(z),F)=\frac{1}{\sqrt{\pi2^{n} n!}}\\
\times \left\{K b_n(w^*(z))-Fe^{\theta
+\frac{\sigma^2}{4\kappa}-G(t^*)} \sum_{m=0}^\infty e^{-\phi(\kappa
m)\int_t^{t^*} a(u)du}\mathcal{L}_{CIR}(\tau,\phi(\kappa m)|z_t)
\frac{\alpha^m}{m!}a_{n,m}(w^*(z))\right\}\notag,
\end{gather}
where $b_n(w)$ and $a_{n,m}(w)$ are given by \eqref{eq:bCoeff} and
\eqref{eq:aCoeff}. The call price is given by the put-call parity.
\end{thm}

\begin{remark}
For options written on the spot price, in contrast to futures
options, we only need the Laplace transform of the time change
instead of the conditional Laplace transform. Furthermore, in this
case $y^*=\ln\left(\frac{K}{F(0,t)}\right)+G(t)$ is independent of
$Z_t$. By setting $t=t^*$ and using
$\int_0^{\infty}\mathcal{L}_{CIR}(t,\phi(\kappa
n)|z_0,z_t)p_{CIR}(t,z_0,z_t)dz_t=\mathcal{L}_{CIR}(t,\phi(\kappa
n)|z_0)$, the put price becomes
$$P(t,K,F)=\sum_{n=0}^\infty e^{-\phi(\kappa n)\int_0^ta(u)du}\mathcal{L}_{CIR}(t,\phi(\kappa n)|z_0)p_n(t,t,w^*,F)\varphi_n(y_0).$$
\end{remark}

\section{Model Implementation and
Calibration Examples}\label{sec:ModelImplementationAndCalibration}

The models introduced in this paper were implemented in C++ on a PC.
Hermite expansions can be efficiently computed using the following
classical recursion for Hermite polynomials (\cite{Lebedev} p.61):
$$
H_0(x)=1,\quad H_1(x)=2x,\quad
H_n(x)=2xH_{n-1}(x)-2(n-1)H_{n-2}(x),\quad n\geqslant 2.
$$
To compute the option pricing formula \eqref{eq:SubOUPutPrice}, we
need to evaluate the coefficients $b_n$ and $a_{n,m}$. From the
equation \eqref{eq:bCoeff}, it is easy to see $b_n$ can be computed
recursively using the recursion for Hermite polynomials. Equation
\eqref{eq:aCoeff} is a closed-form formula for $a_{n,m}$, but it is
not convenient to use from a computational perspective. We have the
following computationally efficient approach for $a_{n,m}$.
\begin{proposition}\label{prop:Compute_acoeff}
The coefficients $a_{n,m}$ satisfy the following:
\begin{align}
a_{0,0}(x)&=\sqrt{\pi}\Phi(\sqrt{2}x),\quad a_{n,n}(x)=2na_{n-1,n-1}(x)-H_{n-1}(x)H_n(x)e^{-x^2},\quad n\geq 1,\label{eq:annCoeffRecusion}\\
a_{n,m}(x)&=\frac{H_n(x)H_{m+1}(x)-H_m(x)H_{n+1}(x)}{2(m-n)}e^{-x^2},\quad
n\neq m, n\geq0, m\geq0\label{eq:anmCoeff}.
\end{align}
\end{proposition}

To evaluate the option pricing formula
\eqref{eq:TI-CIR-SubOUputprice} for the model with stochastic
volatility, we first truncate the integral in $z_t$ at some level
$M$ large enough that the probability of the CIR process to exceed
$M$ at time $t$ is less than the desired error tolerance. We then
use the Simpson rule to discretize the integral on the interval
$[0,M]$. The CIR transition density at each node $z_t(k)$ is
computed by \eqref{eq:CIRTPD}, at each integration node $z_t(k)$ the
value of $y^*(k)$ is found by the bisection algorithm, and the
integrand  is computed similar to the option pricing formula
\eqref{eq:TI-CIR-SubOUputprice} in the SubOU case (with the
distinction that under the time changed SubOU the conditional
Laplace transform \eqref{eq:ConLaplaceCIR} enters the expression in
place of the Laplace transform \eqref{eq:LaplaceCIR} in the SubOU
case).

CPU times generally depend on time to maturity $t$ and the model
parameters.
For short maturities (say, less than two weeks to expiration), one
may have to use infinite-precision arithmetics to achieve required
accuracy in summing up the series. To compute short maturity option
prices we used the GNU MP Bignum library. For longer maturity
options double precision is sufficient. In our numerical experiments
on a PC running Linux (Intel Core 2 Duo CPU at 2.53GHz with 2.00GB
RAM), CPU times ranged from several milliseconds up to hundreds of
milliseconds per option for the SubOU model, depending on the
combination of parameters, and from hundreds of milliseconds up to
several seconds per option for the SubOU model with stochastic
volatility.

\begin{figure}[htbp!]
\subfigure{\includegraphics[scale=0.3]{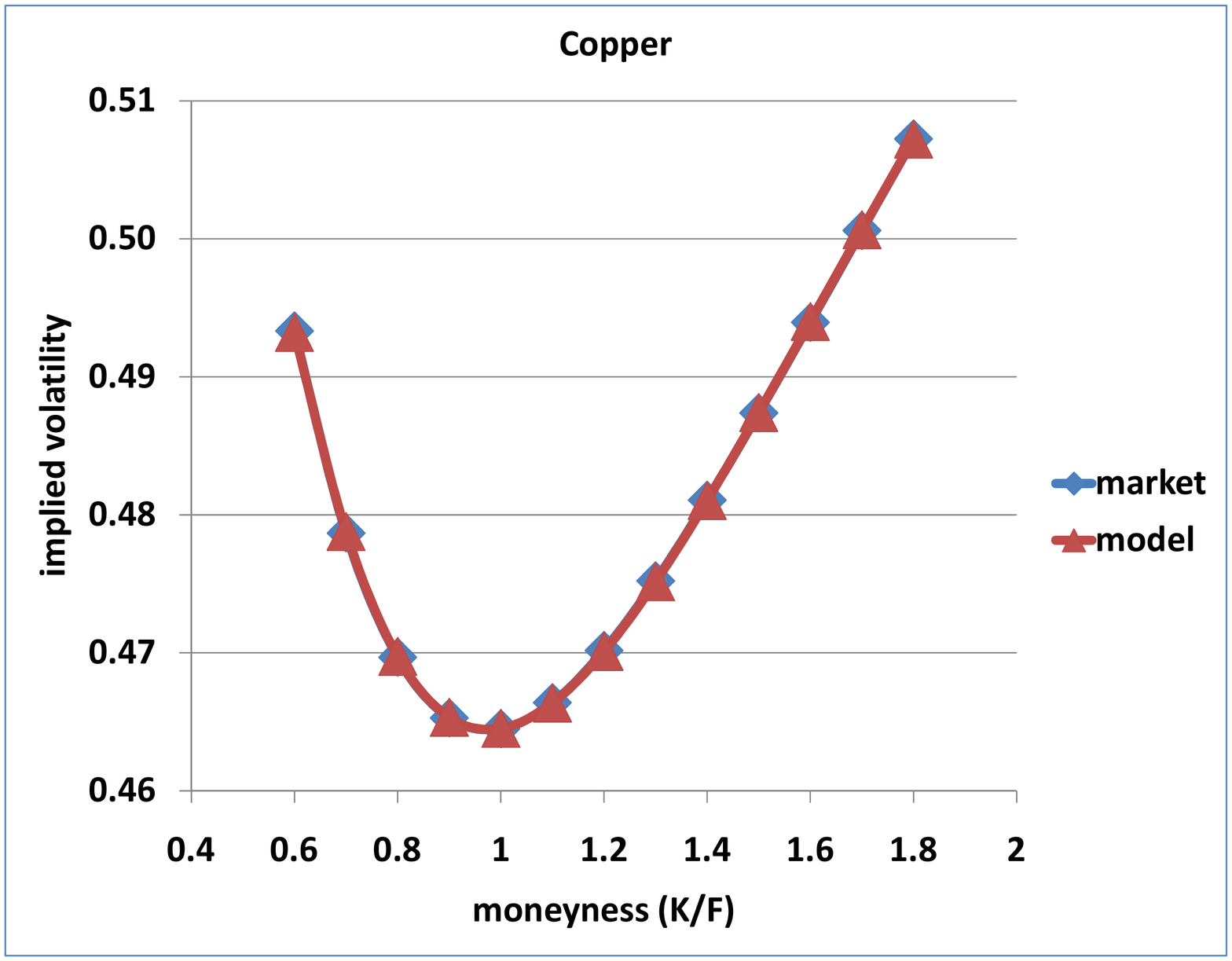}}\hspace{-1cm}\subfigure{\includegraphics[scale=0.3]{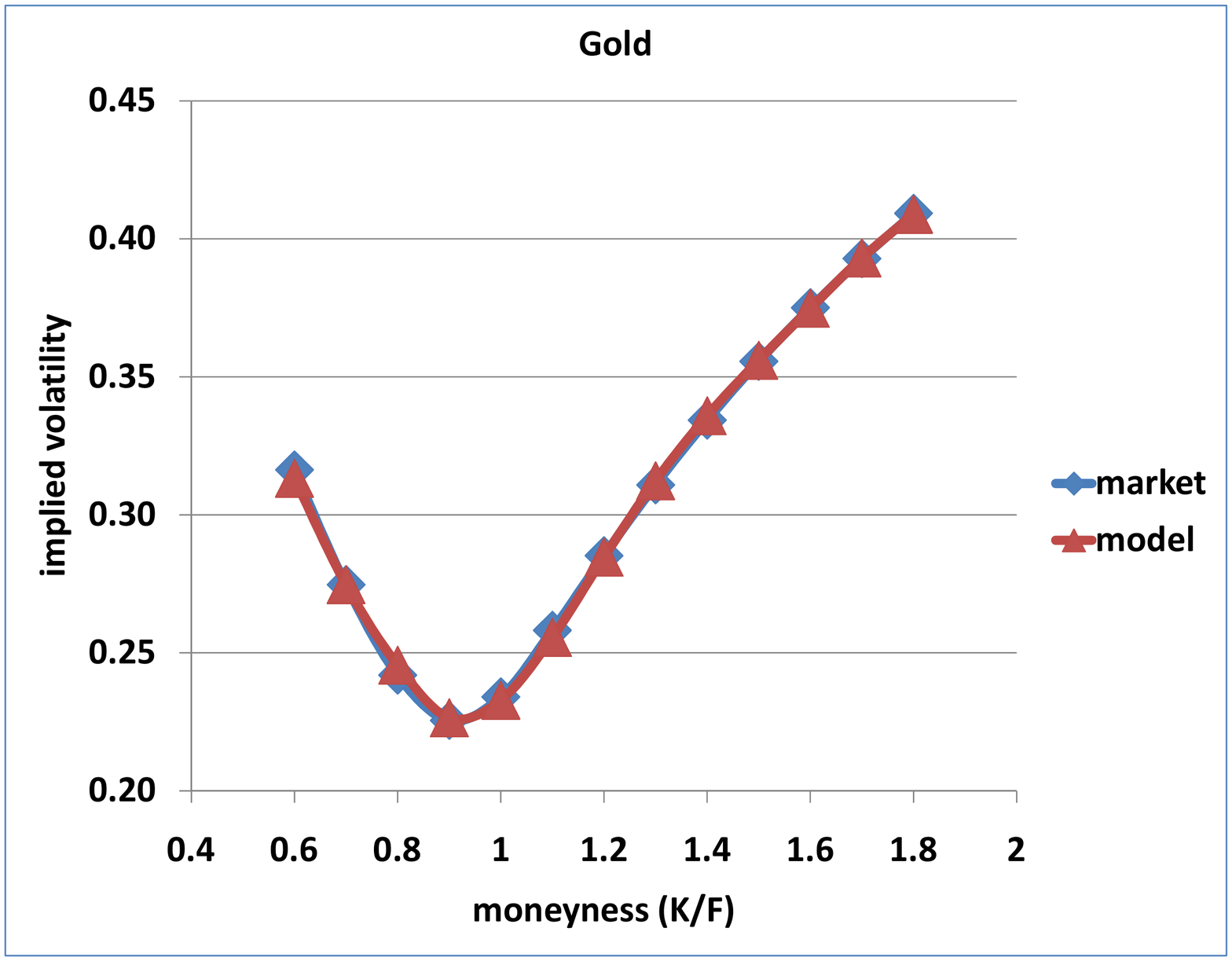}}\\
\subfigure{\includegraphics[scale=0.3]{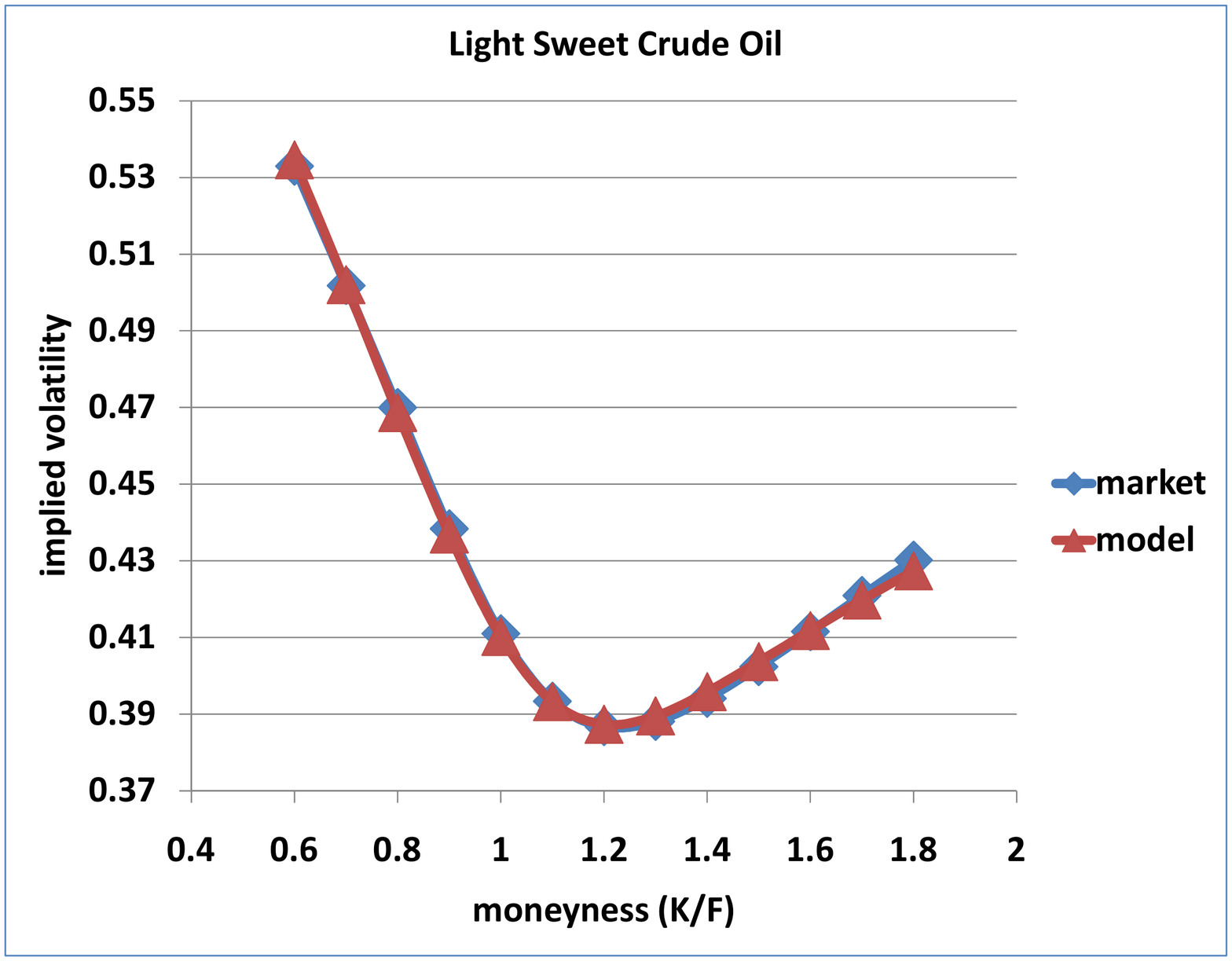}}\hspace{-1cm}\subfigure{\includegraphics[scale=0.3]{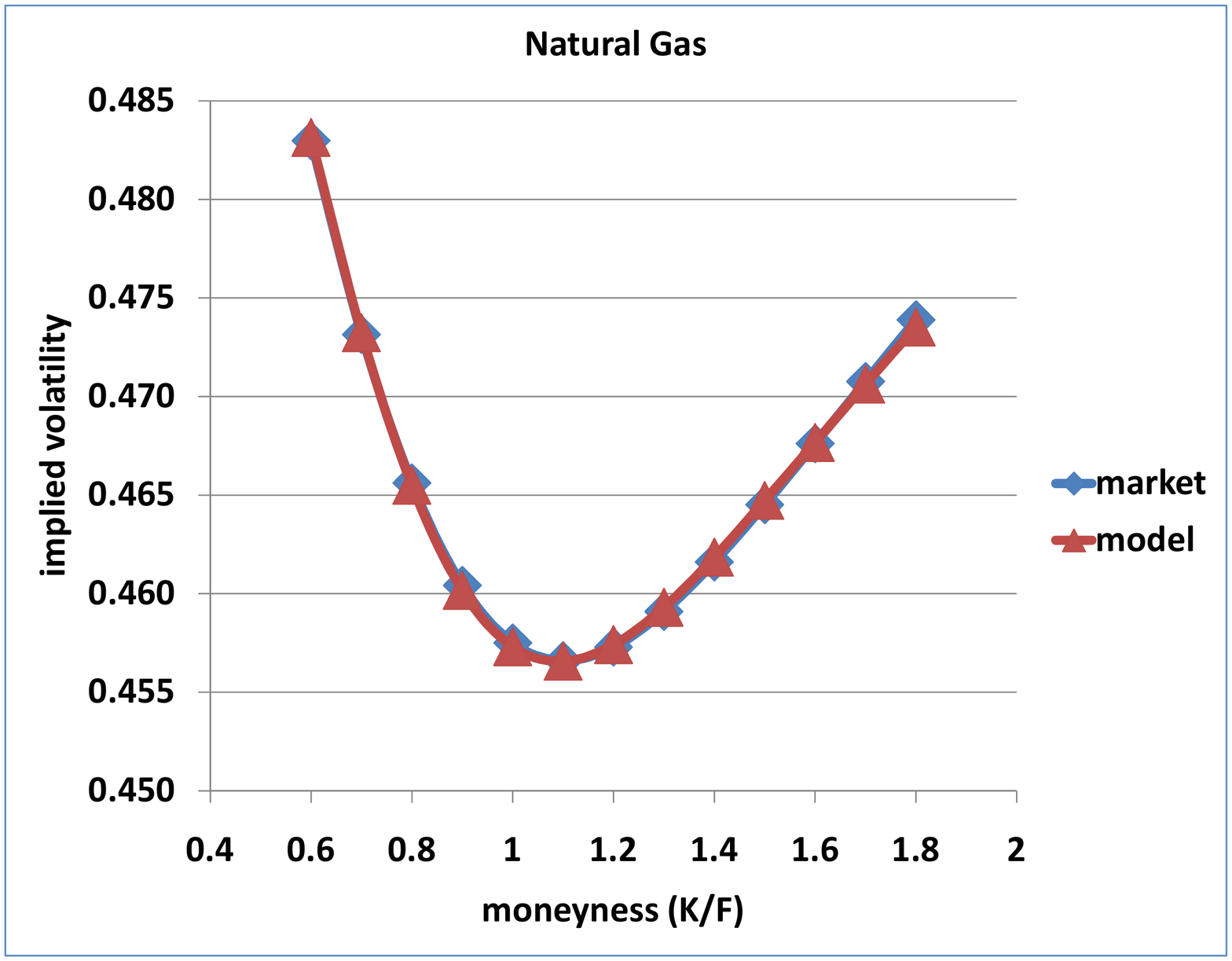}}\\
\subfigure{\includegraphics[scale=0.3]{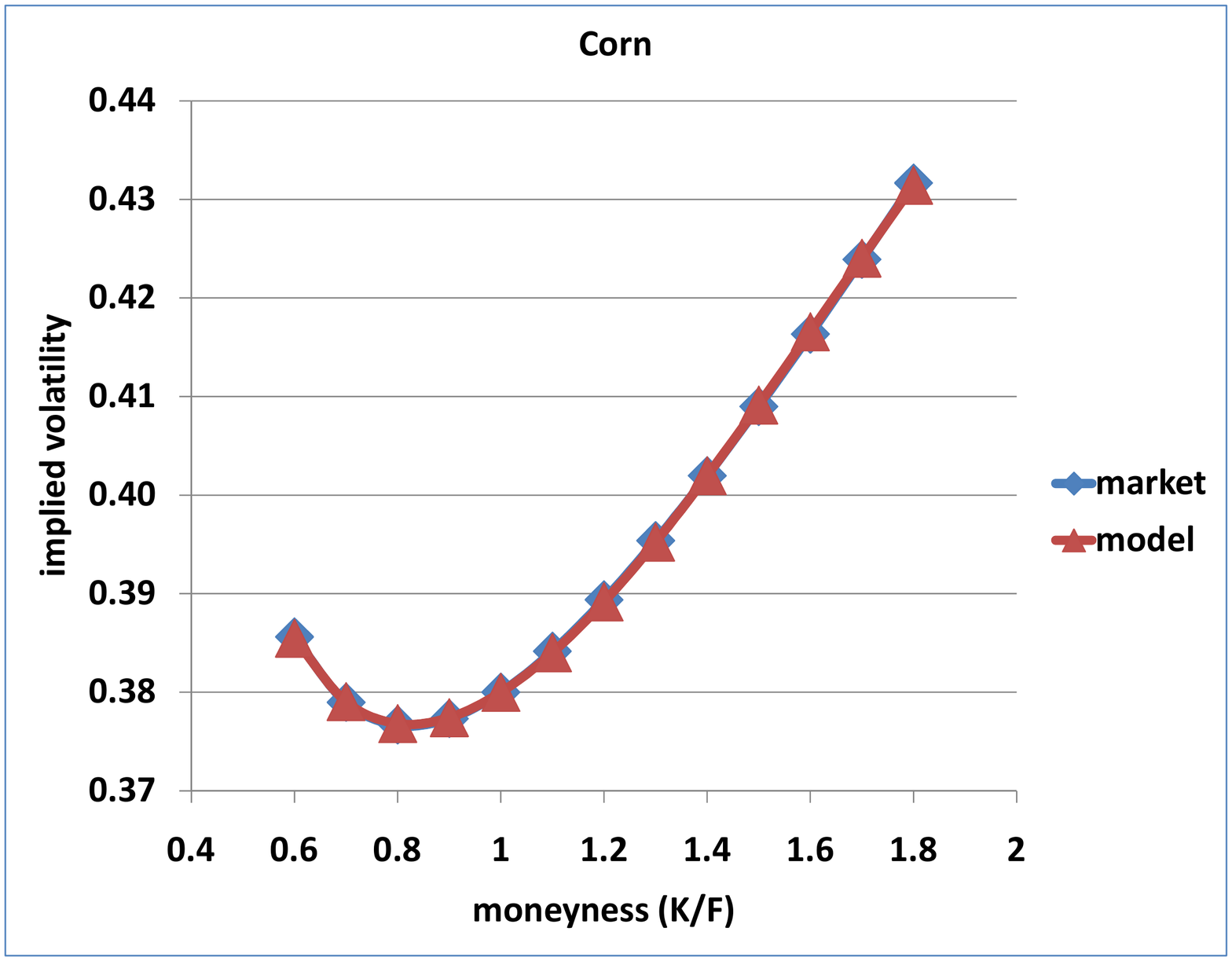}}\hspace{-1cm}\subfigure{\includegraphics[scale=0.3]{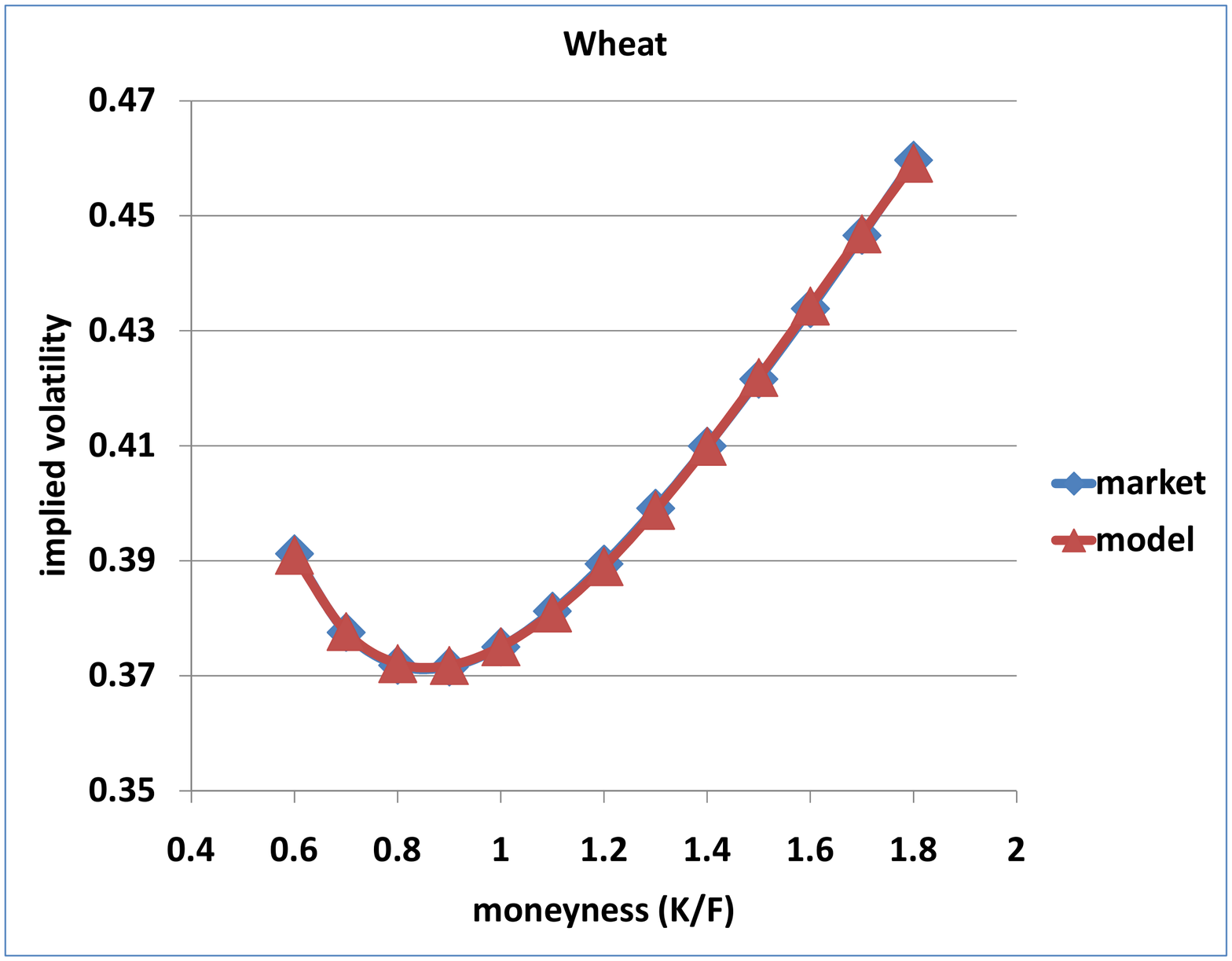}}
\caption{SubOU Model Calibration Results to Implied Volatility
Smiles for Commodities}\label{fig:Calib1M}
\end{figure}

We now present calibration examples of the SubOU model with the IG
subordinator to implied volatility smile curves extracted from
market prices of options on six commodity futures. We have also
calibrated for other commodities, and the results are similar to
what are displayed here. However, due to space constraints, only six
of them are shown. The commodities included two metals (copper,
gold), two energies (crude oil, natural gas), and two agriculturals
(corn and wheat). Market data on implied volatilities for this study
were provided by Morgan Stanley's Commodity Stategies Group and were
extracted from commodity futures options market prices on July 2nd
2009. All options had approximately six months to expiration. The
moneyness defined as the ratio of the option strike price to the
futures price ranged from 0.6 to 1.8 for all commodities. To
calibrate the model to market implied volatilities, we minimized the
sum of squared differences between the market and the model implied
volatilities. There are a total of six parameters in the SubOU
model: three parameters of the background OU process and three
parameters of the inverse Gaussian subordinator with drift. Without
loss of generality, the starting SubOU state $x_0$ was set to zero
(it can always be set to zero by changing $\theta$ to $\theta-x_0$
without affecting the option price). Our calibration results are
presented in Figure \ref{fig:Calib1M}. In these instances the SubOU
model with the IG subordinator provides an excellent fit to
volatility smiles for all eight commodities (well within the bid/ask
spread for each option).

\begin{figure}[htbp!]
\centering\includegraphics[scale=0.4]{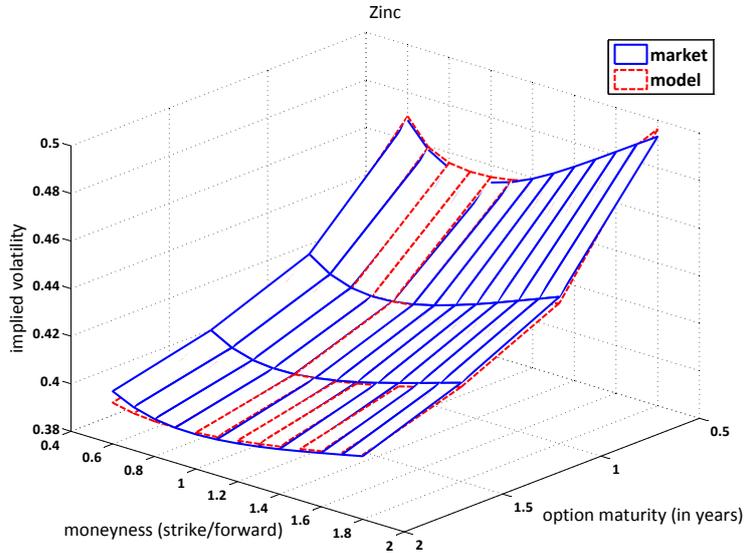}
\caption{Time Changed SubOU Volatility Surface Calibration Results
For Zinc}\label{fig:CalibZinc4M3D}
\end{figure}

While the SubOU model calibrates well to commodity volatility smiles
for a fixed maturity, it may generally lack flexibility to capture
the entire {\em volatility surface} across both the maturity
dimension and the strike (moneyness) dimension. The time changed
SubOU model with stochastic volatility and possible time
inhomogeneity has additional flexibility to capture time dependence
in the shape and steepness of the volatility smile and time
dependence in the at-the-money volatility term structure. In Figure
\ref{fig:CalibZinc4M3D}
we calibrate the
SubOU model with the inverse Gaussian subordinator time
changed with the integral of the CIR process to the implied
volatility surfaces for zinc.
We used four
maturities (6 months, 1 year, 1.5 years and 2 years) in our
calibration. The deterministic activity rate component was taken to
be a piecewise constant function (constant between adjacent futures
maturity dates). The time changed SubOU model
provided an excellent fit to this volatility surface (well within the bid/ask spreads for all options). The
deterministic activity rate allowed us to capture the sharp decay in
the ATM implied volatilities, the IG subordinator
allowed us to capture steep smiles for shorter-dated maturities, and the
CIR stochastic volatility supported the longer-dated smiles. In
contrast, SubOU models without stochastic volatility exhibit faster
flattening of the volatility smile as we go further out in maturity.

\section{Conclusion}
This paper studied a class of subordinate OU processes, their
sample path properties, equivalent measure transformations, and the
spectral representation of their transition semigroup. As an
application, we constructed a new class of commodity models with
mean-reverting jumps based on subordinate OU process. Further time
changing by the integral of a CIR process plus a deterministic
function of time, we induced stochastic volatility and time
inhomogeneity in the models. We obtained analytical solutions for
commodity futures options in terms of Hermite expansions and showed
that the models exhibit the maturity effect and are flexible enough
to capture a wide variety of implied volatility smile patterns
observed in energy, metals, and agricultural commodities futures
options.

We are currently developing computational methods for American-style
futures options in these models. It turns out that the eigenfunction
expansion approach to pricing European options followed in this
paper can be extended to Bermudan-style options with a finite number
of exercise opportunities. Richardson extrapolation can then be used
to obtain solutions for American-style options.

In future work we plan to extend this class of models to
multi-commodity products, such as spread options, and to
path-dependent options such as Asian-style options. An extension to
American-style options is developed in \cite{LiLinetskyOS}. We also
anticipate that subordinate OU processes studied in this paper will
find other applications beyond commodities, such as in interest rate
modeling, volatility modeling, and real options.

\appendix

\section{CIR Processes}\label{append:CIRResults}
Let $\{Z_t,t\geq 0\}$ be a CIR diffusion starting from $Z_0=z>0$ and
solving the SDE
\begin{equation}\label{eq:SDECIR}
dZ_{t}=\kappa\left(\theta-Z_{t}\right)dt+\sigma\sqrt{Z_{t}}dB_{t}.
\end{equation}
Assume the long run level $\theta$, the rate of mean reversion
$\kappa$, and the volatility parameter $\sigma$ satisfy the Feller
condition $d:=\frac{2\theta\kappa}{\sigma^{2}}\geq 1$ to ensure that
the process stays strictly positive (zero is an inaccessible
boundary).

The CIR transition density $p_{CIR}(t,z_0,z)$ is given by
\begin{equation}\label{eq:CIRTPD}
p_{CIR}(t,z_0,z)=\frac{2\kappa}{\sigma^2(1-e^{-\kappa
t})}e^{-\frac{2\kappa(z_0e^{-\kappa t}+z)}{\sigma^2(1-e^{-\kappa
t})}}\left(\frac{z}{z_0e^{-\kappa
t}}\right)^{\frac{d-1}{2}}I_{d-1}\left(\frac{4\kappa\sqrt{z_0ze^{-\kappa
t}}}{\sigma^2(1-e^{-\kappa t})}\right),
\end{equation}
where $I_{d-1}(\cdot)$ is the modified Bessel function of the first
kind of order $d-1$.

The Laplace transform
$\mathcal{L}_{CIR}(t,\lambda|z_0):=\mathbb{E}_{z_0}\left[e^{-\lambda\int_0^tZ_udu}\right]$
is given by the CIR bond pricing formula for the short rate process
$\lambda Z_t$:
\begin{equation}\label{eq:LaplaceCIR}
\mathcal{L}_{CIR}(t,\lambda|z_0)=C(t,\lambda)e^{-B(t,\lambda)z_0},
\end{equation}
where
$C(t,\lambda)=\left(\dfrac{2\gamma(\lambda)e^{(\gamma(\lambda)+\kappa)t/2}}{(\gamma(\lambda)+\kappa)(e^{\gamma(\lambda)
t}-1)+2\gamma(\lambda)}\right)^d$,
$B(t,\lambda)=\dfrac{2\lambda(e^{\gamma(\lambda)
t}-1)}{(\gamma(\lambda)+\kappa)(e^{\gamma(\lambda)t}-1)+2\gamma(\lambda)}$,
and $\gamma(\lambda)=\sqrt{\kappa^2+2\sigma^2 \lambda}$. The
function $\mathcal{L}_{CIR}(t,\lambda|z_0)$ has the following
asymptotic behavior as $\lambda\rightarrow\infty$:
\begin{equation}\label{eq:AsymptoticLaplaceCIR}
\mathcal{L}_{CIR}(t,\lambda|z_0)\sim\exp\left\{-\frac{\kappa\theta}{\sigma}\sqrt{2\lambda}t-\frac{z_0}{\sigma}\sqrt{2\lambda}\right\}.
\end{equation}

The Laplace transform conditional on the state of the process at
time $t$,
$\mathcal{L}_{CIR}(t,\lambda|z_0,z_t):=\mathbb{E}_{z_0}\left[e^{-\lambda\int_0^tZ_udu}\left|\right.Z_t=z_t\right],$
is also known in closed form (\cite{BroadieKaya}):
$$
\mathcal{L}_{CIR}(t,\lambda|z_0,z_t)=\frac{\gamma(\lambda)e^{-0.5(\gamma(\lambda)-\kappa)t}(1-e^{-\kappa
t})}{\kappa (1-e^{-\gamma(\lambda)t})}
$$
\begin{equation}\label{eq:ConLaplaceCIR}
\times\exp\left\{\frac{z_0+z_t}{\sigma^2}\left(\frac{\kappa(1+e^{-\kappa
t})}{1-e^{-\kappa
t}}-\frac{\gamma(\lambda)(1+e^{-\gamma(\lambda)t})}{1-e^{-\gamma(\lambda)t}}\right)\right\}\frac{I_{d-1}\left(\frac{4\gamma(\lambda)\sqrt{z_0z_t}}{\sigma^2}\frac{e^{-0.5\gamma(\lambda)t}}{1-e^{-\gamma(\lambda)t}}\right)}{I_{d-1}\left(\frac{4\kappa\sqrt{z_0z_t}}{\sigma^2}\frac{e^{-0.5\kappa
t}}{1-e^{-\kappa t}}\right)}.
\end{equation}
The function $\mathcal{L}_{CIR}(t,\lambda|z_0,z_t)$ has the
following asymptotic behavior as $\lambda\rightarrow\infty$:
\begin{equation}\label{eq:AsymptoticConLaplaceCIR}
\mathcal{L}_{CIR}(t,\lambda|z_0,z_t)\sim\lambda^{\frac{d}{2}}\exp\left\{-\frac{\kappa\theta}{\sigma}\sqrt{2\lambda}t-\frac{z_0+z_t}{\sigma}\sqrt{2\lambda}\right\}.
\end{equation}

\section{Proofs}\label{append:proofs}

{\bf \emph{Part (2) of Theorem \ref{thm:SubOUSemiMG}}.} Denote the
RHS of $\mathcal{G}^{\phi}$ in Theorem \ref{thm:SubOUgenerator} by
$\mathcal{G}^{\#}$. If $X'$ admits characteristics $(B',C',\Pi')$,
then from It\^{o}'s Formula for semimartingales, for any $f\in
C^2_c(\mathbb{R})$
$$M_t:=f(X'_t)-f(x)-\int_0^t\mathcal{G}^{\#}f(X'_{s-})ds$$
is a local martingale. Since $\mathcal{G}^{\#}f\in C_0(\mathbb{R})$
(the space of continuous functions on $\mathbb{R}$ vanishing at
infinity), $\mathcal{G}^{\#}f$ is bounded. $f(X'_t)-f(x)$ is also
bounded for all $t$. Hence $\mathbb{E}[M^*_t]<\infty$
($M^*_t:=\sup_{s\leq t}|M_s|$) for all $t$, and $M$ is a martingale
by \cite{Protter} Chapter 1 Theorem 51.
Note that from Theorem \ref{thm:SubOUgenerator}, $C^2_c(\mathbb{R})$
is a core of $D(\mathcal{G}^{\phi})$. Hence applying \cite{EthierKurtz}
Chapter 4 Theorem 4.1 to the martingale problem
$\Big((\mathcal{G}^{\phi},C^2_c(\mathbb{R})),\mathbb{P}^x\Big)$ and
Corollary 4.3, it follows that $\mathbb{P}'\circ X'^{-1}=\mathbb{P}^x$ on the
Skorohod space $(\Omega,\mathcal{F}^0)$.\qed

{\bf\emph{Theorem \ref{thm:mean-revertingjumpmeasure}.}} If
$x>\theta$, then for any $y>0$, $|-y+(x-\theta)(1-e^{-\kappa
t})|<|y+(x-\theta)(1-e^{-\kappa t})|.$ From \eqref{eq:OUTPD}, this
implies $p(t,x,x-y)>p(t,x,x+y)$ for any $t>0$. Hence from the
definition of $\pi(x,\cdot)$, $\pi(x,-y)>\pi(x,y)$ for any $y>0$. By
integrating $\pi(x,\cdot)$ on $(-\infty,-y)$ and $(y,\infty)$, we
also get $\Pi(x,(-\infty,-y))>\Pi(x,(y,\infty))$.
The cases with $x<\theta$ and $x=\theta$ are proved similarly.\qed

{\bf\emph{Theorem \ref{thm:EMCSubOU}.}} \emph{Sufficiency}.
By replacing the L\'{e}vy measure used in Remark 33.3 of \cite{Sato}
by our state-dependent L\'{e}vy measure, we can show that the
Hellinger condition
$\int_{y\neq0}\big(\sqrt{\pi'(x,y)}-\sqrt{\pi(x,y)}\big)^2dy<\infty$
implies that
\begin{equation}\label{eq:HellingerConditionEquivalent}
\int_{|y|\leq1}|y|\cdot|\pi'(X_{s-}(\omega),y)-\pi(X_{s-}(\omega),y)|dy<\infty,
\end{equation}
so $h(x)(Y-1)*\Pi$ is finite. We first show that
\begin{align*}
&\int_{[0,\infty)}\int_{|y|\leq1}yp'(u;X_{s-}(\omega),X_{s-}(\omega)+y)dy\nu'(du)-\int_{[0,\infty)}\int_{|y|\leq1}yp(u;X_{s-}(\omega),X_{s-}(\omega)+y)dy\nu(du)\\
&=\int_{|y|\leq1}y\Big[\pi'(X_{s-}(\omega),y)-\pi(X_{s-}(\omega),y)\Big]dy.
\end{align*}
Note that
\begin{align*}
&\int_{[0,\infty)}\int_{|y|\leq 1}y p'(u, x, x+y)dy \nu'(du)-
\int_{[0,\infty)}\int_{|y|\leq 1}y p(u, x, x+y)dy \nu(du) \\
&=\lim_{n\to\infty}\left(\int_{[0,\infty)}\int_{1/n\leq |y|\leq 1}y
p'(u, x, x+y)dy \nu'(du)- \int_{[0,\infty)}\int_{1/n\leq |y|\leq 1}y
p(u, x, x+y)dy \nu(du)
\right)\\
&=\lim_{n\to\infty}\left(\int_{1/n\leq |y|\leq
1}y(\pi'(x,y)-\pi(x,y))dy\right)=\int_{|y|\leq 1}(\pi'(x, y)-\pi(x,
y))dy,
\end{align*}
where the last equality comes from the Dominated Convergence Theorem
since we have \eqref{eq:HellingerConditionEquivalent}. So for all
$\omega$ we have
$B'=B+\gamma\sigma^2\beta\cdot t+h(x)(Y-1)*\Pi$,
$C'=C$,
$\Pi'=Y\cdot\Pi$.

Since $\Pi(\omega,t,dx)=0$, it is clear that $\sigma_{JS}=\infty$,
where $\sigma_{JS}$ is defined in \cite{JacodShiryaev} (JS)
\Rmnum{3}.5.6. The process $H$ defined in JS \Rmnum{3}.5.7
becomes the following in our case:
$$H_t(\omega)=\int_0^t\gamma\sigma^2\beta_s(\omega)^2ds+\int_0^t\int_{y\neq 0}\Big(\sqrt{\pi'(X_{s-}(\omega),y)}-\sqrt{\pi(X_{s-}(\omega),y)}\Big)^2dyds$$

It is clear that the integrand in the above expression is
c\`{a}dl\`{a}g for every $\omega$. This implies that
$H_t(\omega)<\infty$ for every $\omega$ and $t$. Hence the process
$H$ does not jump to infinity as defined in JS \Rmnum{3}.5.8. This
fact together with $\sigma_{JS}=\infty$ implies that Hypothesis
\Rmnum{3}.5.29 of JS holds.

$\mathbb{P}'$ is the unique solution to the martingale problem
$(\sigma(X_0),X|\mathbb{P}_0',B',C',\Pi')$.  As remarked before,
local uniqueness also holds. Note that
$\mathbb{P}'_0\preccurlyeq\mathbb{P}_0$. Now all conditions stated
in JS Theorem \Rmnum{3}.5.34 are satisfied, which implies
$\mathbb{P}'\preccurlyeq\mathbb{P}$ locally.
By interchanging the role of $\mathbb{P}'$ and $\mathbb{P}$ and
similarly defining $\beta'$ and $H'$, we can prove
$\mathbb{P}_0\preccurlyeq \mathbb{P}'_0$ implies
$\mathbb{P}'\preccurlyeq \mathbb{P}$ locally. Hence $\mathbb{P}'\sim
\mathbb{P}$ locally.

\emph{Necessity}. If $\mathbb{P}'\sim\mathbb{P}$ locally, then (i)
holds, and (ii) is implied by JS Theorem \Rmnum{3}.3.24. The
uniqueness of the solution to the martingale problem
$(\sigma(X_0),X|\mathbb{P}_0,B,C,\Pi)$ implies the
$\mathbb{P}$-martingale representation property w.r.t. $X$ (JS
Theorem \Rmnum{3}.4.29), hence JS Theorem \Rmnum{3}.5.19 holds,
which further implies JS \Rmnum{4}.3.32. Now the conditions in JS
Theorem \Rmnum{4}.3.35 are satisfied, and this theorem implies that
the Hellinger process of order $\frac{1}{2}$ (see JS Definition
\Rmnum{4}.1.24) is given by
$$h^\frac{1}{2}_t(\omega)=\frac{1}{8}\int_0^t\gamma\sigma^2\beta_s(\omega)^2ds+\frac{1}{2}\int_0^t\int_{y\neq 0}\Big(\sqrt{\pi'(X_{s-}(\omega),y)}-\sqrt{\pi(X_{s-}(\omega),y)}\Big)^2dyds.$$
JS Theorem \Rmnum{4}.2.1 says that $h^\frac{1}{2}_t(\omega)<\infty$
both $\mathbb{P}$ and $\mathbb{P}'$-a.s., hence there exists
$x_0\in\mathbb{R}$ such that $\int_{y\neq
0}\Big(\sqrt{\pi'(x_0,y)}-\sqrt{\pi(x_0,y)}\Big)^2dy<\infty$. But
one can show that the tail behavior at $y=0$ of
$\sqrt{\pi'(x,y)}-\sqrt{\pi(x,y)}$ does not depend on $x$ (see
Proposition \ref{prop:SubBM Asymptotic Independence of Drift}, whose
proof does not depend on Theorem \ref{thm:EMCSubOU}), so we have
$\int_{y\neq
0}\Big(\sqrt{\pi'(x,y)}-\sqrt{\pi(x,y)}\Big)^2dy<\infty$ for any
$x$.

Therefore, the process $H$ defined in the proof of the sufficiency
part does not jump to infinity. This together with
$\sigma_{JS}=\infty$ allows us to apply JS Corollary \Rmnum{3}.5.22
(ii) which gives the form of the density process. \qed

{\bf \emph{Proposition \ref{prop:asymptotic behavior
SubOU=SubBM}}.} Define $q(s,0,y):=\frac{1}{\sqrt{\pi\sigma^2
s}}\exp{\Big\{-\frac{(y-\kappa(\theta-x)s)^2}{2\sigma^2 s}\Big\}}$,
the transition density of Brownian motion starting at $0$ with
drift $\kappa(\theta-x)$ and volatility $\sigma$. It is easy to see
that
$$\lim_{s\rightarrow0}\frac{p(s,x,x+y)}{q(s,0,y)}=1$$
uniformly for $y$ on any compact interval. We wish to prove that
\begin{equation}\label{eq:limit_equivalence}
\lim_{y\rightarrow0}\frac{\int_{[0,\infty)}p(s,x,x+y)\nu(ds)}{\int_{[0,\infty)}q(s,0,y)\nu(ds)}=1.
\end{equation}
Note that
\begin{equation}\label{eq:limit_eq}
\lim_{y\rightarrow0}\frac{\int_{[0,\delta)}p(s,x,x+y)\nu(ds)}{\int_{[0,\infty)}p(s,x,x+y)\nu(ds)}=1
\end{equation}
for any $\delta>0$. This is because for $s>\delta$, $p(s,x,x+y)$ is
bounded in $s$, and $\int_{[\delta,\infty)}\nu(ds)<\infty$, so
applying the Dominated Convergence Theorem
\begin{align*}
\lim_{y\rightarrow0}\int_{[\delta,\infty)}p(s,x,x+y)\nu(ds)&=\int_{[\delta,\infty)}\lim_{y\rightarrow0}p(s,x,x+y)\nu(ds)\\
&=\int_{[\delta,\infty)}\frac{1}{\sqrt{\frac{\pi\sigma^2}{\kappa}(1-e^{-2\kappa
s})}}\exp{\Big\{-\frac{(\theta-x)^2(1-e^{-\kappa
s})}{\frac{\sigma^2}{\kappa}(1+e^{-\kappa s})}\Big\}}\nu(ds),
\end{align*}
which is finite, and hence
$\lim_{y\rightarrow0}\frac{\int_{[\delta,\infty)}p(s,x,x+y)\nu(ds)}{\int_{[0,\infty)}p(s,x,x+y)\nu(ds)}=0$.
\eqref{eq:limit_eq} is also true when $p(s,x,x+y)$ is replaced by
$q(s,0,y)$ for the same reason.

Fix an interval $[-M,M]$ for $y$. Then for any $\epsilon>0$, there
exists some $\delta>0$, such that for any $y\in[-M,M]$,
$1-\epsilon<\frac{p(s,x,x+y)}{q(s,0,y)}<1+\epsilon$ if $s<\delta$.
Hence
$$1-\epsilon<\frac{\int_{[0,\delta)}p(s,x,x+y)\nu(ds)}{\int_{[0,\delta)}q(s,0,y)\nu(ds)}<1+\epsilon$$
for any $y\in[-M,M]$. Now letting $y\rightarrow0$ we have
$$1-\epsilon\leq\lim_{y\rightarrow0}\frac{\int_{[0,\delta)} p(s,x,x+y)\nu(ds)}{\int_{[0,\delta)}q(s,0,y)\nu(ds)}\leq 1+\epsilon.$$
Equation \eqref{eq:limit_eq} and
$\lim_{y\rightarrow0}\frac{\int_{[0,\delta)}q(s,0,y)\nu(ds)}{\int_{[0,\infty)}q(s,0,y)\nu(ds)}=1$
imply that
$$\lim_{y\rightarrow0}\frac{\int_{[0,\infty)}p(s,x,x+y)\nu(ds)}{\int_{[0,\infty)}q(s,0,y)\nu(ds)}=\lim_{y\rightarrow0}\frac{\int_{[0,\delta)}p(s,x,x+y)\nu(ds)}{\int_{[0,\delta)}q(s,0,y)\nu(ds)}.$$
Hence
$$1-\epsilon\leq\lim_{y\rightarrow0}\frac{\int_{[0,\infty)}p(s,x,x+y)\nu(ds)}{\int_{[0,\infty)}q(s,0,y)\nu(ds)}\leq 1+\epsilon$$
for any $\epsilon$. Now letting $\epsilon\rightarrow0$,
\eqref{eq:limit_equivalence} is proved.\qed

{\bf \emph{Proposition \ref{prop:SubBM Asymptotic
Independence of Drift}}.} Now we prove that the asymptotic of the
L\'{e}vy density $\overline{\pi}(y)$ of a SubBM does not depend on
the drift. Suppose the drift and diffusion coefficients are $\mu$
and $\sigma$ respectively. Then we have
$$
\overline{\pi}(y)=\int_{[0,\infty)}
\frac{1}{\sqrt{2\pi\sigma^2s}}\exp\Big\{-\frac{(y-\mu
s)^2}{2\sigma^2s}\Big\}\nu(ds).
$$
Similar to the proof in Proposition \ref{prop:asymptotic behavior
SubOU=SubBM}, it is straightforward to show that
\begin{equation*}
\lim_{y\rightarrow0}\overline{\pi}(y)=\lim_{y\rightarrow0}\int_{[0,\infty)}
\frac{1}{\sqrt{2\pi\sigma^2s}}\exp\Big\{-\frac{y^2}{2\sigma^2s}\Big\}\nu(ds),
\end{equation*}
which does not depend on $\mu$.\qed

{\bf \emph{Proposition
\ref{prop:SubBMComputeAsymptoticGen}}.} We prove the case with
condition (1) here. The case with condition (2) is proved in
\cite{KimSongVondracek}. We can write:
$$\overline{\pi}(y)=\int_0^{\frac{1}{2\sigma^2\xi}}
\frac{1}{\sqrt{2\pi\sigma^2s}}\exp\Big\{-\frac{y^2}{2\sigma^2s}\Big\}\nu(s)ds+\int_{\frac{1}{2\sigma^2\xi}}^\infty
\frac{1}{\sqrt{2\pi\sigma^2s}}\exp\Big\{-\frac{y^2}{2\sigma^2s}\Big\}\nu(s)ds.$$
Similar to the proof in Proposition \ref{prop:asymptotic behavior
SubOU=SubBM}, the second integral on the RHS is finite as
$y\rightarrow0$, so we only need to be concerned with the first
integral. The rest of the proof is similar to \cite{SongVondracek}.
Let $u=y^2/(2\sigma^2s)$. Then
\begin{align*}
\int_0^{\frac{1}{2\sigma^2\xi}}
\frac{1}{\sqrt{2\pi\sigma^2s}}\exp\Big\{-\frac{y^2}{2\sigma^2s}\Big\}\nu(s)ds&=\frac{|y|}{2\sigma^2\sqrt{\pi}}\int_{\xi
y^2}^\infty u^{-\frac{3}{2}}e^{-u}\nu(\frac{y^2}{2\sigma^2u})du\\
&=\frac{(2\sigma^2)^{\beta-1}}{\sqrt{\pi}|y|^{2\beta-1}\ell(\frac{1}{y^2})}\int_{\xi
y^2}^{\infty}u^{\beta-\frac{3}{2}}e^{-u}\frac{\nu(\frac{y^2}{2\sigma^2u})}{h(y,u)}\frac{\ell(\frac{1}{y^2})}{\ell(\frac{2\sigma^2u}{y^2})}du,
\end{align*}
where
$h(y,u):=\frac{1}{(\frac{y^2}{2\sigma^2u})^\beta\ell(\frac{2\sigma^2u}{y^2})}$.
From assumption \eqref{eq:SubordinatorAsymptoticLevyDensityAt0},
there is a constant $c>0$ such that for all $u>\xi y^2$, we have
$\frac{\nu(\frac{y^2}{2\sigma^2u})}{h(y,u)}<c.$ Note that
$\ell(\frac{1}{y^2})/\ell(\frac{2\sigma^2u}{y^2})=f_{\ell,\xi}(y^2,u)$
for $u>\xi y^2$. So it follows from the assumption that we have
$u^{\beta-\frac{3}{2}}e^{-u}\frac{\nu(\frac{y^2}{2\sigma^2u})}{h(y,u)}\frac{\ell(\frac{1}{y^2})}{\ell(\frac{2\sigma^2u}{y^2})}\leq
cu^{\beta-\frac{3}{2}}e^{-u}g(u).$ By the Dominated Convergence
Theorem,
\begin{align*}
\lim_{y\rightarrow0}\int_{\xi
y^2}^{\infty}u^{\beta-\frac{3}{2}}e^{-u}\frac{\nu(\frac{y^2}{2\sigma^2u})}{h(y,u)}\frac{\ell(\frac{1}{y^2})}{\ell(\frac{2\sigma^2u}{y^2})}du
&=c_0\int_0^\infty
u^{\beta-\frac{3}{2}}e^{-u}du=c_0\Gamma(\beta-\frac{1}{2}).
\end{align*}
So the claim in Proposition \ref{prop:SubBMComputeAsymptoticGen}
follows. \qed

{\bf \emph{Proposition
\ref{prop:SubordinatorBetaAndBGIndex}}.} Let $p=\beta-1$. For any
$0<\alpha<p$ we have that $\beta-\alpha>1$.
$\frac{1}{s^{\beta-\alpha}\ell(\frac{1}{s})}$ is not integrable near
zero because
$\lim_{s\rightarrow0}\frac{1}{s^{\beta-\alpha}\ell(\frac{1}{s})}\Big/\frac{1}{s}=\lim_{s\rightarrow0}\frac{1}{s^{\beta-\alpha-1}\ell(\frac{1}{s})}=\infty$
by \cite{BinghamGoldieTeugels} Proposition 1.3.6.

For any $\alpha>p$ we have that $\beta-\alpha<1$. There is a value
$\gamma$ such that $\beta-\alpha<\gamma<1$. So
$\lim_{s\rightarrow0}\frac{1}{s^{\beta-\alpha}\ell(\frac{1}{s})}\Big/\frac{1}{s^{\gamma}}=\lim_{s\rightarrow0}\frac{1}{s^{\beta-\alpha-\gamma}\ell(\frac{1}{s})}=0,$
again from \cite{BinghamGoldieTeugels} Proposition 1.3.6. Therefore
$\frac{1}{s^{\beta-\alpha}\ell(\frac{1}{s})}$ is integrable near $0$
for any $\alpha>p$.

Together we have the BG index is $\beta-1$. The assertion for the
second part follows from the asymptotic implied by Proposition
\ref{prop:SubBMComputeAsymptoticGen}.\qed

{\bf \emph{Theorem \ref{thm:SubOUGenMeasureChange}}.} The
proof is entirely similar to the proof of the necessity part of
Theorem \ref{thm:EMCSubOU}. First, JS Theorem \Rmnum{3}.3.24 implies
\eqref{eq:SubOUGenMCCharacteristicsRelations},
\eqref{eq:SubOUGenMCBeta} and \eqref{eq:SubOUGenMCRandomMeasure}. To
prove the \eqref{eq:SubOUGenMCHellinger} and the form of the density
process, replace $\int_0^t\int_{y\neq
0}\Big(\sqrt{\pi'(X_{s-}(\omega),y)}-\sqrt{\pi(X_{s-}(\omega),y)}\Big)^2dyds$
by
$\int_0^t\int_{y\neq0}(\sqrt{Y(s,\omega,y)}-1)^2\pi(X_{s-}(\omega),y)dyds$
and the rest remains the same.\qed

{\bf \emph{Theorem 2.7}.} (1) First we notice two facts. (1) On any
compact interval $I\in {\mathbb R}$, there exists a constant $C$
depending on $I$, such that for $n\geq1$ (c.f.
\cite{NikiforovUvarov} p.54 Eq. (28a))
\begin{equation}\label{eq:boundVarphi}
|\varphi_n(x)|\leq C n^{-1/4},\quad x\in I.
\end{equation}
(2) $|f_n|\leq\|f\|$ for all $n$ by the Cauchy-Schwartz inequality.

For $t>0$, on one hand, the RHS of \eqref{eq:OUsp} is bounded by
$|f_0|+C\|f\|\sum_{n=1}^{\infty}e^{-\kappa nt}n^{-\frac{1}{4}}$,
which is finite due to the rapid decay of $e^{-\kappa nt}$. This
expansion converges absolutely for each $x$ and uniformly in $x$ on
compacts, thus it defines a continuous function. On the other hand,
the function $\mathcal{P}_tf(x)$ is infinitely differentiable in
$x$. In fact, if $x$ is replaced with a complex variable $z$,
$\mathcal{P}_tf(z)$ is an entire function (see Theorem 3.1 in
\cite{Thangavelu}). The $L^2$-convergence implies convergence almost
everywhere in this case. To be more precise, let $S(x)$ denotes the
RHS of \eqref{eq:OUsp}, and $S_n(x)$ its $n$-th partial sum.
Convergence of $S_n(x)$ to $\mathcal{P}_tf(x)$ in $L^2$ implies that
there is a subsequence $S_{k_n}(x)$ converging to
$\mathcal{P}_tf(x)$ almost everywhere. But the limit of $S_{k_n}(x)$
is $S(x)$, so $S(x)=\mathcal{P}_tf(x)$ almost everywhere.
Furthermore, since both sides of \eqref{eq:OUsp} are continuous
functions, they must agree at every point. Therefore, for the OU
semigroup, the eigenfunction expansion in \eqref{eq:OUsp} is valid
pointwise for each $f\in L^2(\mathbb{R},\mathfrak{m})$ and $t>0$.

(2) For the SubOU semigroup, when the eigenfunction expansion on the RHS
of \eqref{eq:SubOUsp} converges absolutely for each $x$, the
spectral representation \eqref{eq:SubOUsp} for
$\mathcal{P}_t^{\phi}f(x)$ is valid for each $x$, as the following
calculation can be justified:
\begin{align*}
\mathcal{P}^\phi_tf(x)&=\int_{[0,\infty)}\mathcal{P}_sf(x)q_t(ds)=\int_{[0,\infty)}\sum_{n=0}^{\infty}e^{-\kappa
ns}f_n\varphi_n(x)q_t(ds)\\
&=\sum_{n=0}^{\infty}\int_{[0,\infty)}e^{-\kappa
ns}q_t(ds)f_n\varphi_n(x)=\sum_{n=0}^{\infty}e^{-\phi(\kappa
n)t}f_n\varphi_n(x).
\end{align*}
In the above, we first use the definition of the SubOU semigroup,
then represent $\mathcal{P}_sf(x)$ by the eigenfunction expansion
which also converges pointwise, interchange the summation and
expectation justified by the absolute convergence of the expansion
and the dominated convergence theorem, and use the Laplace transform
of the convolution semigroup.

For $t>0$, either condition (i) or (ii) in Theorem
\ref{thm:PointwiseConvergenceHermiteExpansion} ensures the absolute
convergence of the expansion for each $x$, and thus the
eigenfunction expansion for the SubOU semigroup converges pointwise.
\qed

{\bf \emph{Theorem 2.8}.} Convergence of the expansion in the RHS of
\eqref{eq:OUpdfsp} to the RHS of \eqref{eq:OUTPD} follows from the
well-known Mehler formula for Hermite polynomials (e.g., Thangavelu
(2006) Proposition 2.3). Part (2) follows from the estimate (B.4)
for the eigenfunctions. \qed

{\bf \emph{Theorem 2.9}.} We first notice that for all $n$ and all
real $x$,
$\left|\varphi_n(x)\right|\leq1.0864e^{\frac{\kappa(x-\theta)^2}{2\sigma^2}}$.
This bound is given in \cite{Boyd} and is shown there to be tight.
Therefore we have $\left|\sum_{n=M}^{\infty}e^{-\phi(\kappa
n)t}f_n\varphi_n(x)\right|\leq
1.0864e^{\frac{\kappa(x-\theta)^2}{2\sigma^2}}\sum_{n=M}^{\infty}e^{-\phi(\kappa
n)t}|f_n|$. Using $|f_n|\leq \|f\|$ and assuming that
$\sum_{n=0}^{\infty}e^{-\phi(\kappa n)t}<\infty$ is satisfied for
all $t>0$, we obtain the estimate in Theorem 2.9. \qed

{\bf \emph{Lemma \ref{lemma:expExpansion} and Theorem
\ref{thm:SubOUfuturepricing}}.} First, note that the function
$e^x\in L^2(\mathbb{R},\mathfrak{m})$, so the spectral
representation theorem applies and
\begin{align*}
f_n&=\int_{-\infty}^{\infty}e^x \frac{1}{\sqrt{2^n
n!}}H_n\left(\frac{\sqrt{\kappa}}{\sigma}(x-\theta)\right)\sqrt{\frac{\kappa}{\pi\sigma^2}}e^{-\frac{\kappa(\theta-x)^2}{\sigma^2}}dx\\
&=e^{\theta+\frac{\sigma^2}{4\kappa}}\frac{1}{\sqrt{\pi2^nn!}}\int_{-\infty}^{\infty}e^{-(y-\frac{\sigma}{2\sqrt{\kappa}})^2}H_n(y)dy
=e^{\theta+\frac{\sigma^2}{4\kappa}}\frac{1}{\sqrt{n!}}\left(\frac{\sigma}{\sqrt{2\kappa}}\right)^n,
\end{align*}
where we used the identity
$\int_{-\infty}^{\infty}e^{-(y-z)^2}H_n(y)dy=\sqrt{\pi}(2z)^n$
(\cite{PrudnikovBrychkovMarichev} p.488 No.17 of 2.20.3). It can be
shown by using the estimate of the eigenfunctions
\eqref{eq:boundVarphi} that the Hermite expansion of the exponential
function is absolutely convergent for each $x$, hence condition (i)
in Theorem \ref{thm:PointwiseConvergenceHermiteExpansion} is
satisfied. The results in Theorem \ref{thm:SubOUfuturepricing} are
obtained by applying \eqref{eq:SubOUsp} to $e^x$.\qed

{\bf \emph{Theorem \ref{thm:EMCSubOUvsSubOUDrift}}.} {\em
Necessity.} Let $(B^P,C^P,\Pi^P)$ be the semimartingale
characteristics of the SubOU process with generating tuple
$(\kappa_P,\theta_P,\sigma_P,\gamma_P,\nu_P)$. Then
$(B^P+H,C^P,\Pi^P)$ is the set of characteristics for $X$ under
$\mathbb{P}$. Since $\mathbb{P}$ and $\mathbb{Q}$ are locally
equivalent, Theorem \ref{thm:SubOUGenMeasureChange} implies
condition (2) and (3), and that there exists some deterministic
function $\overline{\beta}$ such
\begin{align*}
B^P_t(\omega)+H(t)&=B_t(\omega)+\gamma\sigma^2\int_0^t\left(\beta_s(\omega)+\overline{\beta}(s)\right)ds\\
&+\int_{[0,t]\times\mathbb{R}}y{1}_{\{|y|\leq1\}}\left(\pi^P(X_{s-}(\omega),y)-\pi(X_{s-}(\omega),y)\right)dyds,
\end{align*}
where
$\beta_s(\omega)=\frac{(\gamma_P\kappa_P\theta_P-\gamma\kappa\theta)-(\gamma_P\kappa_P-\gamma\kappa)X_{s-}(\omega)}{\gamma\sigma^2}{1}_{\{\gamma\neq0\}}$.
Thus if $\gamma>0$, then $H$ is an absolutely continuous function of
time, and $H(0)=0$. If $\gamma=0$, then $H(t)=0$ for all $t$.

{\em Sufficiency.} If $\gamma=0$, then the conclusion is directly
implied by Theorem \ref{thm:EMCSubOU}. If $\gamma>0$, then using
Theorem \ref{thm:EMCSubOU}, we can first find a measure
$\widetilde{\mathbb{P}}$ locally equivalent to $\mathbb{Q}$, and
under $\widetilde{\mathbb{P}}$, $X$ is a SubOU process with
generating tuple $(\kappa_P,\theta_P,\sigma_P,\gamma_P,\nu_P)$. Let
$X^c$ be the continuous local martingale part of $X$ under
$\widetilde{\mathbb{P}}$. Since $H$ is absolutely continuous, we can
define $\lambda(t):=\frac{1}{\gamma_P\sigma_P^2}\frac{dH(t)}{dt}$.
Then define a measure $\mathbb{P}$ by
$\frac{d\mathbb{P}}{d\widetilde{\mathbb{P}}}=\mathscr{E}(\lambda\cdot
X^c)$. This is a Radon-Nikodym density process because the Novikov
condition is satisfied, so the stochastic exponential is a true
martingale. Now $\mathbb{P}$ and $\widetilde{\mathbb{P}}$ are
locally equivalent. Under $\mathbb{P}$, the first component of the
semimartingale characteristics becomes
$B^P_t+\int_0^t\lambda_s\gamma_P\sigma_P^2ds=B^P_t+H(t)$. Thus, $X$
is a SubOU process with the generating tuple
$(\kappa_P,\theta_P,\sigma_P,\gamma_P,\nu_P)$ plus a deterministic
function $H(t)$.\qed

{\bf \emph{Theorem \ref{thm:SubOUOptionPricing}}.} Since
the put payoff is bounded and the measure is Gaussian, it belongs to
$L^2(\mathbb{R},\mathfrak{m})$. The expansion coefficients are
computed as follows:
$$\int_{-\infty}^{\infty}(K-F(x,t,t^*))^+\varphi_n(x)\mathfrak{m}(x)dx=\int_{-\infty}^{\infty}
(K-F(x,t,t^*)){{1}}_{\{x<x^*\}}\varphi_n(x)\mathfrak{m}(x)dx.$$
$$
\int_{-\infty}^{\infty}
K{{1}}_{\{x<x^*\}}\varphi_n(x)\mathfrak{m}(x)dx=\frac{K}{\sqrt{\pi2^nn!}}\int_{-\infty}^{\frac{\sqrt{\kappa}}{\sigma}(x^*-\theta)}H_n(x)e^{-x^2}dx
=\frac{K}{\sqrt{\pi2^nn!}}b_n(w^*).
$$
The integral in \eqref{eq:bCoeff} is given in
\cite{PrudnikovBrychkovMarichev}. For the second integral,
$$
\int_{-\infty}^{\infty}F(x,t,t^*){{1}}_{\{x<x^*\}}\varphi_n(x)\mathfrak{m}(x)dx=\int_{-\infty}^{x^*}Fe^{-G(t^*)}\sum_{m=0}^{\infty}e^{-\phi(\kappa m)\tau}f_m\varphi_m(x)\varphi_n(x)\mathfrak{m}(x)dx
$$
$$
=Fe^{-G(t^*)}\sum_{m=0}^{\infty}e^{-\phi(\kappa
m)\tau}f_m\int_{-\infty}^{x^*}\varphi_m(x)\varphi_n(x)\mathfrak{m}(x)dx
=\frac{1}{\sqrt{\pi2^nn!}}Fe^{\theta +\frac{\sigma^2}{4\kappa}
-G(t^*)} \sum_{m=0}^\infty e^{-\phi(\kappa
m)\tau}\frac{\alpha^m}{m!}a_{n,m}(w^*).
$$
The interchange of integration and summation is justified by the
Dominated Convergence Theorem due to the estimate:
$$\left|\int_{-M}^{x^*}\varphi_m(x)\varphi_n(x)\mathfrak{m}(x)dx\right|\leqslant\int_{-\infty}^{\infty}\left|\varphi_m(x)\varphi_n(x)\right|\mathfrak{m}(x)dx\\
\leqslant||\varphi_m||\cdot||\varphi_n||=1,$$ and
$\sum_{m=0}^{\infty}e^{-\phi(\kappa m)\tau}f_m<\infty$. With some
further simplifications we obtain \eqref{eq:SubOUPutPrice}. The
integral in \eqref{eq:aCoeff} is calculated as follows. Consider the
integral $J_{n,m}^L(x):=\int_{-\infty}^xH_n(z)H_m(z)e^{-z^2}dz$. By
the identity
$H_n(z)H_m(z)=\sum_{k=0}^{min(n,m)}\binom{m}{k}\binom{n}{k}2^kk!H_{n+m-2k}(z)$
(\cite{PrudnikovBrychkovMarichev} p.640 No.11 of 4.5.1), we have
$$J_{n,m}^L(x)
=\sum_{k=0}^{min(n,m)}\binom{m}{k}\binom{n}{k}2^kk!\int_{-\infty}^xH_{n+m-2k}(z)e^{-z^2}dz=\sum_{k=0}^{min(n,m)}\binom{m}{k}\binom{n}{k}2^kk!b_{n+m-2k}(x).
\qed$$

{\bf \emph{Theorem \ref{thm:TI-CIR-SubOUSpectralExpansion}}.}
$$
\mathbb{E}[f(Y_t)|Y_s,Z_s]=\mathbb{E}\Big[\mathbb{E}[f(X_{T_t})|T_t-T_s,Y_s,Z_s]\Big|Y_s,Z_s\Big]
={\mathbb E}\Big[\sum_{n=0}^\infty
e^{-\phi(\kappa n)(T_t-T_s)}f_n\varphi_n(Y_s)\Big|Z_s\Big]
$$
$$
=\sum_{n=0}^\infty {\mathbb E}[e^{-\phi(\kappa n)\int_s^t
(a(u)+Z_u)du}|Z_s]f_n\varphi_n(Y_s)
=\sum_{n=0}^\infty e^{-\phi(\kappa n)\int_s^t a(u)du}
\mathcal{L}_{CIR}\Big(t-s,\phi(\lambda)\Big|Z_s\Big)
f_n\varphi_n(Y_s),
$$
where condition (1) or (2) in Theorem
\ref{thm:TI-CIR-SubOUSpectralExpansion} justify the interchange of
summation and expectation.\qed

{\bf \emph{Proposition \ref{prop:cross-variation}}.} Define
$\tilde{Z}_t=Z^c_{S_t}$, where $S$ is the inverse of $T$. Then
$Z^c_t=\tilde{Z}_{T_t}$. Since the time change $S$ is continuous,
$Z^c$ is adapted to $S$ (see \cite{Jacod} Definition \Rmnum{10}.13
for adaption to a time change), and by \cite{Jacod} Theorem
\Rmnum{10}.16, $\tilde{Z}$ is a continuous local martingale w.r.t.
$(\mathcal{F}_t)_{t\geq0}$. Now
$[Y^c,Z^c]_t=[X^c_T,\tilde{Z}_T]_t=[X^c,\tilde{Z}]_{T_t},$
where the second equality is from \cite{Jacod} Theorem
\Rmnum{10}.17. Since $X$ and $Z$ are independent, $X^c$ and
$\tilde{Z}$ are independent. Because the cross-variation of two
independent continuous local martingale is 0, we have
$[X^c,\tilde{Z}]_t=0$ for all $t$, hence $[X^c,\tilde{Z}]_{T_t}=0$,
and the claim is proved. \qed

{\bf \emph{Theorem \ref{Thm:TI-CIR-SubOUoptionpricing}}.}
Conditioning on the terminal state $Z_t$ of the CIR process, we
have:
\begin{align*}
&\mathbb{E}\left[\left(K-F(Y_t,Z_t,t,t^*)\right)^+\right]
=\int_0^{\infty}\mathbb{E}\left[(K-F(Y_t,z_t,t,t^*))^+\left|Z_t=z_t\right]\right.p_{CIR}(t,z_0,z_t)dz_t\\
&=\int_0^{\infty}\mathbb{E}\left[\left.\sum_{n=0}^{\infty}e^{-\phi(\kappa
n)T_t}p_n(t,t^*,w^*,F)\varphi_n(y_0)\right|Z_t=z_t\right]p_{CIR}(t,z_0,z_t)dz_t\\
&=\int_0^{\infty}\left\{\sum_{n=0}^{\infty}\mathbb{E}\left[\left.e^{-\phi(\kappa
n)T_t}\right|Z_t=z_t\right]p_n(t,t^*,w^*,F)\varphi_n(y_0)\right\}p_{CIR}(t,z_0,z_t)dz_t\\
&=\int_0^{\infty}\left\{\sum_{n=0}^{\infty} e^{-\phi(\kappa
n)\int_0^t a(u)du}{\cal L}_{CIR}(t,\phi(\kappa n)|z_0,z_t)
p_n(t,t^*,w^*,F)\varphi_n(y_0)\right\}p_{CIR}(t,z_0,z_t)dz_t.
\end{align*}
The interchange of expectation and summation is justified by the
assumption.\qed

{\bf {\em Proposition 5.1}.} Using the recursion for
Hermite polynomials, for $n\geq1$, $m\geq0$,
\begin{align*}
&a_{n+1,m+1}(x)=\int_{-\infty}^{x}H_{n+1}(y)H_{m+1}(y)e^{-y^2}dy
=\int_{-\infty}^{x}[2yH_{n}(y)-2nH_{n-1}(y)]H_{m+1}(y)e^{-y^2}dy\\
&=-\int_{-\infty}^{x}H_{n}(y)H_{m+1}(y)de^{-y^2}-2na_{n-1,m+1}(x)\\
&=-H_{n}(y)H_{m+1}(y)e^{-y^2}|_{-\infty}^x+\int_{-\infty}^x[H_{n}'(y)H_{m+1}(y)+H_{n}(y)H_{m+1}'(y)]e^{-y^2}dy
-2na_{n-1,m+1}(x)\\
&=-H_{n}(y)H_{m+1}(y)e^{-y^2}+\int_{-\infty}^x[2nH_{n-1}(y)H_{m+1}(y)+2(m+1)H_{n}(y)H_{m}(y)]e^{-y^2}dy
-2na_{n-1,m+1}(x)\\
&=2(m+1)a_{n,m}(x)-H_{n}(x)H_{m+1}(x)e^{-x^2}.
\end{align*}
It is easy to verify that this recursion is also true for $n=0$.
Therefore we have
\begin{equation}\label{eq:anm}
a_{n+1,m+1}(x)=2(m+1)a_{n,m}(x)-H_{n}(x)H_{m+1}(x)e^{-x^2},\quad
n\geq 0,m\geq 0.
\end{equation}
In particular,
$a_{n,n}(x)=2na_{n-1,n-1}(x)-H_{n-1}(x)H_n(x)e^{-x^2},$ $n\geq 1$.
Noting the symmetry $a_{n,m}(x)=a_{m,n}(x)$, we also obtain
the following by exchanging the role of $n$ and $m$ in
\eqref{eq:anm}:
\begin{equation}\label{eq:amn}
a_{m+1,n+1}(x)=2(n+1)a_{n,m}(x)-H_{m}(x)H_{n+1}(x)e^{-x^2},\quad
m\geq0, n\geq0.
\end{equation}
If $m\neq n$, subtracting \eqref{eq:amn} from \eqref{eq:anm}, we
obtain:
$$a_{n,m}(x)=e^{-x^2}(H_{n}(x)H_{m+1}(x)-H_{m}(x)H_{n+1}(x))/(2(m-n))\ \ (n\neq m,n\geq0,m\geq0).\qed$$

\bibliography{TCOUbib}

\begin{thebibliography}{}

\bibitem[\protect\citeauthoryear{Albanese and Kuznetsov}{Albanese and
  Kuznetsov}{2004}]{AlbaneseKuznetsov}
Albanese, C. and A.~Kuznetsov (2004).
\newblock Unifying the three volatility models.
\newblock {\em Risk\/}~{\em 17\/}(3), 94--98.

\bibitem[\protect\citeauthoryear{Alberver\.{i}o and
  R\"{u}d\.{i}ger}{Alberver\.{i}o and
  R\"{u}d\.{i}ger}{2003}]{AlbeverioRudiger03}
Alberver\.{i}o, S. and B.~R\"{u}d\.{i}ger (2003).
\newblock Infinite-dimensional stochastic differential equations obtained by
  subordination and related {D}irichelet forms.
\newblock {\em Journal of Functional Analysis\/}~{\em 204}, 122--156.

\bibitem[\protect\citeauthoryear{Alberver\.{i}o and
  R\"{u}d\.{i}ger}{Alberver\.{i}o and
  R\"{u}d\.{i}ger}{2005}]{AlbeverioRudiger05}
Alberver\.{i}o, S. and B.~R\"{u}d\.{i}ger (2005).
\newblock Subordination of symmetric quasi-regular {D}irichlet forms.
\newblock {\em Random Operators and Stochastic Equations\/}~{\em 13\/}(1),
  17--38.

\bibitem[\protect\citeauthoryear{Andersen}{Andersen}{2008}]{Andersen}
Andersen, L. (2008).
\newblock Markov models for commodity futures: theory and practice.
\newblock {\em Working Paper, Banc of America Securities\/}.

\bibitem[\protect\citeauthoryear{Andersen, Bollerslev, and Diebold}{Andersen
  et~al.}{2009}]{AndersenBollerslevDiebold}
Andersen, T., T.~Bollerslev, and F.~Diebold (2009).
\newblock Parametric and nonparametric volatility measurement.
\newblock In Y.~A\"{i}t-Sahalia and L.~Hansen (Eds.), {\em Handbook of
  Financial Econometrics}, Volume~1, Chapter~2. North-Holland.

\bibitem[\protect\citeauthoryear{Bakry and Mazet}{Bakry and
  Mazet}{2004}]{BakryMazet}
Bakry, D. and O.~Mazet (2004).
\newblock Characterization of {M}arkov semigroups on $\mathbb{R}$ associated to
  some families of orthogonal polynomials.
\newblock In {\em S\'{e}minaire de Probabilit\'{e}s XXXVII, Lectures Notes in
  Mathematics}, Volume 1832, pp.\  60--80. Springer.

\bibitem[\protect\citeauthoryear{Barndorff-Nielsen}{Barndorff-Nielsen}{1998}]{%
BarndorffNielsen}
Barndorff-Nielsen, O.~E. (1998).
\newblock Processes of {N}ormal {I}nverse {G}aussian type.
\newblock {\em Finance and Stochastics\/}~{\em 2}, 41--68.

\bibitem[\protect\citeauthoryear{Benth and \v{S}altyt\.{e} Benth}{Benth and
  \v{S}altyt\.{e} Benth}{2004}]{2Benth}
Benth, F.~E. and J.~\v{S}altyt\.{e} Benth (2004).
\newblock The {N}ormal {I}nverse {G}aussian distribution and spot price
  modeling in energy markets.
\newblock {\em International Journal of Theoretical and Applied Finance\/}~{\em
  7\/}(2), 177--192.

\bibitem[\protect\citeauthoryear{Bertoin}{Bertoin}{1996}]{Bertoin}
Bertoin, J. (1996).
\newblock {\em L\'{e}vy Processes}.
\newblock Cambidge University Press.

\bibitem[\protect\citeauthoryear{Bessembinder, Coughenour, Seguin, and
  Smoller}{Bessembinder et~al.}{1995}]{BSCC}
Bessembinder, H., J.~F. Coughenour, P.~J. Seguin, and M.~M. Smoller (1995).
\newblock Mean reversion in equilibrium asset prices: evidence from the futures
  term structure.
\newblock {\em The Journal of Finance\/}~{\em 50\/}(1), 361--375.

\bibitem[\protect\citeauthoryear{Bingham, Goldie, and Teugels}{Bingham
  et~al.}{1987}]{BinghamGoldieTeugels}
Bingham, N.~H., C.~M. Goldie, and J.~L. Teugels (1987).
\newblock {\em Regular Variation}.
\newblock Cambridge University Press.

\bibitem[\protect\citeauthoryear{Bochner}{Bochner}{1949}]{Bochner}
Bochner, S. (1949).
\newblock Diffusion equations and stochastic processes.
\newblock {\em Proceedings of the National Academy of Sciences of the United
  States of America\/}~{\em 35}, 368--370.

\bibitem[\protect\citeauthoryear{Boyarchenko and
  Levendorski\u{\.{i}}}{Boyarchenko and
  Levendorski\u{\.{i}}}{2007}]{BoyarchenkoLevendorskii}
Boyarchenko, N. and S.~Levendorski\u{\.{i}} (2007).
\newblock The eigenfunction expansion method in multifactor quadratic term
  structure models.
\newblock {\em Mathematical Finance\/}~{\em 17\/}(4), 503--539.

\bibitem[\protect\citeauthoryear{Boyd}{Boyd}{1984}]{Boyd}
Boyd, J.~P. (1984).
\newblock Asymptotic coefficients of {H}ermite functions series.
\newblock {\em Journal of Computational Physics\/}~{\em 54}, 382--410.

\bibitem[\protect\citeauthoryear{Brigo and Mercurio}{Brigo and
  Mercurio}{2006}]{BrigoMercurio}
Brigo, D. and F.~Mercurio (2006).
\newblock {\em Interest Rate Models---Theory and Practice\/} (2nd ed.).
\newblock Springer.

\bibitem[\protect\citeauthoryear{Broadie and Kaya}{Broadie and
  Kaya}{2006}]{BroadieKaya}
Broadie, M. and O.~Kaya (2006).
\newblock Exact simulation of stochastic volatility and other affine jump
  diffusion processes.
\newblock {\em Operations Research\/}~{\em 54\/}(2), 217--231.

\bibitem[\protect\citeauthoryear{Carr, Geman, Madan, and Yor}{Carr
  et~al.}{2003}]{CarrGemanMadanYorSVLP}
Carr, P., H.~Geman, D.~B. Madan, and M.~Yor (2003).
\newblock Stochastic volatility for {L}\'{e}vy processes.
\newblock {\em Mathematical Finance\/}~{\em 13\/}(3), 345--382.

\bibitem[\protect\citeauthoryear{Carr and Madan}{Carr and
  Madan}{1999}]{CarrMadanFFT}
Carr, P. and D.~Madan (1999).
\newblock Option pricing and the fast {F}ourier transform.
\newblock {\em Journal of Computational Finance\/}~{\em 2\/}(4), 61--73.

\bibitem[\protect\citeauthoryear{Carr and Wu}{Carr and Wu}{2004}]{CarrWu}
Carr, P. and L.~Wu (2004).
\newblock Time changed {L}\'{e}vy processes and option pricing.
\newblock {\em Journal of Financial Economics\/}~{\em 71}, 113--141.

\bibitem[\protect\citeauthoryear{Casassus and Collin-Dufresne}{Casassus and
  Collin-Dufresne}{2005}]{CasassusDufresne}
Casassus, J. and P.~Collin-Dufresne (2005).
\newblock Stochastic convenience yield implied from commodity futures and
  interest rates.
\newblock {\em The Journal of Finance\/}~{\em 60\/}(5), 2283--2331.

\bibitem[\protect\citeauthoryear{Clelow and Strickland}{Clelow and
  Strickland}{1999}]{ClelowStrickland}
Clelow, L. and C.~Strickland (1999).
\newblock Valuing energy options in a one factor model fitted to forward
  prices.
\newblock Technical report, School of Finance and Economics, University of
  Technology, Sydney, Australia.

\bibitem[\protect\citeauthoryear{Crosby}{Crosby}{2008}]{Crosby}
Crosby, J. (2008).
\newblock A multi-factor jump-diffusion model for commodities.
\newblock {\em Quantitative Finance\/}~{\em 8\/}(2), 181--200.

\bibitem[\protect\citeauthoryear{Deng}{Deng}{1999}]{Deng}
Deng, S.~J. (1999).
\newblock Stochastic models of energy commodity prices and their applications:
  mean reversion with jumps and spikes.
\newblock Technical report, POWER.

\bibitem[\protect\citeauthoryear{Duffie, Filipovi\'{c}, and
  Schachermayer}{Duffie et~al.}{2003}]{DuffieFilipovicSchachermayer}
Duffie, D., D.~Filipovi\'{c}, and W.~Schachermayer (2003).
\newblock Affine processes and applications in finance.
\newblock {\em The Annals of Applied Probability\/}~{\em 13\/}(3), 984--1053.

\bibitem[\protect\citeauthoryear{Ethier and Kurtz}{Ethier and
  Kurtz}{1986}]{EthierKurtz}
Ethier, S.~N. and T.~G. Kurtz (1986).
\newblock {\em Markov Processes: Characterization and Convergence}.
\newblock John Wiley \& Sons, Inc.

\bibitem[\protect\citeauthoryear{Eydeland and Geman}{Eydeland and
  Geman}{1998}]{EydelandGeman}
Eydeland, A. and H.~Geman (1998).
\newblock Pricing power derivatives.
\newblock {\em Risk\/}.

\bibitem[\protect\citeauthoryear{Eydeland and Wolyniec}{Eydeland and
  Wolyniec}{2003}]{EydelandWolyniec}
Eydeland, A. and K.~Wolyniec (2003).
\newblock {\em Energy and Power Risk Management}.
\newblock John Wiley \& Sons Inc.

\bibitem[\protect\citeauthoryear{Feng and Linetsky}{Feng and
  Linetsky}{2008}]{FengLinetksy08}
Feng, L. and V.~Linetsky (2008).
\newblock Pricing discretely monitored barrier options and defaultable bonds in
  {L}\'{e}vy process models: A fast {H}ilbert transform approach.
\newblock {\em Mathematical Finance\/}~{\em 18\/}(3), 337--384.

\bibitem[\protect\citeauthoryear{Feng and Linetsky}{Feng and
  Linetsky}{2009}]{FengLinetsky09}
Feng, L. and V.~Linetsky (2009).
\newblock Computing exponential moments of the discrete maximum of a {L}\'{e}vy
  process and lookback options.
\newblock {\em Finance and Stochastics\/}~{\em 13\/}(4), 501--529.

\bibitem[\protect\citeauthoryear{Fukushima, Oshima, and Takeda}{Fukushima
  et~al.}{1994}]{FukushimaOshimaTakeda}
Fukushima, M., Y.~Oshima, and M.~Takeda (1994).
\newblock {\em Dirichlet forms and symmetric Markov processes}.
\newblock W.de Gruyter.

\bibitem[\protect\citeauthoryear{Geman}{Geman}{2005}]{Geman}
Geman, H. (2005).
\newblock {\em Commodities and commodity derivatives: modeling and pricing for
  agriculturals, metals and energy}.
\newblock John Wiley \& Sons Inc.

\bibitem[\protect\citeauthoryear{Geman}{Geman}{2008}]{Geman08}
Geman, H. (2008).
\newblock {\em Risk Management in Commodity Markets : from Shipping to
  Agriculturals and Energy}.
\newblock John Wiley \& Sons Inc.

\bibitem[\protect\citeauthoryear{Geman, Madan, and Yor}{Geman
  et~al.}{2001}]{GemanMadanYor}
Geman, H., D.~B. Madan, and M.~Yor (2001).
\newblock Time changes for {L}\'{e}vy processes.
\newblock {\em Mathemtical Finance\/}~{\em 11\/}(1), 79--96.

\bibitem[\protect\citeauthoryear{Geman and Roncoroni}{Geman and
  Roncoroni}{2005}]{GemanRoncoroni}
Geman, H. and A.~Roncoroni (2005).
\newblock Understanding the fine structure of electricity prices.
\newblock {\em The Journal of Business\/}~{\em 79\/}(3), 1225--1261.

\bibitem[\protect\citeauthoryear{Gorovoi and Linetsky}{Gorovoi and
  Linetsky}{2004}]{GorovoiLinetsky}
Gorovoi, V. and V.~Linetsky (2004).
\newblock {B}lack's model of interest rates as options, eigenfunction expansion
  and {J}apanese interest rates.
\newblock {\em Mathematical Finance\/}~{\em 14}, 49--78.

\bibitem[\protect\citeauthoryear{Haykin}{Haykin}{2001}]{Haykin}
Haykin, S. (Ed.) (2001).
\newblock {\em Kalman Filtering and Neural Networks}.
\newblock Wiley Inter-Science.

\bibitem[\protect\citeauthoryear{Hilliard and Reis}{Hilliard and
  Reis}{1998}]{HilliardReis}
Hilliard, R. and J.~Reis (1998).
\newblock Valuation of commodity futures and options under stochastic
  convenience yileds, interest rates, and jump diffusions in the spot.
\newblock {\em Journal of Financial and Quantitative Analysis\/}~{\em 33\/}(1),
  61--86.

\bibitem[\protect\citeauthoryear{Hilliard and Reis}{Hilliard and
  Reis}{1999}]{HilliardReis99}
Hilliard, R. and J.~Reis (1999).
\newblock {J}ump processes in commodity futures prices and options pricing.
\newblock {\em American Journal of Agricultural Economics\/}~{\em 81\/}(2),
  273--286.

\bibitem[\protect\citeauthoryear{Hull and White}{Hull and
  White}{1993}]{HullWhite}
Hull, J. and A.~White (1993).
\newblock One factor interest rate models and the valuation of interest rate
  derivative securities.
\newblock {\em Journal of Financial and Quantitative Analysis\/}~{\em 28\/}(3),
  235--254.

\bibitem[\protect\citeauthoryear{Jacob}{Jacob}{2001}]{JacobVol1}
Jacob, N. (2001).
\newblock {\em Pseudo-differential operators and {M}arkov processes}, Volume~1.
\newblock Imperial College Press.

\bibitem[\protect\citeauthoryear{Jacob}{Jacob}{2005}]{JacobVol3}
Jacob, N. (2005).
\newblock {\em Pseudo-differential operators and {M}arkov processes}, Volume~3.
\newblock Imperial College Press.

\bibitem[\protect\citeauthoryear{Jacod}{Jacod}{1979}]{Jacod}
Jacod, J. (1979).
\newblock Calcul stochastique et probl\`{e}mes de martingales.
\newblock {\em Lecture Notes in Mathematics\/}~{\em 714}.

\bibitem[\protect\citeauthoryear{Jacod and Shiryaev}{Jacod and
  Shiryaev}{2003}]{JacodShiryaev}
Jacod, J. and A.~Shiryaev (2003).
\newblock {\em Limit Theorems for Stochastic Processes}.
\newblock Springer.

\bibitem[\protect\citeauthoryear{Javaheri, Lautier, and Galli}{Javaheri
  et~al.}{2003}]{JavaheriLautierGalli}
Javaheri, A., D.~Lautier, and A.~Galli (2003).
\newblock Filtering in finance.
\newblock {\em Wilmot Magazine\/}~{\em 2003\/}(3), 67--83.

\bibitem[\protect\citeauthoryear{Kalev and Duong}{Kalev and
  Duong}{2008}]{KalevDuong}
Kalev, P.~S. and H.~N. Duong (2008).
\newblock A test of the {S}amuelson hypothesis using realized range.
\newblock {\em The Journal of Futures Markets\/}~{\em 28\/}(7), 680--696.

\bibitem[\protect\citeauthoryear{Kallsen}{Kallsen}{2006}]{Kallsen}
Kallsen, J. (2006).
\newblock A didactic note on affine stochastic volatility models.
\newblock In Y.~Kabanov, R.~Liptser, and J.~Stoyanov (Eds.), {\em From
  Stochastic Calculus to Mathematical Finance}, pp.\  343--368. Springer.

\bibitem[\protect\citeauthoryear{Kallsen and Shiryaev}{Kallsen and
  Shiryaev}{2002}]{KallsenShiryaev}
Kallsen, J. and A.~N. Shiryaev (2002).
\newblock Time change representation of stochastic integrals.
\newblock {\em Theory of Probability and Its Applications\/}~{\em 46\/}(3),
  522--528.

\bibitem[\protect\citeauthoryear{Karlin and Taylor}{Karlin and
  Taylor}{1981}]{KarlinTaylor}
Karlin, S. and H.~M. Taylor (1981).
\newblock {\em A Second Course in Stochastic Processes}.
\newblock Academic Press.

\bibitem[\protect\citeauthoryear{Kim, Song, and Vondra\v{c}ek}{Kim
  et~al.}{2010}]{KimSongVondracek}
Kim, P., R.~Song, and Z.~Vondra\v{c}ek (2010).
\newblock Two-sided {G}reen function estimates for killed subordinate
  {B}rownian motions.
\newblock {\em Preprint\/}.

\bibitem[\protect\citeauthoryear{Lebedev}{Lebedev}{1972}]{Lebedev}
Lebedev, N.~N. (1972).
\newblock {\em Special Functions and Their Applications}.
\newblock Dover Publications Inc.

\bibitem[\protect\citeauthoryear{Li and Linetsky}{Li and
  Linetsky}{2011}]{LiLinetskyOS}
Li, L. and V.~Linetsky (2011).
\newblock Optimal stopping and early exercise: An eigenfunction expansion
  approach.
\newblock {\em Working Paper, Northwestern University\/}.

\bibitem[\protect\citeauthoryear{Linetsky}{Linetsky}{2004}]{LinetskyIJTAF}
Linetsky, V. (2004).
\newblock The spectral decomposition of the option value.
\newblock {\em International Journal of Theoretical and Applied Finance\/}~{\em
  7\/}(3), 337--384.

\bibitem[\protect\citeauthoryear{Linetsky}{Linetsky}{2007}]{Linetsky}
Linetsky, V. (2007).
\newblock Spectral methods in derivatives pricing.
\newblock In J.~R. Birge and V.~Linetsky (Eds.), {\em Handbook of Financial
  Engineering, Handbooks in Operations Research and Management Sciences},
  Chapter~6. Elsevier.

\bibitem[\protect\citeauthoryear{Madan, Carr, and Chang}{Madan
  et~al.}{1998}]{MadanCarrChang}
Madan, D., P.~Carr, and E.~C. Chang (1998).
\newblock The {V}ariance {G}amma process and option pricing.
\newblock {\em European Finance Review\/}~{\em 2}, 79--105.

\bibitem[\protect\citeauthoryear{McKean}{McKean}{1956}]{McKean}
McKean, H. (1956).
\newblock Elementary solutions for certain parabolic partial differential
  equations.
\newblock {\em Transactions of the American Mathematical Society\/}~{\em 82},
  519--548.

\bibitem[\protect\citeauthoryear{Mendoza, Carr, and Linetsky}{Mendoza
  et~al.}{2010}]{MendozaCarrLinetsky}
Mendoza, R., P.~Carr, and V.~Linetsky (2010).
\newblock Time changed {M}arkov processes in unified credit-equity modeling.
\newblock {\em Mathematical Finance\/}~{\em 20\/}(4), 527--569.

\bibitem[\protect\citeauthoryear{Mendoza and Linetsky}{Mendoza and
  Linetsky}{2010}]{MendozaLinetsky}
Mendoza, R. and V.~Linetsky (2010).
\newblock Pricing equity default swaps under the jump-to-default extended {CEV}
  model.
\newblock {\em Finance and Stochastics\/}~{\em 15\/}(3), 513--540.

\bibitem[\protect\citeauthoryear{Merton}{Merton}{1976}]{Merton}
Merton, R. (1976).
\newblock Option pricing when underlying stock returns are discontinuous.
\newblock {\em Journal of Financial Economics\/}~{\em 3}, 125--144.

\bibitem[\protect\citeauthoryear{Nikiforov and Uvarov}{Nikiforov and
  Uvarov}{1988}]{NikiforovUvarov}
Nikiforov, A.~F. and V.~B. Uvarov (1988).
\newblock {\em Special Functions of Mathematical Physics: A Unified
  Introduction with Applications}.
\newblock Birkh\"{a}user.

\bibitem[\protect\citeauthoryear{Okura}{Okura}{2002}]{Okura}
Okura, H. (2002).
\newblock Recurrence and transience criteria for subordinated symmetric
  {M}arkov processes.
\newblock {\em Forum Mathematicum\/}~{\em 14}, 121--146.

\bibitem[\protect\citeauthoryear{Pindyck}{Pindyck}{2001}]{Pindyck}
Pindyck, R. (2001).
\newblock The dynamics of commodity spot and futures markets: A primer.
\newblock {\em The Energy Journal\/}~{\em 22\/}(3), 1--29.

\bibitem[\protect\citeauthoryear{Protter}{Protter}{2005}]{Protter}
Protter, P.~E. (2005).
\newblock {\em Stochastic Integration and Differential Equations\/} (2nd ed.).
\newblock Springer.

\bibitem[\protect\citeauthoryear{Prudnikov, Brychkov, and Marichev}{Prudnikov
  et~al.}{1986}]{PrudnikovBrychkovMarichev}
Prudnikov, A.~P., Y.~A. Brychkov, and O.~I. Marichev (1986).
\newblock {\em Integrals and Series}, Volume~2.
\newblock Gordon and Breach Science Publishers.

\bibitem[\protect\citeauthoryear{Revuz and Yor}{Revuz and Yor}{1999}]{RevuzYor}
Revuz, D. and M.~Yor (1999).
\newblock {\em Continuous Martingales and Brownian Motion}.
\newblock Springer.

\bibitem[\protect\citeauthoryear{Samuelson}{Samuelson}{1965}]{Samuelson}
Samuelson, P.~A. (1965).
\newblock Proof that properly anticipated prices fluctuate randomly.
\newblock {\em Industrial Management Review\/}~{\em 6}, 41--49.

\bibitem[\protect\citeauthoryear{Sato}{Sato}{1999}]{Sato}
Sato, K. (1999).
\newblock {\em L\'{e}vy Processes and Infinitely Divisible Distributions}.
\newblock Cambidge University Press.

\bibitem[\protect\citeauthoryear{Schilling, Song, and Vondra\v{c}ek}{Schilling
  et~al.}{2010}]{SchillingSongVondracek}
Schilling, R., R.~Song, and Z.~Vondra\v{c}ek (2010).
\newblock {\em Berstein Functions: Theory and Applications}.
\newblock De Gruyter.

\bibitem[\protect\citeauthoryear{Schnurr}{Schnurr}{2009}]{Schnurr}
Schnurr, A. (2009).
\newblock {\em The Symbol of a Markov Semimartingale}.
\newblock Ph.\ D. thesis, Technische Universit\"{a}t Dresden.

\bibitem[\protect\citeauthoryear{Schoutens}{Schoutens}{2000}]{Schoutens}
Schoutens, W. (2000).
\newblock {\em Stochastic Processes and Orthogonal Polynomials}.
\newblock Springer.

\bibitem[\protect\citeauthoryear{Schwartz}{Schwartz}{1997}]{Schwartz97}
Schwartz, E.~S. (1997).
\newblock The stochastic behavior of commodity prices: implications for
  valuation and hedging.
\newblock {\em The Journal of Finance\/}~{\em 52\/}(3), 923--973.

\bibitem[\protect\citeauthoryear{Song and Vondra\v{c}ek}{Song and
  Vondra\v{c}ek}{2009}]{SongVondracek}
Song, R. and Z.~Vondra\v{c}ek (2009).
\newblock Potential theory of subordinate {B}rownian motion.
\newblock {\em Lecture Notes in Mathematics\/}~{\em 1980}.

\bibitem[\protect\citeauthoryear{Thangavelu}{Thangavelu}{2006}]{Thangavelu}
Thangavelu, S. (2006, March).
\newblock Hermite and {L}aguerre semigroups: Some recent developments.
\newblock Technical report, Department of Mathematics, Indian Institute of
  Science.

\bibitem[\protect\citeauthoryear{Vasicek}{Vasicek}{1977}]{Vasicek}
Vasicek, O. (1977).
\newblock An equilibrium characterisation of the term structure.
\newblock {\em Journal of Financial Economics\/}~{\em 5}, 177--188.

\bibitem[\protect\citeauthoryear{Wong}{Wong}{1964}]{Wong}
Wong, E. (1964).
\newblock The construction of a class of stationary {M}arkov processes.
\newblock In R.~Bellman (Ed.), {\em Sixteenth Symposium in Applied
  Mathematics---Stochastic Processes in Mathematical Physics and Engineering},
  pp.\  264--276. American Mathematical Society.

\bibitem[\protect\citeauthoryear{Yan}{Yan}{2002}]{Yan}
Yan, X. (2002).
\newblock Valuation of commodity derivatives in a new multi-factor model.
\newblock {\em Review of Derivatives Research\/}~{\em 5}, 251--271.

\end{thebibliography}
\bibliographystyle{chicago}

\end{document}